\documentclass[preprint,
superscriptaddress,
showpacs,
nofootinbib,
prd,
aps,
floatfix,
]{revtex4-1}

\usepackage{amsmath,amssymb,amsfonts,amsthm}
\usepackage[english]{babel}
\usepackage{url}
\usepackage{bm}
\usepackage[latin1]{inputenc}
\usepackage{graphicx,rotating}
\usepackage[colorlinks=true,linkcolor=black, citecolor=blue]{hyperref}
\usepackage{csvsimple}
\usepackage{booktabs}
\usepackage{multirow, array}
\usepackage{slashed}
\usepackage{xcolor}
\usepackage[export]{adjustbox}

\usepackage{natbib}

\usepackage[normalem]{ulem}

\begin{document}
\title{Impact of reactor neutrino uncertainties on coherent scattering's discovery potential}
\author{Leendert Hayen}
\email[Corresponding author: ]{hayen@lpccaen.in2p3.fr}
\affiliation{LPC Caen, ENSICAEN, Universit\'e de Caen, CNRS/IN2P3, Caen, France}
\affiliation{Department of Physics, North Carolina State University, Raleigh, 27607 North Carolina, USA}

\date{\today}

\begin{abstract}
Nuclear power reactors are the most intense man-made source of antineutrino's and have long been recognized as promising sources for coherent elastic neutrino-nucleus scattering (CE$\nu$NS) studies. Its observation and the spectral shape of the associated recoil spectrum is sensitive to a variety of exotic new physics scenarios and many experimental efforts are underway. Within the context of the reactor antineutrino anomaly, which initially indicated eV-scale sterile neutrino's, the modeling of the reactor antineutrino spectrum has seen a significant evolution in the last decade. Even so, uncertainties remain due to a variety of nuclear structure effects, incomplete information in nuclear databases and fission dynamics complexities. Here, we investigate the effects of these uncertainties on one's ability to accurately distinguish new physics signals. For the scenarios discussed here, we find that reactor spectral uncertainties are similar in magnitude to the projected sensitivities pointing towards a need for $\beta$ spectroscopy measurements below the inverse $\beta$ decay threshold.
\end{abstract}

\maketitle


\section{Introduction}

Despite its theoretical prediction in 1974 \cite{Freedman1974a}, experimental detection of coherent elastic neutrino nuclear scattering (CE$\nu$NS) was demonstrated only in 2017 by the COHERENT collaboration \cite{Akimov2017}. This is largely because of the technical difficulty of detecting the extremely weak recoiling nuclei, in particular since the cross section is many orders of magnitude larger than inverse $\beta$ decay. Following technical advances, attention is turned towards nuclear power reactors and intense sources of antineutrino's to perform CE$\nu$NS studies. Recently, the Dresden-II experiment reported on the first observation of reactor CE$\nu$NS \cite{Colaresi2021, Colaresi2022} and numerous experiments are underway with a variety of detector technologies \cite{Bonet2021, Billard2017, Alekseev2022, Choi2023, Aguilar-Arevalo2022}.

Naturally, predicted cross sections and recoil spectra depend strongly on the antineutrino flux emerging from a nuclear reactor. Predictions of the latter have a long history \cite{Hayes2016}. Their accurate determination is a problem of great complexity, however, as their origin inside the heart of a nuclear reactor consists of the decay of a vast multitude of different decays, each of which depends on the nuclear structure aspects of the initial and final states. The past decade, in particular, has seen a tremendous amount of activity both theoretical and experimental, a lot of which was initially stimulated by a rate mismatch between the measured and predicted total flux, known as the reactor antineutrino anomaly \cite{Mention2011, Huber2011, Mueller2011}. The latter was initially interpreted as a possible sign for a so-called \textit{sterile} neutrino (a particle which transforms as a singlet under all Standard Model symmetry groups, i.e. does not interact except through oscillations) with a mass on the order of $1$ eV$/c^2$ \cite{Abazajian2012, Acero2022}. Precise experimental measurements were performed by several collaborations, including with highly enriched fuels, and current antineutrino spectral measurements have become a precision benchmark for model predictions \cite{Giunti2019}. The latter is in stark contrast to the relatively low statistics measurements of total rates available a decade ago.


Within the context of the using reactor antineutrino's for studies such as coherent antineutrino scattering, the theoretical prediction of the fluxes becomes an indispensable ingredient \cite{Qian2019, Ang2023}.
Specifically, all precisely measured antineutrino spectra thus far are obtained using inverse beta decay (IBD), where an antineutrino is captured onto a proton inside a detector material, leading to a prompt positron and delayed neutron capture signal, or $\bar{\nu}_e + p \to n + e^+$. The latter is possible only when the antineutrino energy exceeds the proton-neutron mass difference and has sufficient amounts of energy to create a positron, which translates into $E_\nu \gtrsim 1.8$ MeV. 
As a consequence, one has to resort to explicit summation calculations which take into account each individual $\beta$ transition using a combination of extensive nuclear databases \cite{Estienne2019}. The 'conversion' approach popular in predicting IBD spectra using experimentally measured composite electron spectra \cite{Hayes2016, Huber2011} is not applicable, as those were performed only over the IBD threshold \cite{Schreckenbach1981, Schreckenbach1985, Haag2014, Kopeikin2021}.

In this article, we therefore aim to discuss the impact and uncertainty of reactor antineutrino flux predictions with an emphasis on its behaviour below the IBD threshold at 1.8 MeV. The paper is constructed as follows: In Sec. \ref{sec:coherent_scattering} we provide a brief summary of CE$\nu$NS cross sections in the Standard Model and Beyond. Section \ref{sec:CEvNS_reactor_neutrinos} reviews the reactor antineutrino production mechanism and the influence of $\beta$ decays on CE$\nu$NS recoil spectra. We report on the impact of CE$\nu$NS discovery potential in Sec. \ref{sec:discovery_impact} and present our conclusions in Sec. \ref{sec:conclusion}.

\section{Coherent scattering in and beyond the Standard Model}
\label{sec:coherent_scattering}
As far as neutrino's go, the cross section for coherent scattering has a number of notable benefits compared to its typical alternative in inverse beta decay. For starters, the inherent cross section is many times larger than that of IBD, and as it scales coherently with the number of nucleons inside the target nucleus, the cross section can be further tuned by changing the target material. Unlike IBD, there is no threshold to be overcome, other than an experimental one for the detection of the recoiling nucleus, meaning that the entire spectrum emerging from a nuclear reactor may be used. This is of particular importance for the latter as the reactor neutrino intensity drops off quickly with increasing energy. It also means that it may be sensitive to Beyond Standard Model physics that largely changes the cross section at low neutrino energies such as the possibility of a neutrino magnetic moment as discussed below. We will commence with the Standard Model cross section and discuss a number of relevant BSM extensions for nuclear CE$\nu$NS.

\subsection{Standard Model cross section}

The tree-level Standard Model differential cross section for CE$\nu$NS on spin-1/2 particles may be written as \cite{Freedman1974a, Lindner2017}
\begin{equation}
    \frac{d\sigma}{dE_\mathrm{rec}} = \frac{G_F^2M}{\pi}Q_W^2\left[1-\frac{ME_\mathrm{rec}}{2E_\nu^2}-\frac{E\mathrm{rec}}{E_\nu}+\frac{1}{2}\left(\frac{E_\mathrm{rec}}{E_\nu}\right)^2\right]F(\bm{q}^2)
    \label{eq:SM_CEvNS_cs}
\end{equation}
where $G_F \approx 1.16$ GeV$^{-2}$, $M$ is the mass of the target nucleus for an incoming antineutrino of energy $E_\nu$, and $F(\bm{q}^2)$ is a nuclear form factor. The latter may be taken to be unity for the relevant energy scales of reactor CE$\nu$NS. Here, $Q_W$ is the Standard Model weak charge
\begin{equation}
    Q_W = g_V^p Z F_Z(\bm{q}^2) + g_V^n N F_N(\bm{q}^2)
\end{equation}
for a target nucleus with $Z$ protons and $N$ neutrons and $\bm{q}^2=2ME_\mathrm{rec}$ is the three-momentum transfer squared. Here, $g_V^p= 1/2-2\sin^2\theta_W$ and $g_V^n = -1/2$, for which we use $\sin^2\theta_W=0.23857$ evaluated at $q=0$ from a renormalization group evolution from the $\overline{MS}$ determination at the $Z$ boson mass \cite{Workman2022, Erler2019}. The form factors $F_{Z, N}(\bm{q}^2)$ are assumed to be equal and, like the nuclear form factor, show no dependence on the momentum transfer for the low-energy antineutrino's emerging from nuclear reactors. 

The kinematics are constrained such that a recoil produced with some kinetic energy $T$ needs an incoming antineutrino of at least $E_{\nu}^\mathrm{min} = (T+\sqrt{T^2-2mT})/2$ where $m$ is the mass of the recoiling particle. Vice versa, the maximum recoil energy for a given antineutrino energy is $T_\mathrm{max} = 2E_\nu^2/(2E_\nu+m)$.

\subsection{Beyond Standard Model phenomenology}
\label{sec:BSM_pheno}
There is a rich phenomenology accessible through CE$\nu$NS, ranging from parametric extensions within the same Lorentz structure as the Standard Model \cite{Valle1987} to neutrino magnetic moments \cite{Vogel1989} and light mediators \cite{Farzan2018, Cerdeno2016}. The combination of elastic scattering off nuclei as well as electrons often allows for more stringent constraints due to different kinematic sensitivities as well as material dependencies. Here, we will treat non-standard neutrino interactions (NSIs), neutrino magnetic moment as well as generalized neutrino interactions. These have been discussed already for the analysis of the Dresden-II data \cite{Coloma2022, Majumdar2022}. We provide only a brief summary of their differential cross sections and treat the effect of spectral uncertainties on their discovery potential in Sec. \ref{sec:discovery_impact}.

\subsubsection{Non-Standard neutrino interactions}
\label{sec:NSI}
Many neutrino mass generation schemes introduce changes similar to the Lorentz and chiral structure of the Standard Model interaction \cite{Valle1987, Boucenna2014}, leading to what are known as non-standard interactions encoded through
\begin{equation}
    \mathcal{L}^\mathrm{NSI} = -2\sqrt{2} G_F \varepsilon_{\alpha\beta}^{qP} (\bar{\nu}_\alpha \gamma_\mu P_L \nu_\beta)(\bar{q}\gamma^\mu P_qq)
\end{equation}
where $\alpha, \beta = e, \mu, \tau$ is the neutrino flavour and $q=u, d$ the quark content, $P = P_{L,R}$ is a chiral projection operator with $P_{L, R} = (1\mp\gamma^5)/2$. Exotic interactions are normalized relative to $G_F$ despite allowing for flavour off-diagonal interactions ($\alpha\neq \beta$). As the Lorentz structure of the interaction is unchanged with respect to the Standard Model, non-zero $\varepsilon$ simply shift the value of $Q_W$. Restricting to vector rather than axial-vector interactions\footnote{Axial-vector CE$\nu$NS interactions are suppressed by a factor $1/A$ with $A=Z+N$ \cite{Barranco2005}.}, one finds \cite{Barranco2005}
\begin{align}
    Q_W^\mathrm{NSI}(\varepsilon) &= (g_V^p+2\varepsilon_{\alpha\alpha}^{uV}+\varepsilon_{\alpha\alpha}^{dV})ZF_Z(\bm{q}^2) + (g_V^n+\varepsilon_{\alpha\alpha}^{uV}+2\varepsilon_{\alpha\alpha}^{dV})ZF_Z(\bm{q}^2)NF_N(\bm{q}^2) \nonumber \\
    &+\sum_{\beta\neq\alpha} (2\varepsilon_{\alpha\beta}^{uV}+\varepsilon_{\alpha\beta}^{dV})ZF_Z(\bm{q}^2) + (\varepsilon_{\alpha\beta}^{uV}+2\varepsilon_{\alpha\beta}^{dV})NF_N(\bm{q}^2)
\end{align}
where $\varepsilon^{qV} = \varepsilon^{qR}+\varepsilon^{qL}$ a vector-like coupling.

\subsubsection{Generalized Neutrino Interactions}
\label{sec:GNI}
A straightforward extension of NSIs is obtained by freeing up the Lorentz structure of the interaction, similarly to the Lee-Yang Hamiltonian of the 1950's in (nuclear) $\beta$ decay \cite{Lee1956}, for which we may write \cite{Lindner2017, Aristizabal-Sierra2018}
\begin{equation}
    \mathcal{L}^{GNI} = \frac{G_F}{\sqrt{2}}\sum_{V, A, S, T, P} [\bar{\nu}\Gamma_X\nu][\bar{q}\Gamma_X(C_X^q+i\gamma^5D_X^q)q]
\end{equation}
with $\Gamma_X = \{1,\gamma^5,\gamma^\mu, \gamma^\mu\gamma^5, \sigma^{\mu\nu}\}$ the operators corresponding to $\{S, P, V, A, T\}$ and $C_X, D_X$ free real parameters. On dimensional grounds the latter may be connected to a BSM scale
\begin{align*}
    C_X, D_X &\propto \left(\frac{M_W}{\Lambda_\mathrm{BSM}}\right)^{n \geq 2} \quad \mathrm{so~ that}  \\
    &\leq 0.01\%\quad \mathrm{means} \quad \Lambda_\mathrm{BSM} \geq 10~\mathrm{TeV}
\end{align*}
Moving to a nucleon system involves the translation of quark to nucleon degrees of freedom, for which we may write $\langle N_F | \bar{q}\Gamma_X q | N_i \rangle = F^X \langle N_f | \bar{N} \Gamma_X N | N_i \rangle$ where $F^X$ is an isoscalar nucleon form factor that may, e.g., be obtained from lattice QCD. Similarly neglecting axial-vector and pseudoscalar interactions as before \cite{Barranco2005} we may connect the nucleon and quark-scale couplings by writing
\begin{align}
    C_S &= ZF_Z(\bm{q}^2)\sum_q C_S^q \frac{m_p}{m_q}f_q^p + NF_N(\bm{q}^2)\sum_qC_S^q\frac{m_n}{m_q}f_q^n \\
    C_T &= ZF_Z(\bm{q}^2)\sum_qC_T^q\delta_q^p + NF_N(\bm{q}^2)\sum_qC_T^q\delta_q^n \\
    C_V &= ZF_Z(\bm{q}^2)(2C_V^u+C_V^d)+NF_N(\bm{q^2})(C_V^u+2C_V^d) 
\end{align}
for hadronic form factors $f_u^p=0.0208$, $f_u^n=0.0189, f_d^p=0.0411$ and $f_d^u=0.0451$, and $\delta_u^p=\delta_d^n=0.54$ and $\delta_d^p=\delta_u^n=-0.23$ \cite{Aristizabal-Sierra2019}.

The GNI Lagrangian results in a richer differential cross section to find \cite{Aristizabal-Sierra2018}
\begin{align}
    \frac{d\sigma}{dE_\mathrm{rec}} = \frac{G_F^2M}{\pi}\Bigg\{&C_S^2\frac{ME_\mathrm{rec}}{8E_\nu^2}+ \left(\frac{C_V}{2}+Q_W\right)^2\left[1-\frac{ME_\mathrm{rec}}{2E_\nu^2}-\frac{E_\mathrm{rec}}{E_\nu}+\left(\frac{E_\mathrm{rec}}{E_\nu}\right)^2\right] \nonumber \\
    &+2C_T^2\left(1-\frac{ME_\mathrm{rec}}{4E_\nu^2}-\frac{E_\mathrm{rec}}{E_\nu}\right)\pm\mathcal{R}\frac{E_\mathrm{rec}}{E_\nu}\Bigg\}
    \label{eq:GNI_diff_cs}
\end{align}
where $\mathcal{R}=C_SC_T/2$ and the minus (plus) sign is for coherent elastic (anti)neutrino scattering. Note that Eq. (\ref{eq:GNI_diff_cs}) assumes flavour-diagonal couplings, unlike those discussed in the previous section.

\subsubsection{Neutrino Magnetic Moment}
\label{sec:neutrino_magnetic_moment}
As a last example, we consider the possible existence of a neutrino magnetic moment \cite{Wong2005, Dvornikov2004}, which has a significantly different kinematic sensitivity. For a neutrino magnetic moment, $\mu_\nu$, one may consider elastic scattering off both electrons and nuclei. In neither case does the amplitude interfere with the Standard Model process so that one is sensitive only to $\mu_\nu^2$ for a differential cross section of \cite{Vogel1989}
\begin{equation}
    \frac{d\sigma_{\mu_\nu}^{CE\nu NS}}{dE_\mathrm{rec}} = Z^2 \left(\frac{\mu_\nu}{\mu_B}\right)^2 \frac{\alpha^2\pi}{m_e^2} \left[\frac{1}{E_\mathrm{rec}}-\frac{1}{E_\nu}\right]|F(\bm{q}^2)|^2
\end{equation}
with $\mu_B$ the Bohr magneton. 


\section{Reactor antineutrino spectra}
\label{sec:reactor_neutrino_production}
\subsection{Overview}
A typical 1~GW nuclear power reactor will produce on the order of $10^{20}$ antineutrino's per second, almost all of which arise from the fission products of four main actinides: $^{235}$U, $^{238}$U, $^{239}$Pu and $^{241}$Pu. Their relative contributions are determined by the initial enrichment of the fuel (i.e. the $^{235}$U/$^{238}$U ratio) and the elapsed time since the start of the burn. In all cases we may write the total antineutrino flux as \cite{Hayes2016}
\begin{equation}
    \frac{d\Phi}{dE_\nu} = \frac{W_\mathrm{th}}{\sum_if_ie_i} \sum_i f_i\left(\frac{dN_i}{dE_\nu}\right)
\end{equation}
where $f_i$ is the relative fission fraction of actinide $i$, $W_\mathrm{th}$ is the total reactor thermal energy, $e_i$ is the effective thermal energy per fission contributed per fission ($202 <e_i <215$ MeV), and $dN_i/dE_\nu$ is the neutrino spectrum originating from each fission actinide's decay chain. Uncertainties on $W_\mathrm{th}$ are estimated to be less than 2\%, while those on $e_i$ are of order 0.25-0.5\% \cite{Hayes2016}. While uncertainties on relative fission fractions can reach several percents, because of the strong overlap in reached final states in the decay chains these do not typically induce large uncertainties in the total antineutrino spectrum. Spectral variations due to uncertainties in individual actinide antineutrino spectra, however, can be significant. These can be broken down into contributions from fission yields and $\beta$ spectrum shapes
\begin{equation}
\frac{dN_i}{dE_\nu} = \sum_n Y_n(Z, A, t) \sum_i b_{n, i}(E_0^i)P_{\nu}(E_\nu, E_0, Z)
\label{eq:actinide_antineutrino_flux}
\end{equation}
where $n$ sums over all reached fission fragments and $i$ sums over all $\beta$ transitions for fission fragment $n$. Here $Y_n$ is the number of $\beta$ decays of the fragment $Z, A$ at time $t$, which converges to the cumulative fission yield. Relative branching ratios of individual $\beta$ transitions with electron endpoint energy $E_0^i$ are written as $b_{n, i}(E_0^i)$ and are normalized such that $\sum_i b_{n, i} = 1$. The $\beta$ spectrum shape for each transition is denoted $P_{\nu}$ and is normalized such that its total integral equals unity. Uncertainties on all elements of Eq. (\ref{eq:actinide_antineutrino_flux}) contribute substantially to those on the total antineutrino flux, and many recent Total Absorption Gamma Spectroscopy (TAGS) measurements have significantly shifted previously determined values of $b_{n, i}$ \cite{Fijakowska2017, Guadilla2019, Zakari-Issoufou2015}. Uncertainties in $\beta$ spectrum shapes originate from nuclear structure complexities, and the vast number of individual transitions (around 8000) prevents one from performing many-body calculations for but a few nuclei. While the normalization of individual spectra guarantees spectrum shape differences do not change the \textit{total} antineutrino flux, it significantly modifies the fraction of which exceeds 1.8 MeV for an IBD reaction to occur, as well as average energies used for, e.g., decay heat studies \cite{Fallot2019}. Similarly, due to the dependence on $E_\nu$ in CE$\nu$NS cross sections (see Eq. (\ref{eq:SM_CEvNS_cs})), total rates and spectrum shapes will depend on an accurate knowledge on individual transition shapes.

\subsection{$\beta$ spectrum shapes and uncertainties}
Individual $\beta$ spectrum shapes may be written as follows \cite{Hayen2018}
\begin{align}
P_\nu(E_\nu, E_0, Z) &\propto \sqrt{(E_0-E_\nu)^2-m_e^2}E_\nu^2(E_0-E_\nu) F(Z, E_0-E_\nu) C(Z, E_\nu) K(Z, E_\nu)\nonumber \\
&\times(1+\frac{\alpha}{2\pi}h(E_\nu, E_0))
\end{align}
where $F(Z, E_e)$ is the traditional Fermi function, $C$ is often referred to as the shape factor and contains nuclear structure information, $K$ is a collection of smaller correction factors and $[1+\alpha/(2\pi)h]$ is a radiative correction due to real and virtual photon emission. The latter is written as follows \cite{Sirlin2011}
\begin{align}
    h(E_\nu, E_0) &= 3\ln\left(\frac{m_p}{m_e}\right) + \frac{23}{4} - \frac{8}{\hat{\beta}}\mathrm{Li}_2\left(\frac{2\hat{\beta}}{1+\hat{\beta}}\right)+8\left(\frac{\tanh^{-1}\hat{\beta}}{\hat{\beta}}-1\right)\ln\left(\frac{2(E_0-E_\nu)\hat{\beta}}{m_e}\right) \nonumber \\
    &+4\frac{\tanh^{-1}\hat{\beta}}{\hat{\beta}}\left[\frac{7+3\hat{\beta}^2}{8}-2\tanh^{-1}\hat{\beta}\right]
\end{align}
where $\hat{\beta} = \sqrt{(E_0-E_\nu)^2-m_e^2}/(E_0-E_\nu)$ and is the only correction in the antineutrino spectrum shape that cannot be obtained by taking the electron spectrum shape and inverting it.

\begin{figure}[ht]
    \centering
    \includegraphics[width=0.8\textwidth]{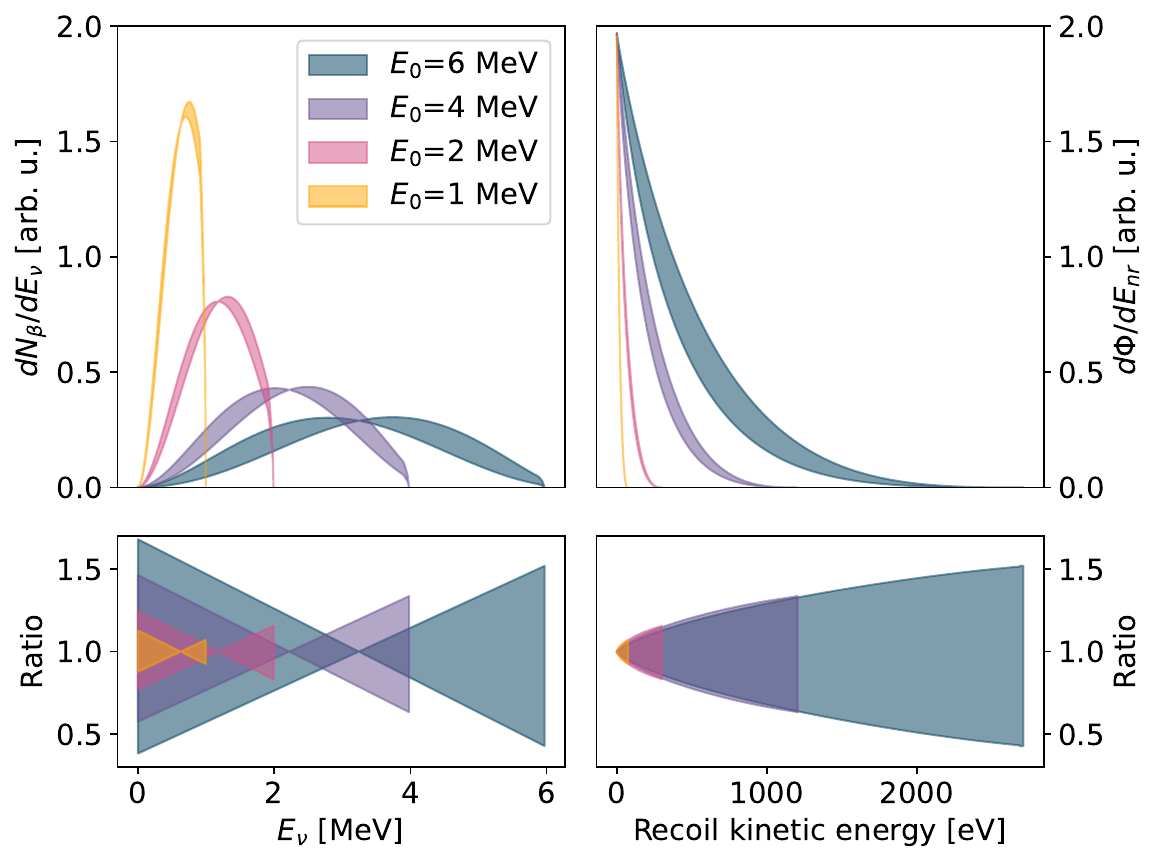}
    \caption{Absolute and relative spectral changes for both antineutrino and recoil energy spectra due to a change in slope of the shape factor of the individual $\beta$ transition. (Left) Normalized antineutrino spectra for different endpoints following changes to the nuclear shape factor within a range of slopes, shown in the bottom left. (Right) Recoil energy spectra following CE$\nu$NS on Silicon atoms taking the individual $\beta$ spectra on the left as input, with relative uncertainties shown on the bottom right.}
    \label{fig:single_spect_changes}
\end{figure}

The largest source of uncertainty in the spectrum shape calculation comes from the shape factor. The latter depends heavily on the spin-parity change of the transition and can change slope significantly depending on nuclear structure details. Specific calculations have been performed for a limited number of nuclei in the past \cite{Hayen2019, Hayen2019b}, and we provide a short summary of the theoretical framework and anticipated changes in the Appendix. For the purposes of this study it is sufficient to consider the effect of generic changes to the shape factor and its corresponding effect on the anticipated recoil energy spectrum following CE$\nu$NS. This is shown in Fig. \ref{fig:single_spect_changes} for Silicon as the target nucleus.

Because of the increasing cross section with increasing $E_\nu$, relative differences are largest near the endpoint in recoil kinetic energy. While the linear distortions to the $\beta$ spectrum are transformed into non-linear changes to the recoil CE$\nu$NS spectrum, relative changes near the endpoint of both spectra are identical as expected. While the anticipated shape factor changes are often non-linear (see Refs. \cite{Hayen2019, Hayen2019b}), the envelope taken in Fig. \ref{fig:single_spect_changes} is similar to what is found through explicit calculations. 

\section{CE$\nu$NS with reactor neutrinos}
\label{sec:CEvNS_reactor_neutrinos}

Compared to the initial demonstration of CE$\nu$NS where one used neutrino's originating from decay-at-rest pions \cite{Akimov2017}, the total antineutrino energy emerging from nuclear reactors is much lower. In order to detect the corresponding nuclear recoil energies one therefore needs a different type of detector technology. The Dresden-II experiment observed a statistically significant number of reactor CE$\nu$NS events using point-type contact Germanium detectors \cite{Colaresi2021}, while many other experiments are investigating both semiconductor and superconducting technologies. The former are more easily scaled up to increase the target mass whereas the latter typically allow for much smaller energy thresholds, thereby keeping complementarity between both techniques. In general, we may write the experimental CE$\nu$NS differential rate as follows
\begin{equation}
    \frac{dR}{dE_\mathrm{m}}\Bigg|_\mathrm{exp} = N_\mathrm{target} \int_{0}^{\infty} dE_\mathrm{rec}~ \mathcal{E}(E_m) \mathcal{R}(E_m, E_\mathrm{rec}) \int^{E^\mathrm{max}_\nu}_{E^\mathrm{min}_\nu} dE_\nu \frac{d\Phi}{dE_\nu} \frac{d\sigma}{dE_\mathrm{rec}}
\end{equation}
where $N_\mathrm{target}$ is the total number of target nuclei, $\mathcal{R}(E_m, E_\mathrm{rec})$ is the detector response function resolving in a measured energy $E_m$ due to a deposited energy $E_\mathrm{rec}$, and $\mathcal{E}(E_m)$ is a detection efficiency function. Many effects, including Fano noise, quenching, electronic noise, etc., will determine the final spectrum shape \cite{Bonhomme2022, Khan2019}. For the purposes of this work, however, we will consider an idealized detection setup as a best case scenario, meaning we take $\mathcal{R}(E_m, E_{\mathrm{rec}}) = \delta(E_m-E_\mathrm{rec})$ and a step function response for the detection efficiency above $E_\mathrm{thr}$, $\mathcal{E}(E_m) = \theta(E_m-E_\mathrm{thr})$.

\begin{figure}[ht]
    \centering
    \includegraphics[width=0.8\textwidth]{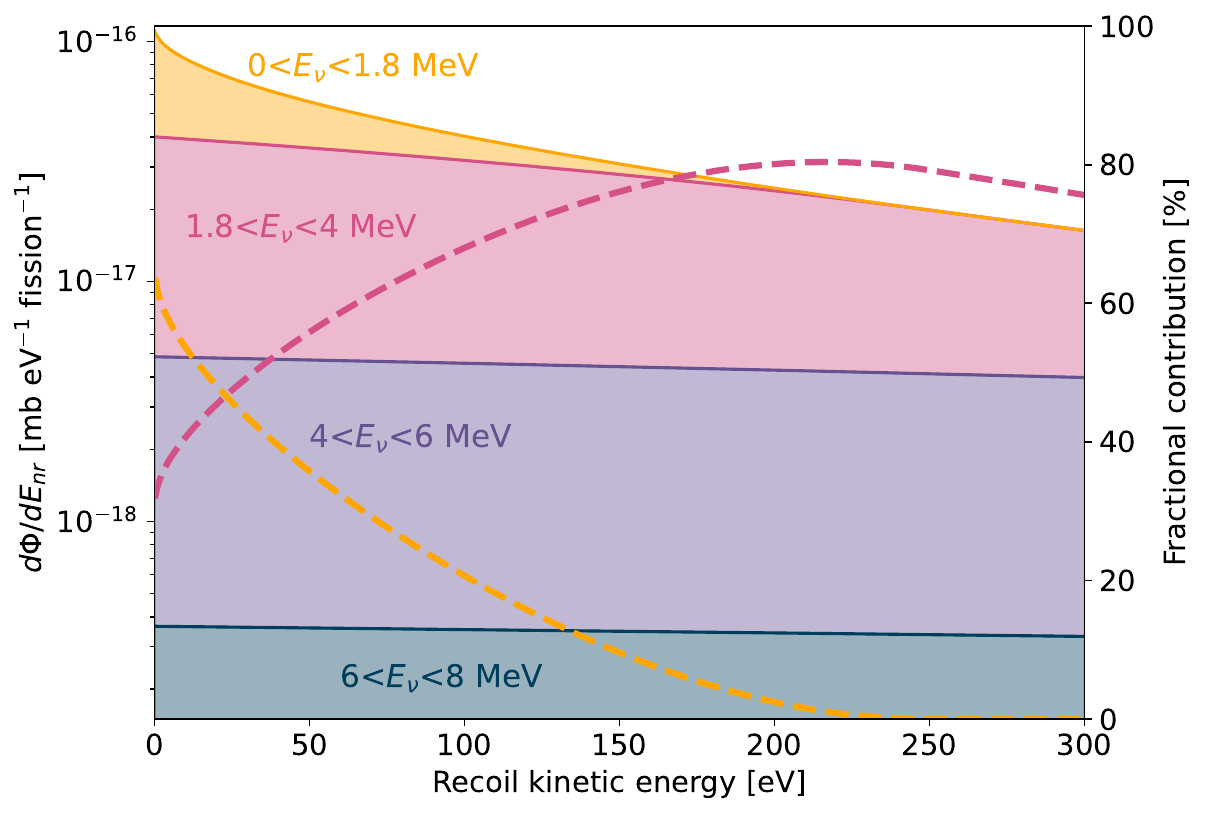}
    \caption{Stackplot for the differential CE$\nu$NS cross section for Silicon atoms using a typical reactor antineutrino spectrum segmented as a function of the antineutrino energy. The dashed lines show the relative contribution to the total differential cross section for segments where $0<E_\nu < 1.8$ MeV and $1.8 < E_\nu <4$ MeV.}
    \label{fig:spectrum_contrib}
\end{figure}

Using such an idealized detector, it is instructive to see which parts of the antineutrino spectrum contribute at different recoil energies. In particular, as experimental measurements of antineutrino spectra have currently only been performed using IBD, no experimental spectral information is available for $E_\nu < 1.8$ MeV. Instead, only \textit{ab initio}, or database driven spectral calculations can provide information for the lowest recoil energies. Much of this spectral content comes from nuclear $\beta$ transitions with endpoints between $2$ and $4$ MeV and often comprise of transitions to nuclear excited states for which nuclear structure information is less available and generally harder to confidently calculate. Figure \ref{fig:spectrum_contrib} shows the recoil kinetic energy distribution segmented as a function of the antineutrino energy. 

For the lowest kinetic energies, a little over half of the recoil spectrum is determined by reactor antineutrino's with energies below the IBD threshold and for which no experimental data is present. The crossing point occurs at relatively low energies compared to the threshold energy of many current devices, at around 30 eV. Most of the experimental recoil spectrum accessible to, e.g., the Dresden-II experiment ($E_\mathrm{thr} \sim 200$ eV) comes from the reactor antineutrino spectrum for $1.8 < E_\nu < 4$ MeV, while superconducting technologies can typically reach threshold energies of a few (tens) of eV.

\begin{figure}[ht]
    \centering
    \includegraphics[width=0.8\textwidth]{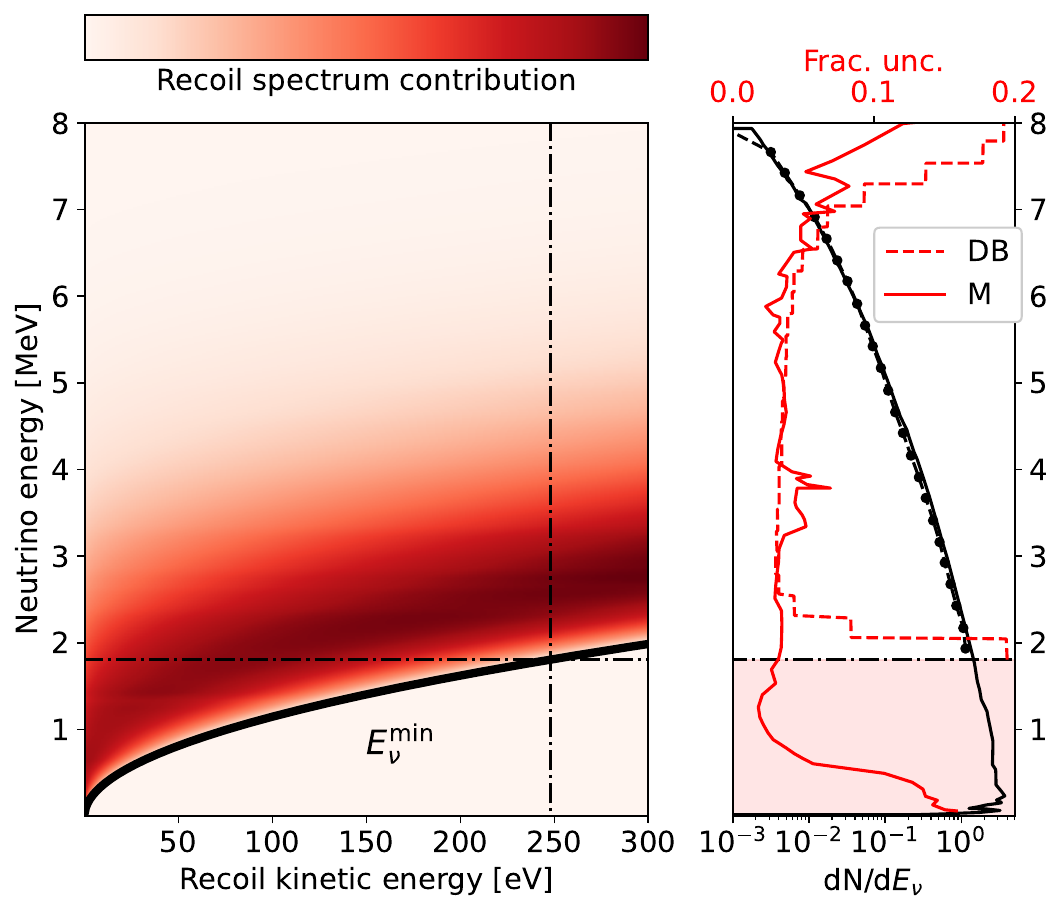}
    \caption{Spectral contribution to the detected differential cross section for Silicon using an experimental and Daya Bay-like modeled antineutrino spectrum \cite{DayaBaycollaboration2021}, together with their respective errors. The blank line on the left plot shows the minimum antineutrino energy required to generate a recoiling particle with a certain kinetic energy. The maximum of the spectral contribution at each recoil kinetic energy occurs for energies slightly above $E_\nu^\mathrm{min}$ due to the compensating effect of a strongly decreasing antineutrino flux but increasing CE$\nu$NS differential cross section with increasing $E_\nu$. The dotted lines on the left plot determine the point for which the recoil spectrum obtains no more contributions from the antineutrino spectrum below the IBD threshold.}
    \label{fig:spectral_contrib_errors}
\end{figure}

Figure \ref{fig:spectral_contrib_errors} shows a more fine-grained image of the spectral contribution to the CE$\nu$NS differential cross section depending on the antineutrino energy using the deconvoluted Daya Bay antineutrino spectrum \cite{DayaBaycollaboration2021} and an \textit{ab initio} approach with their respective uncertainties \cite{Letourneau2023}. The latter increases significantly for the experimental Daya Bay spectrum as one approaches the IBD threshold, whereas the theoretical estimate remains at an uncertainty of a few percent down to about 800 keV. The theoretical uncertainties in this case were determined through variation of a number of parameters, including poorly known shape factors, branching ratio's, etc, but even so obtains disagreements with experimental data larger than either uncertainty estimate. This is therefore to be taken as a lower bound rather than an absolute determination. 

Finally, due to the low amount of anticipated CE$\nu$NS events in current generation detector technologies, it is interesting to see how $\beta$ spectral changes influence the total number of detected CE$\nu$NS events depending on an experimental threshold. For this we consider two possibilities for the shape factor: That of an allowed decay, in which to first order we treat $C\approx 1$, and that of a first-unique forbidden transition, in which case we may write
\begin{equation}
    C(Z, E_e) \propto (E_0-E_e)^2+\lambda_2(Z, E_e)(E_e^2-m_e^2) + \Delta_{NS}
    \label{eq:C_UFF}
\end{equation}
where $\lambda_2(Z, E_e)$ is an $\mathcal{O}(1)$ Coulomb function \cite{Behrens1969, Hayen2019b}, $\Delta_{NS}$ are percent-level nuclear structure corrections \cite{Behrens1982} and we have have omitted the prefactor of the dominant matrix element due to the spectral normalization

\begin{figure}[ht]
    \centering
    \includegraphics[width=0.8\textwidth]{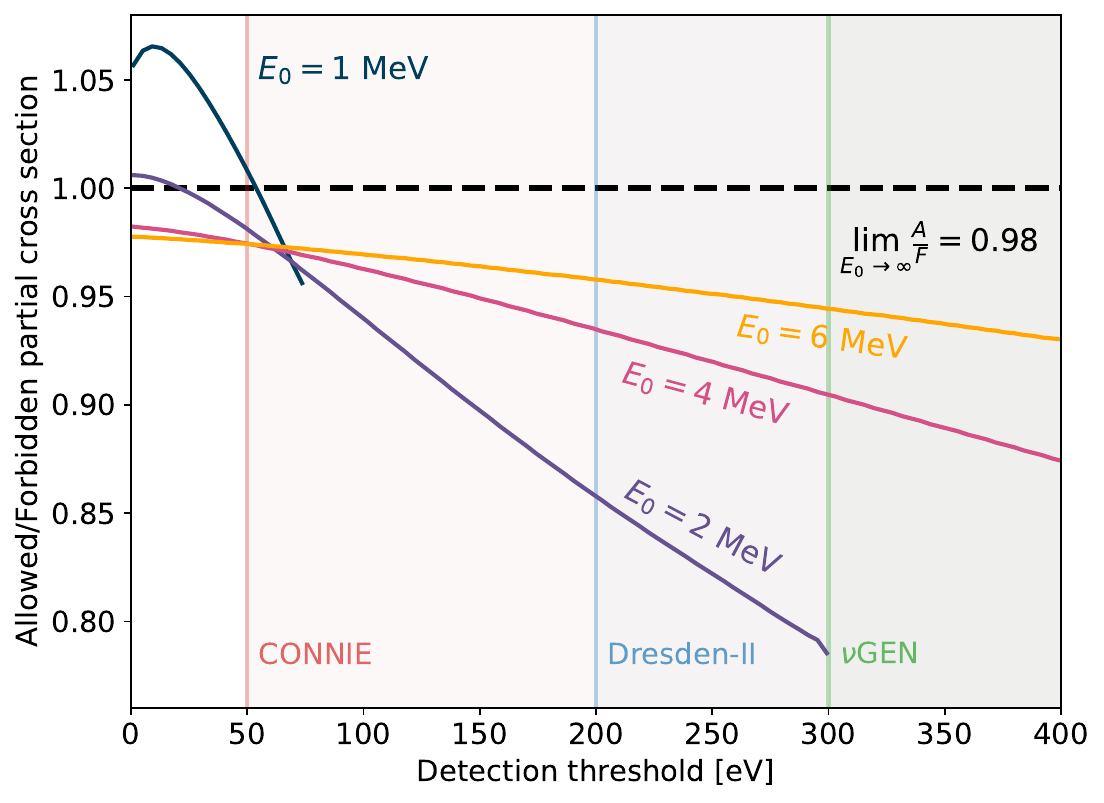}
    \caption{Ratio of integrated counts as a function of recoil threshold energy for Silicon target atoms where a single $\beta$ transition has either an allowed ($C\approx 1$) or unique first-forbidden ($C$ from Eq. \ref{eq:C_UFF}) shape factor for different $\beta$ decay endpoints. Also shown are proposed experimental thresholds from the CONNIE \cite{Nasteva2021, Aguilar-Arevalo2022}, Dresden-II \cite{Colaresi2022} and $\nu$GEN experiments \cite{Alekseev2022}. See text for discussion.}
    \label{fig:ratio_A_F_thresholds}
\end{figure}

Figure \ref{fig:ratio_A_F_thresholds} shows the ratio of \textit{integrated} counts depending on the transition endpoint and the lower detection threshold. A single transition with a 2 MeV $\beta$ endpoint can, for example, result in a 15\% lower detected total number of counts depending on whether the transition was allowed unique first-forbidden. It is interesting to note that, despite the $\beta$ spectra being separately normalized to unity, the relative number of detected counts integrated over the full spectrum does not converge to unity as the endpoint tends towards infinity (and therefore the relative fraction of events that exceed the detection threshold reaches 100\%). This is another consequence of the energy-dependence of the CE$\nu$NS cross section. As such, experiments using total integrated number of counts in order to, for example, search for exotic physics through variations in $Q_W$, should be extremely careful to properly understand the initial spectral content of the antineutrino source.

\section{Discovery impact}
\label{sec:discovery_impact}
The use of an intense man-made antineutrino source for CE$\nu$NS detection has generated a significant amount of research towards potential discoveries of exotic, Beyond the Standard Model physics. In Sec. \ref{sec:coherent_scattering} we described the differential cross sections of a number of different physics extensions. The question is then how the reactor antineutrino spectral uncertainties discussed in the previous section compare to the sensitivity projections, if they turn out to be comparable, where effort should be placed in the coming years.

\subsection{Neutrino magnetic moment and nonstandard interactions}

As a simple scenario, we investigate how the current spectral uncertainties due to the reactor antineutrino predictions translate into bounds on a potential neutrino magnetic moment and a change in $Q_W$. The procedure is as follows: In order to estimate the antineutrino spectral uncertainty we, bin-by-bin, take the maximum of the estimated model uncertainty and the actual discrepancy between model and experimental data. For the latter we take the Daya Bay data set, as in Fig. \ref{fig:spectral_contrib_errors}, since these have been measured to the highest precision. Then, the total relative uncertainty is transformed into an absolute uncertainty for which we calculate the resulting the recoil energy spectrum. The latter is then taken to be Gaussian noise around the original recoil spectrum without bin-to-bin correlation. The resulting spectrum is then fit for a change to $Q_W$, denoted $\delta Q_W$, and a neutrino magnetic moment normalized by the Bohr magneton, $\mu_\nu/\mu_B$. This procedure is done $N$ times to sufficiently sample the added noise.

\begin{figure}[ht]
    \centering
    \includegraphics[width=0.8\textwidth]{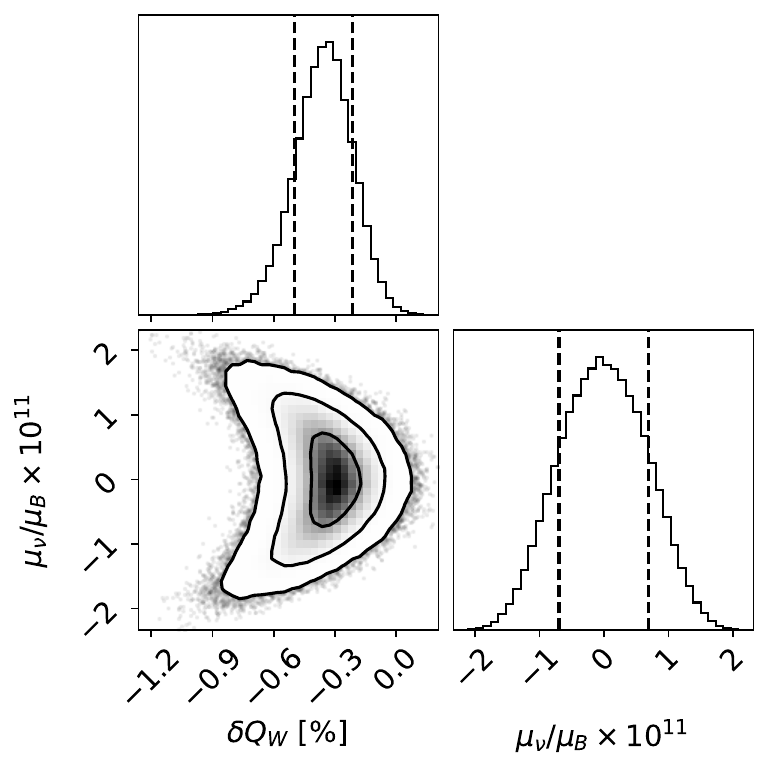}
    \caption{Two-dimensional representation of the uncertainties induced in a change to $Q_W$, denoted $\delta Q_W$, and a neutrino magnetic moment, denoted $\mu_\nu$, due to reactor antineutrino spectral uncertainties in reactor CE$\nu$NS events. Curves in the lower left plot correspond to the equivalent 1,2 and 3 standard deviation quantiles for a two-dimensional Gaussian.}
    \label{fig:fit_unc_results}
\end{figure}

A two-dimensional representation of the resulting uncertainty distributions and their correlation is shown in Fig. \ref{fig:fit_unc_results}. A combination of the different kinematic signature ($[E_\mathrm{rec}^{-1}-E_\nu^{-1}]$ for a neutrino magnetic moment, compared to $ME_\mathrm{rec}/E_\nu^2$ for SM CE$\nu$NS), and the quadratic sensitivity to $\mu_\nu$ results in a non-Gaussian correlation between the two parameters. Current reactor antineutrino spectral uncertainties induce systematic uncertainties in potential NSI contributions at the few percent level, while for neutrino magnetic moments one finds uncertainties on the order of a few parts in $10^{-11}$ for $\mu_\nu/\mu_B$.

\begin{figure}[ht]
    \centering
    \includegraphics[width=0.8\textwidth]{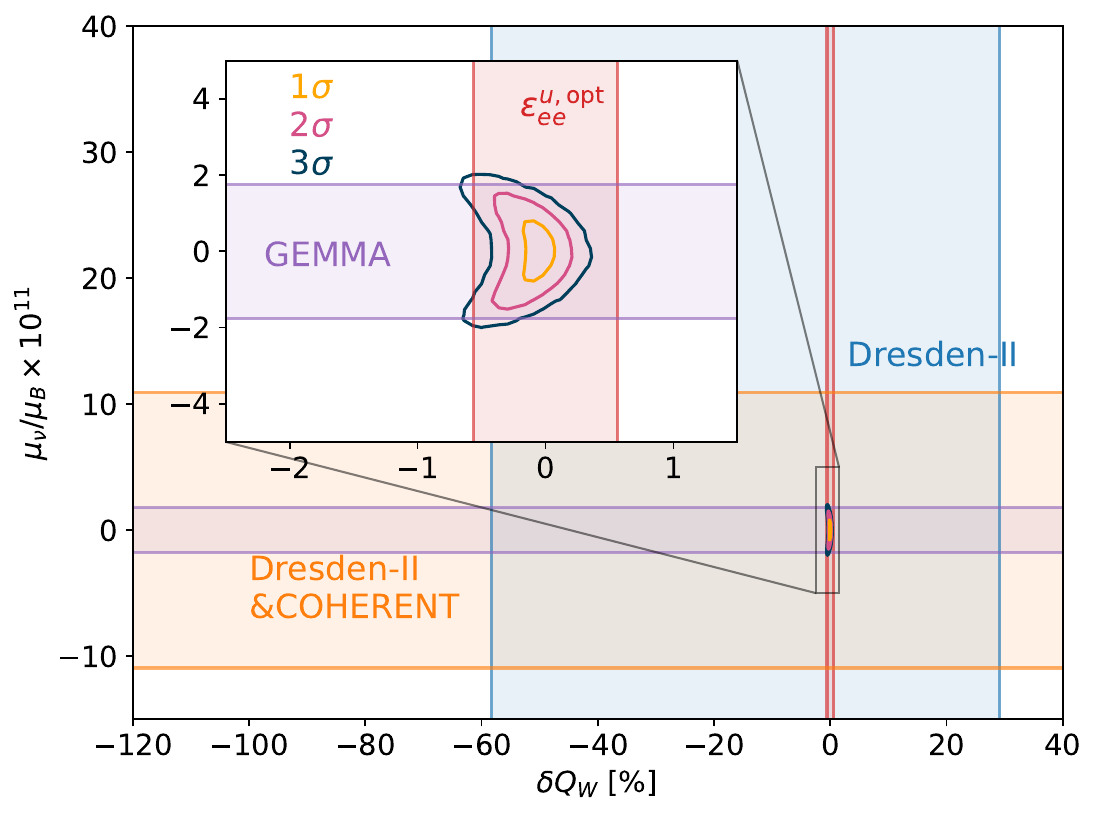}
    \caption{Comparison of uncertainties induced on a neutrino magnetic moment and shift from the Standard Model value for $Q_W$ compared to current and projected sensitivities. The Dresden-II and Coherent analysis results are taken from Ref. \cite{Coloma2022}, the GEMMA experiment from Ref. \cite{Beda2013}, while the projected sensitivity for $\varepsilon_{ee}$ is from Ref. \cite{Lindner2017}.}
    \label{fig:limits_dQw_mu_bounds}
\end{figure}

We may compare how these uncertainties compare to current global constraints \cite{Agostini2017a, Lattimer1988} on these parameters as well as projected sensitivities for reactor CE$\nu$NS. These results are shown in Fig. \ref{fig:limits_dQw_mu_bounds}. We compare against recent analyses of the reactor CE$\nu$NS data by the Dresden-II collaboration and those from pion decay-at-rest from Coherent \cite{Coloma2022}, as well as an analysis from the GEMMA experiment \cite{Beda2013} and projected sensitivities by Lindner \textit{et al.} \cite{Lindner2017}. While the current limits from CE$\nu$NS experiments are very wide, the current uncertainty in $\mu_\nu$ from reactor spectral uncertainties is on the same order of magnitude as those found by GEMMA. For NSI changes to $Q_W$, the reactor spectral uncertainties are smaller than those for the projected sensitivity and therefore do not pose a problem. Current spectral uncertainties are therefore only anticipated to be a problem when CE$\nu$NS tests of non-neutrino $\mu_\nu$ approach a relative precision of $\mu_\nu/\mu_B \sim 2 \cdot 10^{-11}$. 

\subsection{Generalized Neutrino Interactions}

\begin{figure}[ht]
    \centering
    \includegraphics[width=0.8\textwidth]{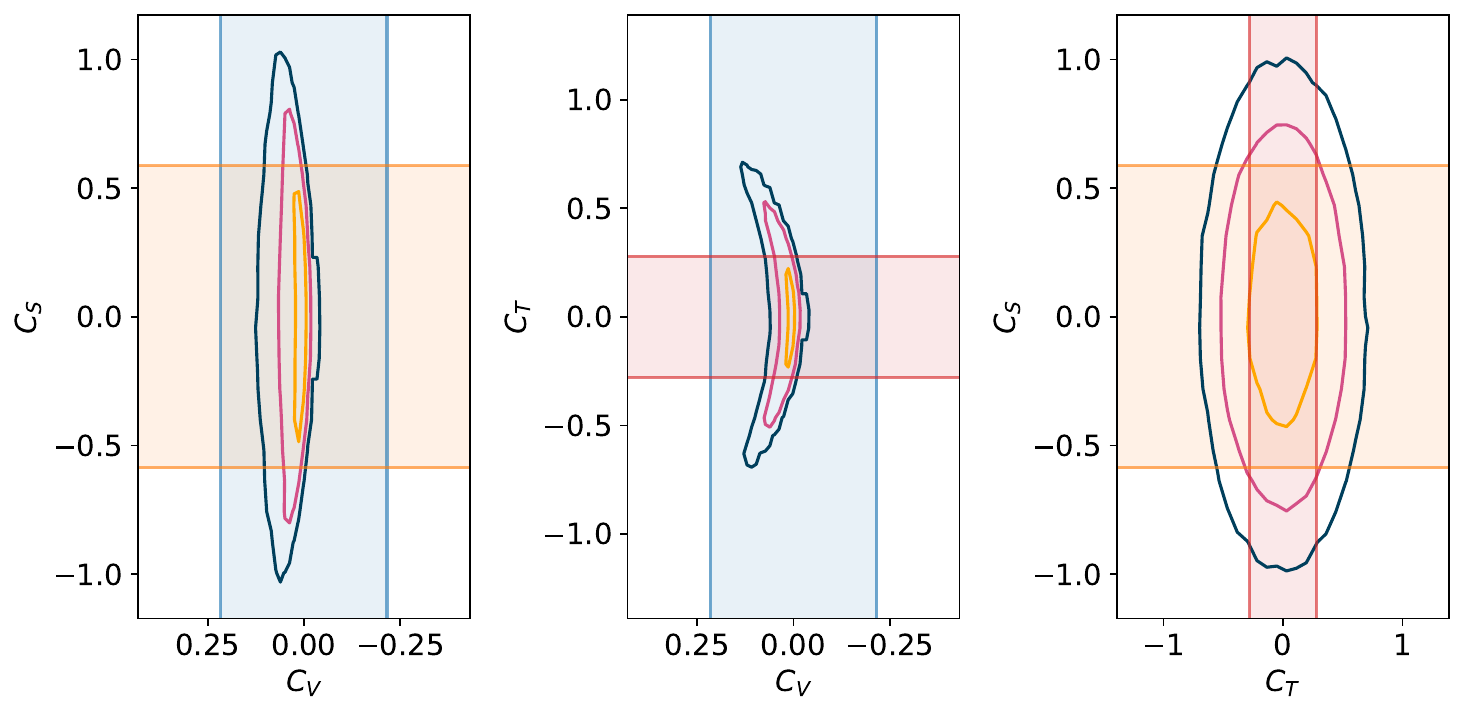}
    \caption{Reactor spectral uncertainties for 1, 2, and 3 $\sigma$ compared to the projected sensitivities of Ref. \cite{Lindner2017} for each $C_i$.}
    \label{fig:limits_GNI}
\end{figure}

Opening up the Lorentz structure of the CE$\nu$NS interaction provides the richer spectral structure of Eq. (\ref{eq:GNI_diff_cs}). Due to the latter, correlations between different $C_i$ are typically limited and uncertainties are approximately Gaussian. In Fig. \ref{fig:limits_GNI} we therefore immediately show the resultant 1, 2, and 3 $\sigma$ uncertainties compared to the projected sensitivities of Ref. \cite{Lindner2017}.

Changes to the vector charge, $C_V$, are substantially less sensitive to reactor spectral uncertainties than either $C_S$ or $C_T$. In case of the former, the projected sensitivity is a factor two larger than the 3$\sigma$ uncertainty resulting from reactor spectral uncertainties and is therefore not affected in the foreseeable future. This is consistent with the results of the previous section as the sensitivity decreases when allowing for more free parameters in the fit.

Exotic scalar and tensor interactions, on the other hand, are significantly more affected as the different kinetic sensitivity captures all spectral distortions that do not resemble the usual vector interaction. The induced uncertainties due to the reactor antineutrino spectrum are of similar magnitude to the projected sensitivities. For exotic tensor interactions this is at the 1$\sigma$ level, while for scalar this is around 1.7$\sigma$. 

\section{Conclusion and outlook}
\label{sec:conclusion}

Nuclear reactors are the most intense man-made sources of antineutrinos and form an attractive option for studies on coherent elastic (anti)neutrino nuclear scattering. The observation of the latter is sensitive to a plethora of different exotic physics scenarios out of reach in many other observables. The prediction of reactor antineutrino energy spectra has undergone a significant evolution in the last decade, however, and while many sources of uncertainty have been investigated the lowest energies can be obtained solely from theory. We have investigated the resultant uncertainties in the recoil spectra and shown the effects of changes to the $\beta$ spectrum shape due to nuclear structure effects. Following this, we performed an analysis using uncertainties in aggregate reactor spectra and the systematic uncertainties for three different exotic physics scenarios: the existence of a neutrino magnetic moment, nonstandard neutrino interactions and generalized neutrino interactions.

We have shown how reactor antineutrino spectral uncertainties are at the same order of magnitude as the projected sensitivities by Lindner \textit{et al.} for all but changes to $Q_W$, the weak vector charge. Systematic uncertainties on a neutrino magnetic moment are of similar magnitude as those obtained by the GEMMA experiment which provides the current most precise upper limit. A substantial part of this uncertainty arises from the antineutrino spectrum below the inverse $\beta$ decay threshold and is not experimentally constrained. These results depend on a number of experimental parameters, however, such as the target material and detection threshold. Largely speaking, however, we argue that in order for the projected sensitivities to be reached further $\beta$ spectroscopy studies are necessary.

\bibliography{library}

\begin{thebibliography}{57}%
\makeatletter
\providecommand \@ifxundefined [1]{%
 \@ifx{#1\undefined}
}%
\providecommand \@ifnum [1]{%
 \ifnum #1\expandafter \@firstoftwo
 \else \expandafter \@secondoftwo
 \fi
}%
\providecommand \@ifx [1]{%
 \ifx #1\expandafter \@firstoftwo
 \else \expandafter \@secondoftwo
 \fi
}%
\providecommand \natexlab [1]{#1}%
\providecommand \enquote  [1]{``#1''}%
\providecommand \bibnamefont  [1]{#1}%
\providecommand \bibfnamefont [1]{#1}%
\providecommand \citenamefont [1]{#1}%
\providecommand \href@noop [0]{\@secondoftwo}%
\providecommand \href [0]{\begingroup \@sanitize@url \@href}%
\providecommand \@href[1]{\@@startlink{#1}\@@href}%
\providecommand \@@href[1]{\endgroup#1\@@endlink}%
\providecommand \@sanitize@url [0]{\catcode `\\12\catcode `\$12\catcode `\&12\catcode `\#12\catcode `\^12\catcode `\_12\catcode `\%12\relax}%
\providecommand \@@startlink[1]{}%
\providecommand \@@endlink[0]{}%
\providecommand \url  [0]{\begingroup\@sanitize@url \@url }%
\providecommand \@url [1]{\endgroup\@href {#1}{\urlprefix }}%
\providecommand \urlprefix  [0]{URL }%
\providecommand \Eprint [0]{\href }%
\providecommand \doibase [0]{http://dx.doi.org/}%
\providecommand \selectlanguage [0]{\@gobble}%
\providecommand \bibinfo  [0]{\@secondoftwo}%
\providecommand \bibfield  [0]{\@secondoftwo}%
\providecommand \translation [1]{[#1]}%
\providecommand \BibitemOpen [0]{}%
\providecommand \bibitemStop [0]{}%
\providecommand \bibitemNoStop [0]{.\EOS\space}%
\providecommand \EOS [0]{\spacefactor3000\relax}%
\providecommand \BibitemShut  [1]{\csname bibitem#1\endcsname}%
\let\auto@bib@innerbib\@empty
\bibitem [{\citenamefont {Freedman}(1974)}]{Freedman1974a}%
  \BibitemOpen
  \bibfield  {author} {\bibinfo {author} {\bibfnamefont {D.~Z.}\ \bibnamefont {Freedman}},\ }\href {\doibase 10.1103/PhysRevD.9.1389} {\bibfield  {journal} {\bibinfo  {journal} {Physical Review D}\ }\textbf {\bibinfo {volume} {9}},\ \bibinfo {pages} {1389} (\bibinfo {year} {1974})}\BibitemShut {NoStop}%
\bibitem [{\citenamefont {Akimov}\ \emph {et~al.}(2017)\citenamefont {Akimov}, \citenamefont {Albert}, \citenamefont {An}, \citenamefont {Awe}, \citenamefont {Barbeau}, \citenamefont {Becker}, \citenamefont {Belov}, \citenamefont {Brown}, \citenamefont {Bolozdynya}, \citenamefont {Cabrera-Palmer}, \citenamefont {Cervantes}, \citenamefont {Collar}, \citenamefont {Cooper}, \citenamefont {Cooper}, \citenamefont {Cuesta}, \citenamefont {Dean}, \citenamefont {Detwiler}, \citenamefont {Eberhardt}, \citenamefont {Efremenko}, \citenamefont {Elliott}, \citenamefont {Erkela}, \citenamefont {Fabris}, \citenamefont {Febbraro}, \citenamefont {Fields}, \citenamefont {Fox}, \citenamefont {Fu}, \citenamefont {Galindo-Uribarri}, \citenamefont {Green}, \citenamefont {Hai}, \citenamefont {Heath}, \citenamefont {Hedges}, \citenamefont {Hornback}, \citenamefont {Hossbach}, \citenamefont {Iverson}, \citenamefont {Kaufman}, \citenamefont {Ki}, \citenamefont {Klein}, \citenamefont {Khromov}, \citenamefont {Konovalov}, \citenamefont
  {Kremer}, \citenamefont {Kumpan}, \citenamefont {Leadbetter}, \citenamefont {Li}, \citenamefont {Lu}, \citenamefont {Mann}, \citenamefont {Markoff}, \citenamefont {Miller}, \citenamefont {Moreno}, \citenamefont {Mueller}, \citenamefont {Newby}, \citenamefont {Orrell}, \citenamefont {Overman}, \citenamefont {Parno}, \citenamefont {Penttila}, \citenamefont {Perumpilly}, \citenamefont {Ray}, \citenamefont {Raybern}, \citenamefont {Reyna}, \citenamefont {Rich}, \citenamefont {Rimal}, \citenamefont {Rudik}, \citenamefont {Scholberg}, \citenamefont {Scholz}, \citenamefont {Sinev}, \citenamefont {Snow}, \citenamefont {Sosnovtsev}, \citenamefont {Shakirov}, \citenamefont {Suchyta}, \citenamefont {Suh}, \citenamefont {Tayloe}, \citenamefont {Thornton}, \citenamefont {Tolstukhin}, \citenamefont {Vanderwerp}, \citenamefont {Varner}, \citenamefont {Virtue}, \citenamefont {Wan}, \citenamefont {Yoo}, \citenamefont {Yu}, \citenamefont {Zawada}, \citenamefont {Zettlemoyer},\ and\ \citenamefont {Zderic}}]{Akimov2017}%
  \BibitemOpen
  \bibfield  {author} {\bibinfo {author} {\bibfnamefont {D.}~\bibnamefont {Akimov}}, \bibinfo {author} {\bibfnamefont {J.~B.}\ \bibnamefont {Albert}}, \bibinfo {author} {\bibfnamefont {P.}~\bibnamefont {An}}, \bibinfo {author} {\bibfnamefont {C.}~\bibnamefont {Awe}}, \bibinfo {author} {\bibfnamefont {P.~S.}\ \bibnamefont {Barbeau}}, \bibinfo {author} {\bibfnamefont {B.}~\bibnamefont {Becker}}, \bibinfo {author} {\bibfnamefont {V.}~\bibnamefont {Belov}}, \bibinfo {author} {\bibfnamefont {A.}~\bibnamefont {Brown}}, \bibinfo {author} {\bibfnamefont {A.}~\bibnamefont {Bolozdynya}}, \bibinfo {author} {\bibfnamefont {B.}~\bibnamefont {Cabrera-Palmer}}, \bibinfo {author} {\bibfnamefont {M.}~\bibnamefont {Cervantes}}, \bibinfo {author} {\bibfnamefont {J.~I.}\ \bibnamefont {Collar}}, \bibinfo {author} {\bibfnamefont {R.~J.}\ \bibnamefont {Cooper}}, \bibinfo {author} {\bibfnamefont {R.~L.}\ \bibnamefont {Cooper}}, \bibinfo {author} {\bibfnamefont {C.}~\bibnamefont {Cuesta}}, \bibinfo {author} {\bibfnamefont {D.~J.}\
  \bibnamefont {Dean}}, \bibinfo {author} {\bibfnamefont {J.~A.}\ \bibnamefont {Detwiler}}, \bibinfo {author} {\bibfnamefont {A.}~\bibnamefont {Eberhardt}}, \bibinfo {author} {\bibfnamefont {Y.}~\bibnamefont {Efremenko}}, \bibinfo {author} {\bibfnamefont {S.~R.}\ \bibnamefont {Elliott}}, \bibinfo {author} {\bibfnamefont {E.~M.}\ \bibnamefont {Erkela}}, \bibinfo {author} {\bibfnamefont {L.}~\bibnamefont {Fabris}}, \bibinfo {author} {\bibfnamefont {M.}~\bibnamefont {Febbraro}}, \bibinfo {author} {\bibfnamefont {N.~E.}\ \bibnamefont {Fields}}, \bibinfo {author} {\bibfnamefont {W.}~\bibnamefont {Fox}}, \bibinfo {author} {\bibfnamefont {Z.}~\bibnamefont {Fu}}, \bibinfo {author} {\bibfnamefont {A.}~\bibnamefont {Galindo-Uribarri}}, \bibinfo {author} {\bibfnamefont {M.~P.}\ \bibnamefont {Green}}, \bibinfo {author} {\bibfnamefont {M.}~\bibnamefont {Hai}}, \bibinfo {author} {\bibfnamefont {M.~R.}\ \bibnamefont {Heath}}, \bibinfo {author} {\bibfnamefont {S.}~\bibnamefont {Hedges}}, \bibinfo {author} {\bibfnamefont
  {D.}~\bibnamefont {Hornback}}, \bibinfo {author} {\bibfnamefont {T.~W.}\ \bibnamefont {Hossbach}}, \bibinfo {author} {\bibfnamefont {E.~B.}\ \bibnamefont {Iverson}}, \bibinfo {author} {\bibfnamefont {L.~J.}\ \bibnamefont {Kaufman}}, \bibinfo {author} {\bibfnamefont {S.}~\bibnamefont {Ki}}, \bibinfo {author} {\bibfnamefont {S.~R.}\ \bibnamefont {Klein}}, \bibinfo {author} {\bibfnamefont {A.}~\bibnamefont {Khromov}}, \bibinfo {author} {\bibfnamefont {A.}~\bibnamefont {Konovalov}}, \bibinfo {author} {\bibfnamefont {M.}~\bibnamefont {Kremer}}, \bibinfo {author} {\bibfnamefont {A.}~\bibnamefont {Kumpan}}, \bibinfo {author} {\bibfnamefont {C.}~\bibnamefont {Leadbetter}}, \bibinfo {author} {\bibfnamefont {L.}~\bibnamefont {Li}}, \bibinfo {author} {\bibfnamefont {W.}~\bibnamefont {Lu}}, \bibinfo {author} {\bibfnamefont {K.}~\bibnamefont {Mann}}, \bibinfo {author} {\bibfnamefont {D.~M.}\ \bibnamefont {Markoff}}, \bibinfo {author} {\bibfnamefont {K.}~\bibnamefont {Miller}}, \bibinfo {author} {\bibfnamefont
  {H.}~\bibnamefont {Moreno}}, \bibinfo {author} {\bibfnamefont {P.~E.}\ \bibnamefont {Mueller}}, \bibinfo {author} {\bibfnamefont {J.}~\bibnamefont {Newby}}, \bibinfo {author} {\bibfnamefont {J.~L.}\ \bibnamefont {Orrell}}, \bibinfo {author} {\bibfnamefont {C.~T.}\ \bibnamefont {Overman}}, \bibinfo {author} {\bibfnamefont {D.~S.}\ \bibnamefont {Parno}}, \bibinfo {author} {\bibfnamefont {S.}~\bibnamefont {Penttila}}, \bibinfo {author} {\bibfnamefont {G.}~\bibnamefont {Perumpilly}}, \bibinfo {author} {\bibfnamefont {H.}~\bibnamefont {Ray}}, \bibinfo {author} {\bibfnamefont {J.}~\bibnamefont {Raybern}}, \bibinfo {author} {\bibfnamefont {D.}~\bibnamefont {Reyna}}, \bibinfo {author} {\bibfnamefont {G.~C.}\ \bibnamefont {Rich}}, \bibinfo {author} {\bibfnamefont {D.}~\bibnamefont {Rimal}}, \bibinfo {author} {\bibfnamefont {D.}~\bibnamefont {Rudik}}, \bibinfo {author} {\bibfnamefont {K.}~\bibnamefont {Scholberg}}, \bibinfo {author} {\bibfnamefont {B.~J.}\ \bibnamefont {Scholz}}, \bibinfo {author} {\bibfnamefont
  {G.}~\bibnamefont {Sinev}}, \bibinfo {author} {\bibfnamefont {W.~M.}\ \bibnamefont {Snow}}, \bibinfo {author} {\bibfnamefont {V.}~\bibnamefont {Sosnovtsev}}, \bibinfo {author} {\bibfnamefont {A.}~\bibnamefont {Shakirov}}, \bibinfo {author} {\bibfnamefont {S.}~\bibnamefont {Suchyta}}, \bibinfo {author} {\bibfnamefont {B.}~\bibnamefont {Suh}}, \bibinfo {author} {\bibfnamefont {R.}~\bibnamefont {Tayloe}}, \bibinfo {author} {\bibfnamefont {R.~T.}\ \bibnamefont {Thornton}}, \bibinfo {author} {\bibfnamefont {I.}~\bibnamefont {Tolstukhin}}, \bibinfo {author} {\bibfnamefont {J.}~\bibnamefont {Vanderwerp}}, \bibinfo {author} {\bibfnamefont {R.~L.}\ \bibnamefont {Varner}}, \bibinfo {author} {\bibfnamefont {C.~J.}\ \bibnamefont {Virtue}}, \bibinfo {author} {\bibfnamefont {Z.}~\bibnamefont {Wan}}, \bibinfo {author} {\bibfnamefont {J.}~\bibnamefont {Yoo}}, \bibinfo {author} {\bibfnamefont {C.~H.}\ \bibnamefont {Yu}}, \bibinfo {author} {\bibfnamefont {A.}~\bibnamefont {Zawada}}, \bibinfo {author} {\bibfnamefont
  {J.}~\bibnamefont {Zettlemoyer}}, \ and\ \bibinfo {author} {\bibfnamefont {A.~M.}\ \bibnamefont {Zderic}},\ }\href {\doibase 10.1126/science.aao0990} {\bibfield  {journal} {\bibinfo  {journal} {Science}\ }\textbf {\bibinfo {volume} {357}},\ \bibinfo {pages} {1123} (\bibinfo {year} {2017})},\ \Eprint {http://arxiv.org/abs/1708.01294} {arXiv:1708.01294} \BibitemShut {NoStop}%
\bibitem [{\citenamefont {Colaresi}\ \emph {et~al.}(2021)\citenamefont {Colaresi}, \citenamefont {Collar}, \citenamefont {Hossbach}, \citenamefont {Kavner}, \citenamefont {Lewis}, \citenamefont {Robinson},\ and\ \citenamefont {Yocum}}]{Colaresi2021}%
  \BibitemOpen
  \bibfield  {author} {\bibinfo {author} {\bibfnamefont {J.}~\bibnamefont {Colaresi}}, \bibinfo {author} {\bibfnamefont {J.}~\bibnamefont {Collar}}, \bibinfo {author} {\bibfnamefont {T.}~\bibnamefont {Hossbach}}, \bibinfo {author} {\bibfnamefont {A.}~\bibnamefont {Kavner}}, \bibinfo {author} {\bibfnamefont {C.}~\bibnamefont {Lewis}}, \bibinfo {author} {\bibfnamefont {A.}~\bibnamefont {Robinson}}, \ and\ \bibinfo {author} {\bibfnamefont {K.}~\bibnamefont {Yocum}},\ }\href {\doibase 10.1103/PhysRevD.104.072003} {\bibfield  {journal} {\bibinfo  {journal} {Physical Review D}\ }\textbf {\bibinfo {volume} {104}},\ \bibinfo {pages} {072003} (\bibinfo {year} {2021})}\BibitemShut {NoStop}%
\bibitem [{\citenamefont {Colaresi}\ \emph {et~al.}(2022)\citenamefont {Colaresi}, \citenamefont {Collar}, \citenamefont {Hossbach}, \citenamefont {Lewis},\ and\ \citenamefont {Yocum}}]{Colaresi2022}%
  \BibitemOpen
  \bibfield  {author} {\bibinfo {author} {\bibfnamefont {J.}~\bibnamefont {Colaresi}}, \bibinfo {author} {\bibfnamefont {J.}~\bibnamefont {Collar}}, \bibinfo {author} {\bibfnamefont {T.}~\bibnamefont {Hossbach}}, \bibinfo {author} {\bibfnamefont {C.}~\bibnamefont {Lewis}}, \ and\ \bibinfo {author} {\bibfnamefont {K.}~\bibnamefont {Yocum}},\ }\href {\doibase 10.1103/PhysRevLett.129.211802} {\bibfield  {journal} {\bibinfo  {journal} {Physical Review Letters}\ }\textbf {\bibinfo {volume} {129}},\ \bibinfo {pages} {211802} (\bibinfo {year} {2022})}\BibitemShut {NoStop}%
\bibitem [{\citenamefont {Bonet}\ \emph {et~al.}(2021)\citenamefont {Bonet}, \citenamefont {Bonhomme}, \citenamefont {Buck}, \citenamefont {Fülber}, \citenamefont {Hakenmüller}, \citenamefont {Heusser}, \citenamefont {Hugle}, \citenamefont {Lindner}, \citenamefont {Maneschg}, \citenamefont {Rink}, \citenamefont {Strecker},\ and\ \citenamefont {Wink}}]{Bonet2021}%
  \BibitemOpen
  \bibfield  {author} {\bibinfo {author} {\bibfnamefont {H.}~\bibnamefont {Bonet}}, \bibinfo {author} {\bibfnamefont {A.}~\bibnamefont {Bonhomme}}, \bibinfo {author} {\bibfnamefont {C.}~\bibnamefont {Buck}}, \bibinfo {author} {\bibfnamefont {K.}~\bibnamefont {Fülber}}, \bibinfo {author} {\bibfnamefont {J.}~\bibnamefont {Hakenmüller}}, \bibinfo {author} {\bibfnamefont {G.}~\bibnamefont {Heusser}}, \bibinfo {author} {\bibfnamefont {T.}~\bibnamefont {Hugle}}, \bibinfo {author} {\bibfnamefont {M.}~\bibnamefont {Lindner}}, \bibinfo {author} {\bibfnamefont {W.}~\bibnamefont {Maneschg}}, \bibinfo {author} {\bibfnamefont {T.}~\bibnamefont {Rink}}, \bibinfo {author} {\bibfnamefont {H.}~\bibnamefont {Strecker}}, \ and\ \bibinfo {author} {\bibfnamefont {R.}~\bibnamefont {Wink}},\ }\href {\doibase 10.1103/PhysRevLett.126.041804} {\bibfield  {journal} {\bibinfo  {journal} {Physical Review Letters}\ }\textbf {\bibinfo {volume} {126}},\ \bibinfo {pages} {041804} (\bibinfo {year} {2021})}\BibitemShut {NoStop}%
\bibitem [{\citenamefont {Billard}\ \emph {et~al.}(2017)\citenamefont {Billard}, \citenamefont {Carr}, \citenamefont {Dawson}, \citenamefont {Figueroa-Feliciano}, \citenamefont {Formaggio}, \citenamefont {Gascon}, \citenamefont {Heine}, \citenamefont {Jesus}, \citenamefont {Johnston}, \citenamefont {Lasserre}, \citenamefont {Leder}, \citenamefont {Palladino}, \citenamefont {Sibille}, \citenamefont {Vivier},\ and\ \citenamefont {Winslow}}]{Billard2017}%
  \BibitemOpen
  \bibfield  {author} {\bibinfo {author} {\bibfnamefont {J.}~\bibnamefont {Billard}}, \bibinfo {author} {\bibfnamefont {R.}~\bibnamefont {Carr}}, \bibinfo {author} {\bibfnamefont {J.}~\bibnamefont {Dawson}}, \bibinfo {author} {\bibfnamefont {E.}~\bibnamefont {Figueroa-Feliciano}}, \bibinfo {author} {\bibfnamefont {J.~A.}\ \bibnamefont {Formaggio}}, \bibinfo {author} {\bibfnamefont {J.}~\bibnamefont {Gascon}}, \bibinfo {author} {\bibfnamefont {S.~T.}\ \bibnamefont {Heine}}, \bibinfo {author} {\bibfnamefont {M.~D.}\ \bibnamefont {Jesus}}, \bibinfo {author} {\bibfnamefont {J.}~\bibnamefont {Johnston}}, \bibinfo {author} {\bibfnamefont {T.}~\bibnamefont {Lasserre}}, \bibinfo {author} {\bibfnamefont {A.}~\bibnamefont {Leder}}, \bibinfo {author} {\bibfnamefont {K.~J.}\ \bibnamefont {Palladino}}, \bibinfo {author} {\bibfnamefont {V.}~\bibnamefont {Sibille}}, \bibinfo {author} {\bibfnamefont {M.}~\bibnamefont {Vivier}}, \ and\ \bibinfo {author} {\bibfnamefont {L.}~\bibnamefont {Winslow}},\ }\href {\doibase
  10.1088/1361-6471/aa83d0} {\bibfield  {journal} {\bibinfo  {journal} {Journal of Physics G: Nuclear and Particle Physics}\ }\textbf {\bibinfo {volume} {44}},\ \bibinfo {pages} {105101} (\bibinfo {year} {2017})}\BibitemShut {NoStop}%
\bibitem [{\citenamefont {Alekseev}\ \emph {et~al.}(2022)\citenamefont {Alekseev}, \citenamefont {Balej}, \citenamefont {Belov}, \citenamefont {Evseev}, \citenamefont {Filosofov}, \citenamefont {Fomina}, \citenamefont {Hons}, \citenamefont {Karaivanov}, \citenamefont {Kazartsev}, \citenamefont {Khushvaktov}, \citenamefont {Kuznetsov}, \citenamefont {Lubashevskiy}, \citenamefont {Medvedev}, \citenamefont {Ponomarev}, \citenamefont {Rakhimov}, \citenamefont {Shakhov}, \citenamefont {Shevchik}, \citenamefont {Shirchenko}, \citenamefont {Smolek}, \citenamefont {Rozov}, \citenamefont {Rozova}, \citenamefont {Vasilyev}, \citenamefont {Yakushev},\ and\ \citenamefont {Zhitnikov}}]{Alekseev2022}%
  \BibitemOpen
  \bibfield  {author} {\bibinfo {author} {\bibfnamefont {I.}~\bibnamefont {Alekseev}}, \bibinfo {author} {\bibfnamefont {K.}~\bibnamefont {Balej}}, \bibinfo {author} {\bibfnamefont {V.}~\bibnamefont {Belov}}, \bibinfo {author} {\bibfnamefont {S.}~\bibnamefont {Evseev}}, \bibinfo {author} {\bibfnamefont {D.}~\bibnamefont {Filosofov}}, \bibinfo {author} {\bibfnamefont {M.}~\bibnamefont {Fomina}}, \bibinfo {author} {\bibfnamefont {Z.}~\bibnamefont {Hons}}, \bibinfo {author} {\bibfnamefont {D.}~\bibnamefont {Karaivanov}}, \bibinfo {author} {\bibfnamefont {S.}~\bibnamefont {Kazartsev}}, \bibinfo {author} {\bibfnamefont {J.}~\bibnamefont {Khushvaktov}}, \bibinfo {author} {\bibfnamefont {A.}~\bibnamefont {Kuznetsov}}, \bibinfo {author} {\bibfnamefont {A.}~\bibnamefont {Lubashevskiy}}, \bibinfo {author} {\bibfnamefont {D.}~\bibnamefont {Medvedev}}, \bibinfo {author} {\bibfnamefont {D.}~\bibnamefont {Ponomarev}}, \bibinfo {author} {\bibfnamefont {A.}~\bibnamefont {Rakhimov}}, \bibinfo {author} {\bibfnamefont
  {K.}~\bibnamefont {Shakhov}}, \bibinfo {author} {\bibfnamefont {E.}~\bibnamefont {Shevchik}}, \bibinfo {author} {\bibfnamefont {M.}~\bibnamefont {Shirchenko}}, \bibinfo {author} {\bibfnamefont {K.}~\bibnamefont {Smolek}}, \bibinfo {author} {\bibfnamefont {S.}~\bibnamefont {Rozov}}, \bibinfo {author} {\bibfnamefont {I.}~\bibnamefont {Rozova}}, \bibinfo {author} {\bibfnamefont {S.}~\bibnamefont {Vasilyev}}, \bibinfo {author} {\bibfnamefont {E.}~\bibnamefont {Yakushev}}, \ and\ \bibinfo {author} {\bibfnamefont {I.}~\bibnamefont {Zhitnikov}},\ }\href {\doibase 10.1103/PhysRevD.106.L051101} {\bibfield  {journal} {\bibinfo  {journal} {Physical Review D}\ }\textbf {\bibinfo {volume} {106}},\ \bibinfo {pages} {L051101} (\bibinfo {year} {2022})}\BibitemShut {NoStop}%
\bibitem [{\citenamefont {Choi}\ \emph {et~al.}(2023)\citenamefont {Choi}, \citenamefont {Jeon}, \citenamefont {Kim}, \citenamefont {Kim}, \citenamefont {Kim}, \citenamefont {Kim}, \citenamefont {Kim}, \citenamefont {Ko}, \citenamefont {Koh}, \citenamefont {Ha}, \citenamefont {Park}, \citenamefont {Lee}, \citenamefont {Lee}, \citenamefont {Lee}, \citenamefont {Lee}, \citenamefont {Lee},\ and\ \citenamefont {Oh}}]{Choi2023}%
  \BibitemOpen
  \bibfield  {author} {\bibinfo {author} {\bibfnamefont {J.~J.}\ \bibnamefont {Choi}}, \bibinfo {author} {\bibfnamefont {E.~J.}\ \bibnamefont {Jeon}}, \bibinfo {author} {\bibfnamefont {J.~Y.}\ \bibnamefont {Kim}}, \bibinfo {author} {\bibfnamefont {K.~W.}\ \bibnamefont {Kim}}, \bibinfo {author} {\bibfnamefont {S.~H.}\ \bibnamefont {Kim}}, \bibinfo {author} {\bibfnamefont {S.~K.}\ \bibnamefont {Kim}}, \bibinfo {author} {\bibfnamefont {Y.~D.}\ \bibnamefont {Kim}}, \bibinfo {author} {\bibfnamefont {Y.~J.}\ \bibnamefont {Ko}}, \bibinfo {author} {\bibfnamefont {B.~C.}\ \bibnamefont {Koh}}, \bibinfo {author} {\bibfnamefont {C.}~\bibnamefont {Ha}}, \bibinfo {author} {\bibfnamefont {B.~J.}\ \bibnamefont {Park}}, \bibinfo {author} {\bibfnamefont {S.~H.}\ \bibnamefont {Lee}}, \bibinfo {author} {\bibfnamefont {I.~S.}\ \bibnamefont {Lee}}, \bibinfo {author} {\bibfnamefont {H.}~\bibnamefont {Lee}}, \bibinfo {author} {\bibfnamefont {H.~S.}\ \bibnamefont {Lee}}, \bibinfo {author} {\bibfnamefont {J.}~\bibnamefont {Lee}}, \
  and\ \bibinfo {author} {\bibfnamefont {Y.~M.}\ \bibnamefont {Oh}},\ }\href {\doibase 10.1140/epjc/s10052-023-11352-x} {\bibfield  {journal} {\bibinfo  {journal} {The European Physical Journal C}\ }\textbf {\bibinfo {volume} {83}},\ \bibinfo {pages} {226} (\bibinfo {year} {2023})}\BibitemShut {NoStop}%
\bibitem [{\citenamefont {Aguilar-Arevalo}\ \emph {et~al.}(2022)\citenamefont {Aguilar-Arevalo}, \citenamefont {Bernal}, \citenamefont {Bertou}, \citenamefont {Bonifazi}, \citenamefont {Cancelo}, \citenamefont {de~Carvalho}, \citenamefont {Cervantes-Vergara}, \citenamefont {Chavez}, \citenamefont {Correa}, \citenamefont {D'Olivo}, \citenamefont {dos Anjos}, \citenamefont {Estrada}, \citenamefont {Neto}, \citenamefont {Moroni}, \citenamefont {Foguel}, \citenamefont {Ford}, \citenamefont {Barbuscio}, \citenamefont {Cuevas}, \citenamefont {Hernandez}, \citenamefont {Izraelevitch}, \citenamefont {Kilminster}, \citenamefont {Kuk}, \citenamefont {Lima}, \citenamefont {Makler}, \citenamefont {Montero}, \citenamefont {Mendes}, \citenamefont {Molina}, \citenamefont {Mota}, \citenamefont {Nasteva}, \citenamefont {Paolini}, \citenamefont {Rodrigues}, \citenamefont {Sarkis}, \citenamefont {Haro}, \citenamefont {Stalder},\ and\ \citenamefont {Tiffenberg}}]{Aguilar-Arevalo2022}%
  \BibitemOpen
  \bibfield  {author} {\bibinfo {author} {\bibfnamefont {A.}~\bibnamefont {Aguilar-Arevalo}}, \bibinfo {author} {\bibfnamefont {J.}~\bibnamefont {Bernal}}, \bibinfo {author} {\bibfnamefont {X.}~\bibnamefont {Bertou}}, \bibinfo {author} {\bibfnamefont {C.}~\bibnamefont {Bonifazi}}, \bibinfo {author} {\bibfnamefont {G.}~\bibnamefont {Cancelo}}, \bibinfo {author} {\bibfnamefont {V.~G. P.~B.}\ \bibnamefont {de~Carvalho}}, \bibinfo {author} {\bibfnamefont {B.~A.}\ \bibnamefont {Cervantes-Vergara}}, \bibinfo {author} {\bibfnamefont {C.}~\bibnamefont {Chavez}}, \bibinfo {author} {\bibfnamefont {G.~C.}\ \bibnamefont {Correa}}, \bibinfo {author} {\bibfnamefont {J.~C.}\ \bibnamefont {D'Olivo}}, \bibinfo {author} {\bibfnamefont {J.~C.}\ \bibnamefont {dos Anjos}}, \bibinfo {author} {\bibfnamefont {J.}~\bibnamefont {Estrada}}, \bibinfo {author} {\bibfnamefont {A.~R.~F.}\ \bibnamefont {Neto}}, \bibinfo {author} {\bibfnamefont {G.~F.}\ \bibnamefont {Moroni}}, \bibinfo {author} {\bibfnamefont {A.}~\bibnamefont {Foguel}},
  \bibinfo {author} {\bibfnamefont {R.}~\bibnamefont {Ford}}, \bibinfo {author} {\bibfnamefont {J.~G.}\ \bibnamefont {Barbuscio}}, \bibinfo {author} {\bibfnamefont {J.~G.}\ \bibnamefont {Cuevas}}, \bibinfo {author} {\bibfnamefont {S.}~\bibnamefont {Hernandez}}, \bibinfo {author} {\bibfnamefont {F.}~\bibnamefont {Izraelevitch}}, \bibinfo {author} {\bibfnamefont {B.}~\bibnamefont {Kilminster}}, \bibinfo {author} {\bibfnamefont {K.}~\bibnamefont {Kuk}}, \bibinfo {author} {\bibfnamefont {H.~P.}\ \bibnamefont {Lima}}, \bibinfo {author} {\bibfnamefont {M.}~\bibnamefont {Makler}}, \bibinfo {author} {\bibfnamefont {M.~M.}\ \bibnamefont {Montero}}, \bibinfo {author} {\bibfnamefont {L.~H.}\ \bibnamefont {Mendes}}, \bibinfo {author} {\bibfnamefont {J.}~\bibnamefont {Molina}}, \bibinfo {author} {\bibfnamefont {P.}~\bibnamefont {Mota}}, \bibinfo {author} {\bibfnamefont {I.}~\bibnamefont {Nasteva}}, \bibinfo {author} {\bibfnamefont {E.}~\bibnamefont {Paolini}}, \bibinfo {author} {\bibfnamefont {D.}~\bibnamefont
  {Rodrigues}}, \bibinfo {author} {\bibfnamefont {Y.}~\bibnamefont {Sarkis}}, \bibinfo {author} {\bibfnamefont {M.~S.}\ \bibnamefont {Haro}}, \bibinfo {author} {\bibfnamefont {D.}~\bibnamefont {Stalder}}, \ and\ \bibinfo {author} {\bibfnamefont {J.}~\bibnamefont {Tiffenberg}},\ }\href {\doibase 10.1007/JHEP05(2022)017} {\bibfield  {journal} {\bibinfo  {journal} {Journal of High Energy Physics}\ }\textbf {\bibinfo {volume} {2022}},\ \bibinfo {pages} {17} (\bibinfo {year} {2022})}\BibitemShut {NoStop}%
\bibitem [{\citenamefont {Hayes}\ and\ \citenamefont {Vogel}(2016)}]{Hayes2016}%
  \BibitemOpen
  \bibfield  {author} {\bibinfo {author} {\bibfnamefont {A.~C.}\ \bibnamefont {Hayes}}\ and\ \bibinfo {author} {\bibfnamefont {P.}~\bibnamefont {Vogel}},\ }\href {\doibase 10.1146/annurev-nucl-102115-044826} {\bibfield  {journal} {\bibinfo  {journal} {Annual Review of Nuclear and Particle Science}\ }\textbf {\bibinfo {volume} {66}},\ \bibinfo {pages} {219} (\bibinfo {year} {2016})},\ \Eprint {http://arxiv.org/abs/1605.02047} {arXiv:1605.02047} \BibitemShut {NoStop}%
\bibitem [{\citenamefont {Mention}\ \emph {et~al.}(2011)\citenamefont {Mention}, \citenamefont {Fechner}, \citenamefont {Lasserre}, \citenamefont {Mueller}, \citenamefont {Lhuillier}, \citenamefont {Cribier},\ and\ \citenamefont {Letourneau}}]{Mention2011}%
  \BibitemOpen
  \bibfield  {author} {\bibinfo {author} {\bibfnamefont {G.}~\bibnamefont {Mention}}, \bibinfo {author} {\bibfnamefont {M.}~\bibnamefont {Fechner}}, \bibinfo {author} {\bibfnamefont {T.}~\bibnamefont {Lasserre}}, \bibinfo {author} {\bibfnamefont {T.~A.}\ \bibnamefont {Mueller}}, \bibinfo {author} {\bibfnamefont {D.}~\bibnamefont {Lhuillier}}, \bibinfo {author} {\bibfnamefont {M.}~\bibnamefont {Cribier}}, \ and\ \bibinfo {author} {\bibfnamefont {A.}~\bibnamefont {Letourneau}},\ }\href {\doibase 10.1103/PhysRevD.83.073006} {\bibfield  {journal} {\bibinfo  {journal} {Physical Review D}\ }\textbf {\bibinfo {volume} {83}},\ \bibinfo {pages} {073006} (\bibinfo {year} {2011})},\ \Eprint {http://arxiv.org/abs/1101.2755} {arXiv:1101.2755} \BibitemShut {NoStop}%
\bibitem [{\citenamefont {Huber}(2011)}]{Huber2011}%
  \BibitemOpen
  \bibfield  {author} {\bibinfo {author} {\bibfnamefont {P.}~\bibnamefont {Huber}},\ }\href {\doibase 10.1103/PhysRevC.84.024617} {\bibfield  {journal} {\bibinfo  {journal} {Physical Review C}\ }\textbf {\bibinfo {volume} {84}},\ \bibinfo {pages} {24617} (\bibinfo {year} {2011})}\BibitemShut {NoStop}%
\bibitem [{\citenamefont {Mueller}\ \emph {et~al.}(2011)\citenamefont {Mueller}, \citenamefont {Lhuillier}, \citenamefont {Fallot}, \citenamefont {Letourneau}, \citenamefont {Cormon}, \citenamefont {Fechner}, \citenamefont {Giot}, \citenamefont {Lasserre}, \citenamefont {Martino}, \citenamefont {Mention}, \citenamefont {Porta},\ and\ \citenamefont {Yermia}}]{Mueller2011}%
  \BibitemOpen
  \bibfield  {author} {\bibinfo {author} {\bibfnamefont {T.~A.}\ \bibnamefont {Mueller}}, \bibinfo {author} {\bibfnamefont {D.}~\bibnamefont {Lhuillier}}, \bibinfo {author} {\bibfnamefont {M.}~\bibnamefont {Fallot}}, \bibinfo {author} {\bibfnamefont {A.}~\bibnamefont {Letourneau}}, \bibinfo {author} {\bibfnamefont {S.}~\bibnamefont {Cormon}}, \bibinfo {author} {\bibfnamefont {M.}~\bibnamefont {Fechner}}, \bibinfo {author} {\bibfnamefont {L.}~\bibnamefont {Giot}}, \bibinfo {author} {\bibfnamefont {T.}~\bibnamefont {Lasserre}}, \bibinfo {author} {\bibfnamefont {J.}~\bibnamefont {Martino}}, \bibinfo {author} {\bibfnamefont {G.}~\bibnamefont {Mention}}, \bibinfo {author} {\bibfnamefont {A.}~\bibnamefont {Porta}}, \ and\ \bibinfo {author} {\bibfnamefont {F.}~\bibnamefont {Yermia}},\ }\href {\doibase 10.1103/PhysRevC.83.054615} {\bibfield  {journal} {\bibinfo  {journal} {Physical Review C - Nuclear Physics}\ }\textbf {\bibinfo {volume} {83}},\ \bibinfo {pages} {054615} (\bibinfo {year} {2011})},\ \Eprint
  {http://arxiv.org/abs/1101.2663} {arXiv:1101.2663} \BibitemShut {NoStop}%
\bibitem [{\citenamefont {Abazajian}\ \emph {et~al.}(2012)\citenamefont {Abazajian}, \citenamefont {Acero}, \citenamefont {Agarwalla}, \citenamefont {Aguilar-Arevalo}, \citenamefont {Albright}, \citenamefont {Antusch}, \citenamefont {Arguelles}, \citenamefont {Balantekin}, \citenamefont {Barenboim}, \citenamefont {Barger}, \citenamefont {Bernardini}, \citenamefont {Bezrukov}, \citenamefont {Bjaelde}, \citenamefont {Bogacz}, \citenamefont {Bowden}, \citenamefont {Boyarsky}, \citenamefont {Bravar}, \citenamefont {Berguno}, \citenamefont {Brice}, \citenamefont {Bross}, \citenamefont {Caccianiga}, \citenamefont {Cavanna}, \citenamefont {Chun}, \citenamefont {Cleveland}, \citenamefont {Collin}, \citenamefont {Coloma}, \citenamefont {Conrad}, \citenamefont {Cribier}, \citenamefont {Cucoanes}, \citenamefont {D'Olivo}, \citenamefont {Das}, \citenamefont {de~Gouvea}, \citenamefont {Derbin}, \citenamefont {Dharmapalan}, \citenamefont {Diaz}, \citenamefont {Ding}, \citenamefont {Djurcic}, \citenamefont {Donini},
  \citenamefont {Duchesneau}, \citenamefont {Ejiri}, \citenamefont {Elliott}, \citenamefont {Ernst}, \citenamefont {Esmaili}, \citenamefont {Evans}, \citenamefont {Fernandez-Martinez}, \citenamefont {Figueroa-Feliciano}, \citenamefont {Fleming}, \citenamefont {Formaggio}, \citenamefont {Franco}, \citenamefont {Gaffiot}, \citenamefont {Gandhi}, \citenamefont {Gao}, \citenamefont {Garvey}, \citenamefont {Gavrin}, \citenamefont {Ghoshal}, \citenamefont {Gibin}, \citenamefont {Giunti}, \citenamefont {Gninenko}, \citenamefont {Gorbachev}, \citenamefont {Gorbunov}, \citenamefont {Guenette}, \citenamefont {Guglielmi}, \citenamefont {Halzen}, \citenamefont {Hamann}, \citenamefont {Hannestad}, \citenamefont {Haxton}, \citenamefont {Heeger}, \citenamefont {Henning}, \citenamefont {Hernandez}, \citenamefont {Huber}, \citenamefont {Huelsnitz}, \citenamefont {Ianni}, \citenamefont {Ibragimova}, \citenamefont {Karadzhov}, \citenamefont {Karagiorgi}, \citenamefont {Keefer}, \citenamefont {Kim}, \citenamefont {Kopp},
  \citenamefont {Kornoukhov}, \citenamefont {Kusenko}, \citenamefont {Kyberd}, \citenamefont {Langacker}, \citenamefont {Lasserre}, \citenamefont {Laveder}, \citenamefont {Letourneau}, \citenamefont {Lhuillier}, \citenamefont {Li}, \citenamefont {Lindner}, \citenamefont {Link}, \citenamefont {Littlejohn}, \citenamefont {Lombardi}, \citenamefont {Long}, \citenamefont {Lopez-Pavon}, \citenamefont {Louis}, \citenamefont {Ludhova}, \citenamefont {Lykken}, \citenamefont {Machado}, \citenamefont {Maltoni}, \citenamefont {Mann}, \citenamefont {Marfatia}, \citenamefont {Mariani}, \citenamefont {Matveev}, \citenamefont {Mavromatos}, \citenamefont {Melchiorri}, \citenamefont {Meloni}, \citenamefont {Mena}, \citenamefont {Mention}, \citenamefont {Merle}, \citenamefont {Meroni}, \citenamefont {Mezzetto}, \citenamefont {Mills}, \citenamefont {Minic}, \citenamefont {Miramonti}, \citenamefont {Mohapatra}, \citenamefont {Mohapatra}, \citenamefont {Montanari}, \citenamefont {Mori}, \citenamefont {Mueller}, \citenamefont
  {Mumm}, \citenamefont {Muratova}, \citenamefont {Nelson}, \citenamefont {Nico}, \citenamefont {Noah}, \citenamefont {Nowak}, \citenamefont {Smirnov}, \citenamefont {Obolensky}, \citenamefont {Pakvasa}, \citenamefont {Palamara}, \citenamefont {Pallavicini}, \citenamefont {Pascoli}, \citenamefont {Patrizii}, \citenamefont {Pavlovic}, \citenamefont {Peres}, \citenamefont {Pessard}, \citenamefont {Pietropaolo}, \citenamefont {Pitt}, \citenamefont {Popovic}, \citenamefont {Pradler}, \citenamefont {Ranucci}, \citenamefont {Ray}, \citenamefont {Razzaque}, \citenamefont {Rebel}, \citenamefont {Robertson}, \citenamefont {Rodejohann}, \citenamefont {Rountree}, \citenamefont {Rubbia}, \citenamefont {Ruchayskiy}, \citenamefont {Sala}, \citenamefont {Scholberg}, \citenamefont {Schwetz}, \citenamefont {Shaevitz}, \citenamefont {Shaposhnikov}, \citenamefont {Shrock}, \citenamefont {Simone}, \citenamefont {Skorokhvatov}, \citenamefont {Sorel}, \citenamefont {Sousa}, \citenamefont {Spergel}, \citenamefont {Spitz},
  \citenamefont {Stanco}, \citenamefont {Stancu}, \citenamefont {Suzuki}, \citenamefont {Takeuchi}, \citenamefont {Tamborra}, \citenamefont {Tang}, \citenamefont {Testera}, \citenamefont {Tian}, \citenamefont {Tonazzo}, \citenamefont {Tunnell}, \citenamefont {{Van de Water}}, \citenamefont {Verde}, \citenamefont {Veretenkin}, \citenamefont {Vignoli}, \citenamefont {Vivier}, \citenamefont {Vogelaar}, \citenamefont {Wascko}, \citenamefont {Wilkerson}, \citenamefont {Winter}, \citenamefont {Wong}, \citenamefont {Yanagida}, \citenamefont {Yasuda}, \citenamefont {Yeh}, \citenamefont {Yermia}, \citenamefont {Yokley}, \citenamefont {Zeller}, \citenamefont {Zhan},\ and\ \citenamefont {Zhang}}]{Abazajian2012}%
  \BibitemOpen
  \bibfield  {author} {\bibinfo {author} {\bibfnamefont {K.~N.}\ \bibnamefont {Abazajian}}, \bibinfo {author} {\bibfnamefont {M.~A.}\ \bibnamefont {Acero}}, \bibinfo {author} {\bibfnamefont {S.~K.}\ \bibnamefont {Agarwalla}}, \bibinfo {author} {\bibfnamefont {A.~A.}\ \bibnamefont {Aguilar-Arevalo}}, \bibinfo {author} {\bibfnamefont {C.~H.}\ \bibnamefont {Albright}}, \bibinfo {author} {\bibfnamefont {S.}~\bibnamefont {Antusch}}, \bibinfo {author} {\bibfnamefont {C.~A.}\ \bibnamefont {Arguelles}}, \bibinfo {author} {\bibfnamefont {A.~B.}\ \bibnamefont {Balantekin}}, \bibinfo {author} {\bibfnamefont {G.}~\bibnamefont {Barenboim}}, \bibinfo {author} {\bibfnamefont {V.}~\bibnamefont {Barger}}, \bibinfo {author} {\bibfnamefont {P.}~\bibnamefont {Bernardini}}, \bibinfo {author} {\bibfnamefont {F.}~\bibnamefont {Bezrukov}}, \bibinfo {author} {\bibfnamefont {O.~E.}\ \bibnamefont {Bjaelde}}, \bibinfo {author} {\bibfnamefont {S.~A.}\ \bibnamefont {Bogacz}}, \bibinfo {author} {\bibfnamefont {N.~S.}\ \bibnamefont
  {Bowden}}, \bibinfo {author} {\bibfnamefont {A.}~\bibnamefont {Boyarsky}}, \bibinfo {author} {\bibfnamefont {A.}~\bibnamefont {Bravar}}, \bibinfo {author} {\bibfnamefont {D.~B.}\ \bibnamefont {Berguno}}, \bibinfo {author} {\bibfnamefont {S.~J.}\ \bibnamefont {Brice}}, \bibinfo {author} {\bibfnamefont {A.~D.}\ \bibnamefont {Bross}}, \bibinfo {author} {\bibfnamefont {B.}~\bibnamefont {Caccianiga}}, \bibinfo {author} {\bibfnamefont {F.}~\bibnamefont {Cavanna}}, \bibinfo {author} {\bibfnamefont {E.~J.}\ \bibnamefont {Chun}}, \bibinfo {author} {\bibfnamefont {B.~T.}\ \bibnamefont {Cleveland}}, \bibinfo {author} {\bibfnamefont {A.~P.}\ \bibnamefont {Collin}}, \bibinfo {author} {\bibfnamefont {P.}~\bibnamefont {Coloma}}, \bibinfo {author} {\bibfnamefont {J.~M.}\ \bibnamefont {Conrad}}, \bibinfo {author} {\bibfnamefont {M.}~\bibnamefont {Cribier}}, \bibinfo {author} {\bibfnamefont {A.~S.}\ \bibnamefont {Cucoanes}}, \bibinfo {author} {\bibfnamefont {J.~C.}\ \bibnamefont {D'Olivo}}, \bibinfo {author} {\bibfnamefont
  {S.}~\bibnamefont {Das}}, \bibinfo {author} {\bibfnamefont {A.}~\bibnamefont {de~Gouvea}}, \bibinfo {author} {\bibfnamefont {A.~V.}\ \bibnamefont {Derbin}}, \bibinfo {author} {\bibfnamefont {R.}~\bibnamefont {Dharmapalan}}, \bibinfo {author} {\bibfnamefont {J.~S.}\ \bibnamefont {Diaz}}, \bibinfo {author} {\bibfnamefont {X.~J.}\ \bibnamefont {Ding}}, \bibinfo {author} {\bibfnamefont {Z.}~\bibnamefont {Djurcic}}, \bibinfo {author} {\bibfnamefont {A.}~\bibnamefont {Donini}}, \bibinfo {author} {\bibfnamefont {D.}~\bibnamefont {Duchesneau}}, \bibinfo {author} {\bibfnamefont {H.}~\bibnamefont {Ejiri}}, \bibinfo {author} {\bibfnamefont {S.~R.}\ \bibnamefont {Elliott}}, \bibinfo {author} {\bibfnamefont {D.~J.}\ \bibnamefont {Ernst}}, \bibinfo {author} {\bibfnamefont {A.}~\bibnamefont {Esmaili}}, \bibinfo {author} {\bibfnamefont {J.~J.}\ \bibnamefont {Evans}}, \bibinfo {author} {\bibfnamefont {E.}~\bibnamefont {Fernandez-Martinez}}, \bibinfo {author} {\bibfnamefont {E.}~\bibnamefont {Figueroa-Feliciano}}, \bibinfo
  {author} {\bibfnamefont {B.~T.}\ \bibnamefont {Fleming}}, \bibinfo {author} {\bibfnamefont {J.~A.}\ \bibnamefont {Formaggio}}, \bibinfo {author} {\bibfnamefont {D.}~\bibnamefont {Franco}}, \bibinfo {author} {\bibfnamefont {J.}~\bibnamefont {Gaffiot}}, \bibinfo {author} {\bibfnamefont {R.}~\bibnamefont {Gandhi}}, \bibinfo {author} {\bibfnamefont {Y.}~\bibnamefont {Gao}}, \bibinfo {author} {\bibfnamefont {G.~T.}\ \bibnamefont {Garvey}}, \bibinfo {author} {\bibfnamefont {V.~N.}\ \bibnamefont {Gavrin}}, \bibinfo {author} {\bibfnamefont {P.}~\bibnamefont {Ghoshal}}, \bibinfo {author} {\bibfnamefont {D.}~\bibnamefont {Gibin}}, \bibinfo {author} {\bibfnamefont {C.}~\bibnamefont {Giunti}}, \bibinfo {author} {\bibfnamefont {S.~N.}\ \bibnamefont {Gninenko}}, \bibinfo {author} {\bibfnamefont {V.~V.}\ \bibnamefont {Gorbachev}}, \bibinfo {author} {\bibfnamefont {D.~S.}\ \bibnamefont {Gorbunov}}, \bibinfo {author} {\bibfnamefont {R.}~\bibnamefont {Guenette}}, \bibinfo {author} {\bibfnamefont {A.}~\bibnamefont
  {Guglielmi}}, \bibinfo {author} {\bibfnamefont {F.}~\bibnamefont {Halzen}}, \bibinfo {author} {\bibfnamefont {J.}~\bibnamefont {Hamann}}, \bibinfo {author} {\bibfnamefont {S.}~\bibnamefont {Hannestad}}, \bibinfo {author} {\bibfnamefont {W.}~\bibnamefont {Haxton}}, \bibinfo {author} {\bibfnamefont {K.~M.}\ \bibnamefont {Heeger}}, \bibinfo {author} {\bibfnamefont {R.}~\bibnamefont {Henning}}, \bibinfo {author} {\bibfnamefont {P.}~\bibnamefont {Hernandez}}, \bibinfo {author} {\bibfnamefont {P.}~\bibnamefont {Huber}}, \bibinfo {author} {\bibfnamefont {W.}~\bibnamefont {Huelsnitz}}, \bibinfo {author} {\bibfnamefont {A.}~\bibnamefont {Ianni}}, \bibinfo {author} {\bibfnamefont {T.~V.}\ \bibnamefont {Ibragimova}}, \bibinfo {author} {\bibfnamefont {Y.}~\bibnamefont {Karadzhov}}, \bibinfo {author} {\bibfnamefont {G.}~\bibnamefont {Karagiorgi}}, \bibinfo {author} {\bibfnamefont {G.}~\bibnamefont {Keefer}}, \bibinfo {author} {\bibfnamefont {Y.~D.}\ \bibnamefont {Kim}}, \bibinfo {author} {\bibfnamefont {J.}~\bibnamefont
  {Kopp}}, \bibinfo {author} {\bibfnamefont {V.~N.}\ \bibnamefont {Kornoukhov}}, \bibinfo {author} {\bibfnamefont {A.}~\bibnamefont {Kusenko}}, \bibinfo {author} {\bibfnamefont {P.}~\bibnamefont {Kyberd}}, \bibinfo {author} {\bibfnamefont {P.}~\bibnamefont {Langacker}}, \bibinfo {author} {\bibfnamefont {T.}~\bibnamefont {Lasserre}}, \bibinfo {author} {\bibfnamefont {M.}~\bibnamefont {Laveder}}, \bibinfo {author} {\bibfnamefont {A.}~\bibnamefont {Letourneau}}, \bibinfo {author} {\bibfnamefont {D.}~\bibnamefont {Lhuillier}}, \bibinfo {author} {\bibfnamefont {Y.~F.}\ \bibnamefont {Li}}, \bibinfo {author} {\bibfnamefont {M.}~\bibnamefont {Lindner}}, \bibinfo {author} {\bibfnamefont {J.~M.}\ \bibnamefont {Link}}, \bibinfo {author} {\bibfnamefont {B.~L.}\ \bibnamefont {Littlejohn}}, \bibinfo {author} {\bibfnamefont {P.}~\bibnamefont {Lombardi}}, \bibinfo {author} {\bibfnamefont {K.}~\bibnamefont {Long}}, \bibinfo {author} {\bibfnamefont {J.}~\bibnamefont {Lopez-Pavon}}, \bibinfo {author} {\bibfnamefont {W.~C.}\
  \bibnamefont {Louis}}, \bibinfo {author} {\bibfnamefont {L.}~\bibnamefont {Ludhova}}, \bibinfo {author} {\bibfnamefont {J.~D.}\ \bibnamefont {Lykken}}, \bibinfo {author} {\bibfnamefont {P.~A.~N.}\ \bibnamefont {Machado}}, \bibinfo {author} {\bibfnamefont {M.}~\bibnamefont {Maltoni}}, \bibinfo {author} {\bibfnamefont {W.~A.}\ \bibnamefont {Mann}}, \bibinfo {author} {\bibfnamefont {D.}~\bibnamefont {Marfatia}}, \bibinfo {author} {\bibfnamefont {C.}~\bibnamefont {Mariani}}, \bibinfo {author} {\bibfnamefont {V.~A.}\ \bibnamefont {Matveev}}, \bibinfo {author} {\bibfnamefont {N.~E.}\ \bibnamefont {Mavromatos}}, \bibinfo {author} {\bibfnamefont {A.}~\bibnamefont {Melchiorri}}, \bibinfo {author} {\bibfnamefont {D.}~\bibnamefont {Meloni}}, \bibinfo {author} {\bibfnamefont {O.}~\bibnamefont {Mena}}, \bibinfo {author} {\bibfnamefont {G.}~\bibnamefont {Mention}}, \bibinfo {author} {\bibfnamefont {A.}~\bibnamefont {Merle}}, \bibinfo {author} {\bibfnamefont {E.}~\bibnamefont {Meroni}}, \bibinfo {author} {\bibfnamefont
  {M.}~\bibnamefont {Mezzetto}}, \bibinfo {author} {\bibfnamefont {G.~B.}\ \bibnamefont {Mills}}, \bibinfo {author} {\bibfnamefont {D.}~\bibnamefont {Minic}}, \bibinfo {author} {\bibfnamefont {L.}~\bibnamefont {Miramonti}}, \bibinfo {author} {\bibfnamefont {D.}~\bibnamefont {Mohapatra}}, \bibinfo {author} {\bibfnamefont {R.~N.}\ \bibnamefont {Mohapatra}}, \bibinfo {author} {\bibfnamefont {C.}~\bibnamefont {Montanari}}, \bibinfo {author} {\bibfnamefont {Y.}~\bibnamefont {Mori}}, \bibinfo {author} {\bibfnamefont {T.~A.}\ \bibnamefont {Mueller}}, \bibinfo {author} {\bibfnamefont {H.~P.}\ \bibnamefont {Mumm}}, \bibinfo {author} {\bibfnamefont {V.}~\bibnamefont {Muratova}}, \bibinfo {author} {\bibfnamefont {A.~E.}\ \bibnamefont {Nelson}}, \bibinfo {author} {\bibfnamefont {J.~S.}\ \bibnamefont {Nico}}, \bibinfo {author} {\bibfnamefont {E.}~\bibnamefont {Noah}}, \bibinfo {author} {\bibfnamefont {J.}~\bibnamefont {Nowak}}, \bibinfo {author} {\bibfnamefont {O.~Y.}\ \bibnamefont {Smirnov}}, \bibinfo {author}
  {\bibfnamefont {M.}~\bibnamefont {Obolensky}}, \bibinfo {author} {\bibfnamefont {S.}~\bibnamefont {Pakvasa}}, \bibinfo {author} {\bibfnamefont {O.}~\bibnamefont {Palamara}}, \bibinfo {author} {\bibfnamefont {M.}~\bibnamefont {Pallavicini}}, \bibinfo {author} {\bibfnamefont {S.}~\bibnamefont {Pascoli}}, \bibinfo {author} {\bibfnamefont {L.}~\bibnamefont {Patrizii}}, \bibinfo {author} {\bibfnamefont {Z.}~\bibnamefont {Pavlovic}}, \bibinfo {author} {\bibfnamefont {O.~L.~G.}\ \bibnamefont {Peres}}, \bibinfo {author} {\bibfnamefont {H.}~\bibnamefont {Pessard}}, \bibinfo {author} {\bibfnamefont {F.}~\bibnamefont {Pietropaolo}}, \bibinfo {author} {\bibfnamefont {M.~L.}\ \bibnamefont {Pitt}}, \bibinfo {author} {\bibfnamefont {M.}~\bibnamefont {Popovic}}, \bibinfo {author} {\bibfnamefont {J.}~\bibnamefont {Pradler}}, \bibinfo {author} {\bibfnamefont {G.}~\bibnamefont {Ranucci}}, \bibinfo {author} {\bibfnamefont {H.}~\bibnamefont {Ray}}, \bibinfo {author} {\bibfnamefont {S.}~\bibnamefont {Razzaque}}, \bibinfo
  {author} {\bibfnamefont {B.}~\bibnamefont {Rebel}}, \bibinfo {author} {\bibfnamefont {R.~G.~H.}\ \bibnamefont {Robertson}}, \bibinfo {author} {\bibfnamefont {W.}~\bibnamefont {Rodejohann}}, \bibinfo {author} {\bibfnamefont {S.~D.}\ \bibnamefont {Rountree}}, \bibinfo {author} {\bibfnamefont {C.}~\bibnamefont {Rubbia}}, \bibinfo {author} {\bibfnamefont {O.}~\bibnamefont {Ruchayskiy}}, \bibinfo {author} {\bibfnamefont {P.~R.}\ \bibnamefont {Sala}}, \bibinfo {author} {\bibfnamefont {K.}~\bibnamefont {Scholberg}}, \bibinfo {author} {\bibfnamefont {T.}~\bibnamefont {Schwetz}}, \bibinfo {author} {\bibfnamefont {M.~H.}\ \bibnamefont {Shaevitz}}, \bibinfo {author} {\bibfnamefont {M.}~\bibnamefont {Shaposhnikov}}, \bibinfo {author} {\bibfnamefont {R.}~\bibnamefont {Shrock}}, \bibinfo {author} {\bibfnamefont {S.}~\bibnamefont {Simone}}, \bibinfo {author} {\bibfnamefont {M.}~\bibnamefont {Skorokhvatov}}, \bibinfo {author} {\bibfnamefont {M.}~\bibnamefont {Sorel}}, \bibinfo {author} {\bibfnamefont {A.}~\bibnamefont
  {Sousa}}, \bibinfo {author} {\bibfnamefont {D.~N.}\ \bibnamefont {Spergel}}, \bibinfo {author} {\bibfnamefont {J.}~\bibnamefont {Spitz}}, \bibinfo {author} {\bibfnamefont {L.}~\bibnamefont {Stanco}}, \bibinfo {author} {\bibfnamefont {I.}~\bibnamefont {Stancu}}, \bibinfo {author} {\bibfnamefont {A.}~\bibnamefont {Suzuki}}, \bibinfo {author} {\bibfnamefont {T.}~\bibnamefont {Takeuchi}}, \bibinfo {author} {\bibfnamefont {I.}~\bibnamefont {Tamborra}}, \bibinfo {author} {\bibfnamefont {J.}~\bibnamefont {Tang}}, \bibinfo {author} {\bibfnamefont {G.}~\bibnamefont {Testera}}, \bibinfo {author} {\bibfnamefont {X.~C.}\ \bibnamefont {Tian}}, \bibinfo {author} {\bibfnamefont {A.}~\bibnamefont {Tonazzo}}, \bibinfo {author} {\bibfnamefont {C.~D.}\ \bibnamefont {Tunnell}}, \bibinfo {author} {\bibfnamefont {R.~G.}\ \bibnamefont {{Van de Water}}}, \bibinfo {author} {\bibfnamefont {L.}~\bibnamefont {Verde}}, \bibinfo {author} {\bibfnamefont {E.~P.}\ \bibnamefont {Veretenkin}}, \bibinfo {author} {\bibfnamefont
  {C.}~\bibnamefont {Vignoli}}, \bibinfo {author} {\bibfnamefont {M.}~\bibnamefont {Vivier}}, \bibinfo {author} {\bibfnamefont {R.~B.}\ \bibnamefont {Vogelaar}}, \bibinfo {author} {\bibfnamefont {M.~O.}\ \bibnamefont {Wascko}}, \bibinfo {author} {\bibfnamefont {J.~F.}\ \bibnamefont {Wilkerson}}, \bibinfo {author} {\bibfnamefont {W.}~\bibnamefont {Winter}}, \bibinfo {author} {\bibfnamefont {Y.~Y.~Y.}\ \bibnamefont {Wong}}, \bibinfo {author} {\bibfnamefont {T.~T.}\ \bibnamefont {Yanagida}}, \bibinfo {author} {\bibfnamefont {O.}~\bibnamefont {Yasuda}}, \bibinfo {author} {\bibfnamefont {M.}~\bibnamefont {Yeh}}, \bibinfo {author} {\bibfnamefont {F.}~\bibnamefont {Yermia}}, \bibinfo {author} {\bibfnamefont {Z.~W.}\ \bibnamefont {Yokley}}, \bibinfo {author} {\bibfnamefont {G.~P.}\ \bibnamefont {Zeller}}, \bibinfo {author} {\bibfnamefont {L.}~\bibnamefont {Zhan}}, \ and\ \bibinfo {author} {\bibfnamefont {H.}~\bibnamefont {Zhang}},\ }\href {http://arxiv.org/abs/1204.5379} {\  (\bibinfo {year} {2012})},\ \Eprint
  {http://arxiv.org/abs/1204.5379} {arXiv:1204.5379} \BibitemShut {NoStop}%
\bibitem [{\citenamefont {Acero}\ \emph {et~al.}(2022)\citenamefont {Acero}, \citenamefont {Arg{\"{u}}elles}, \citenamefont {Hostert}, \citenamefont {Kalra}, \citenamefont {Karagiorgi}, \citenamefont {Kelly}, \citenamefont {Littlejohn}, \citenamefont {Machado}, \citenamefont {Pettus}, \citenamefont {Toups}, \citenamefont {Ross-Lonergan}, \citenamefont {Sousa}, \citenamefont {Surukuchi}, \citenamefont {Wong}, \citenamefont {Abdallah}, \citenamefont {Abdullahi}, \citenamefont {Akutsu}, \citenamefont {Alvarez-Ruso}, \citenamefont {Alves}, \citenamefont {Aurisano}, \citenamefont {Balantekin}, \citenamefont {Berryman}, \citenamefont {Bert{\'{o}}lez-Mart{\'{i}}nez}, \citenamefont {Brunner}, \citenamefont {Blennow}, \citenamefont {Bolognesi}, \citenamefont {Borusinski}, \citenamefont {Cianci}, \citenamefont {Collin}, \citenamefont {Conrad}, \citenamefont {Crow}, \citenamefont {Denton}, \citenamefont {Duvall}, \citenamefont {Fern{\'{a}}ndez-Martinez}, \citenamefont {Fong}, \citenamefont {Foppiani}, \citenamefont
  {Forero}, \citenamefont {Friend}, \citenamefont {Garc{\'{i}}a-Soto}, \citenamefont {Giganti}, \citenamefont {Giunti}, \citenamefont {Gandhi}, \citenamefont {Ghosh}, \citenamefont {Hardin}, \citenamefont {Heeger}, \citenamefont {Ishitsuka}, \citenamefont {Izmaylov}, \citenamefont {Jones}, \citenamefont {Jordan}, \citenamefont {Kamp}, \citenamefont {Katori}, \citenamefont {Kim}, \citenamefont {Koerner}, \citenamefont {Lamoureux}, \citenamefont {Lasserre}, \citenamefont {Leach}, \citenamefont {Learned}, \citenamefont {Li}, \citenamefont {Link}, \citenamefont {Louis}, \citenamefont {Mahn}, \citenamefont {Meyers}, \citenamefont {Maricic}, \citenamefont {Marko}, \citenamefont {Maruyama}, \citenamefont {Mertens}, \citenamefont {Minakata}, \citenamefont {Mocioiu}, \citenamefont {Mooney}, \citenamefont {Moulai}, \citenamefont {Nunokawa}, \citenamefont {Ochoa-Ricoux}, \citenamefont {Oh}, \citenamefont {Ohlsson}, \citenamefont {P{\"{a}}s}, \citenamefont {Pershey}, \citenamefont {Robertson}, \citenamefont
  {Rosauro-Alcaraz}, \citenamefont {Rott}, \citenamefont {Roy}, \citenamefont {Salvado}, \citenamefont {Scott}, \citenamefont {Seo}, \citenamefont {Shaevitz}, \citenamefont {Smiley}, \citenamefont {Spitz}, \citenamefont {Stachurska}, \citenamefont {Thakore}, \citenamefont {Ternes}, \citenamefont {Thompson}, \citenamefont {Tseng}, \citenamefont {Vogelaar}, \citenamefont {Weiss}, \citenamefont {Wendell}, \citenamefont {Wright}, \citenamefont {Xin}, \citenamefont {Yang}, \citenamefont {Yoo}, \citenamefont {Zennamo}, \citenamefont {Zettlemoyer}, \citenamefont {Zornoza}, \citenamefont {Ahmad}, \citenamefont {Basto-Gonzalez}, \citenamefont {Bowden}, \citenamefont {Ca{\~{n}}as}, \citenamefont {Caratelli}, \citenamefont {Chang}, \citenamefont {Chen}, \citenamefont {Classen}, \citenamefont {Convery}, \citenamefont {Davies}, \citenamefont {Dennis}, \citenamefont {Djurcic}, \citenamefont {Dorrill}, \citenamefont {Du}, \citenamefont {Evans}, \citenamefont {Fahrendholz}, \citenamefont {Formaggio}, \citenamefont {Foust},
  \citenamefont {Gatti}, \citenamefont {Garcia-Gamez}, \citenamefont {Gariazzo}, \citenamefont {Gehrlein}, \citenamefont {Grant}, \citenamefont {Gomes}, \citenamefont {Hansell}, \citenamefont {Halzen}, \citenamefont {Ho}, \citenamefont {Zink}, \citenamefont {Jones}, \citenamefont {Kunkle}, \citenamefont {Li}, \citenamefont {Li}, \citenamefont {Luo}, \citenamefont {Malyshkin}, \citenamefont {Massaro}, \citenamefont {Mastbaum}, \citenamefont {Mohanta}, \citenamefont {Mumm}, \citenamefont {Nebot-Guinot}, \citenamefont {Neilson}, \citenamefont {Ni}, \citenamefont {Nieves}, \citenamefont {Gann}, \citenamefont {Pandey}, \citenamefont {Pascoli}, \citenamefont {Qian}, \citenamefont {Rajaoalisoa}, \citenamefont {Roca}, \citenamefont {Roskovec}, \citenamefont {Saul-Sala}, \citenamefont {Salda{\~{n}}a}, \citenamefont {Scholberg}, \citenamefont {Shakya}, \citenamefont {Slocum}, \citenamefont {Snider}, \citenamefont {Steiger}, \citenamefont {Steklain}, \citenamefont {Stock}, \citenamefont {Sutanto}, \citenamefont
  {Takhistov}, \citenamefont {Tsai}, \citenamefont {Tsai}, \citenamefont {Venegas-Vargas}, \citenamefont {Wallbank}, \citenamefont {Wang}, \citenamefont {Weatherly}, \citenamefont {Westerdale}, \citenamefont {Worcester}, \citenamefont {Wu}, \citenamefont {Yang},\ and\ \citenamefont {Zamorano}}]{Acero2022}%
  \BibitemOpen
  \bibfield  {author} {\bibinfo {author} {\bibfnamefont {M.~A.}\ \bibnamefont {Acero}}, \bibinfo {author} {\bibfnamefont {C.~A.}\ \bibnamefont {Arg{\"{u}}elles}}, \bibinfo {author} {\bibfnamefont {M.}~\bibnamefont {Hostert}}, \bibinfo {author} {\bibfnamefont {D.}~\bibnamefont {Kalra}}, \bibinfo {author} {\bibfnamefont {G.}~\bibnamefont {Karagiorgi}}, \bibinfo {author} {\bibfnamefont {K.~J.}\ \bibnamefont {Kelly}}, \bibinfo {author} {\bibfnamefont {B.}~\bibnamefont {Littlejohn}}, \bibinfo {author} {\bibfnamefont {P.}~\bibnamefont {Machado}}, \bibinfo {author} {\bibfnamefont {W.}~\bibnamefont {Pettus}}, \bibinfo {author} {\bibfnamefont {M.}~\bibnamefont {Toups}}, \bibinfo {author} {\bibfnamefont {M.}~\bibnamefont {Ross-Lonergan}}, \bibinfo {author} {\bibfnamefont {A.}~\bibnamefont {Sousa}}, \bibinfo {author} {\bibfnamefont {P.~T.}\ \bibnamefont {Surukuchi}}, \bibinfo {author} {\bibfnamefont {Y.~Y.~Y.}\ \bibnamefont {Wong}}, \bibinfo {author} {\bibfnamefont {W.}~\bibnamefont {Abdallah}}, \bibinfo {author}
  {\bibfnamefont {A.~M.}\ \bibnamefont {Abdullahi}}, \bibinfo {author} {\bibfnamefont {R.}~\bibnamefont {Akutsu}}, \bibinfo {author} {\bibfnamefont {L.}~\bibnamefont {Alvarez-Ruso}}, \bibinfo {author} {\bibfnamefont {D.~S.~M.}\ \bibnamefont {Alves}}, \bibinfo {author} {\bibfnamefont {A.}~\bibnamefont {Aurisano}}, \bibinfo {author} {\bibfnamefont {A.~B.}\ \bibnamefont {Balantekin}}, \bibinfo {author} {\bibfnamefont {J.~M.}\ \bibnamefont {Berryman}}, \bibinfo {author} {\bibfnamefont {T.}~\bibnamefont {Bert{\'{o}}lez-Mart{\'{i}}nez}}, \bibinfo {author} {\bibfnamefont {J.}~\bibnamefont {Brunner}}, \bibinfo {author} {\bibfnamefont {M.}~\bibnamefont {Blennow}}, \bibinfo {author} {\bibfnamefont {S.}~\bibnamefont {Bolognesi}}, \bibinfo {author} {\bibfnamefont {M.}~\bibnamefont {Borusinski}}, \bibinfo {author} {\bibfnamefont {D.}~\bibnamefont {Cianci}}, \bibinfo {author} {\bibfnamefont {G.}~\bibnamefont {Collin}}, \bibinfo {author} {\bibfnamefont {J.~M.}\ \bibnamefont {Conrad}}, \bibinfo {author} {\bibfnamefont
  {B.}~\bibnamefont {Crow}}, \bibinfo {author} {\bibfnamefont {P.~B.}\ \bibnamefont {Denton}}, \bibinfo {author} {\bibfnamefont {M.}~\bibnamefont {Duvall}}, \bibinfo {author} {\bibfnamefont {E.}~\bibnamefont {Fern{\'{a}}ndez-Martinez}}, \bibinfo {author} {\bibfnamefont {C.~S.}\ \bibnamefont {Fong}}, \bibinfo {author} {\bibfnamefont {N.}~\bibnamefont {Foppiani}}, \bibinfo {author} {\bibfnamefont {D.~V.}\ \bibnamefont {Forero}}, \bibinfo {author} {\bibfnamefont {M.}~\bibnamefont {Friend}}, \bibinfo {author} {\bibfnamefont {A.}~\bibnamefont {Garc{\'{i}}a-Soto}}, \bibinfo {author} {\bibfnamefont {C.}~\bibnamefont {Giganti}}, \bibinfo {author} {\bibfnamefont {C.}~\bibnamefont {Giunti}}, \bibinfo {author} {\bibfnamefont {R.}~\bibnamefont {Gandhi}}, \bibinfo {author} {\bibfnamefont {M.}~\bibnamefont {Ghosh}}, \bibinfo {author} {\bibfnamefont {J.}~\bibnamefont {Hardin}}, \bibinfo {author} {\bibfnamefont {K.~M.}\ \bibnamefont {Heeger}}, \bibinfo {author} {\bibfnamefont {M.}~\bibnamefont {Ishitsuka}}, \bibinfo {author}
  {\bibfnamefont {A.}~\bibnamefont {Izmaylov}}, \bibinfo {author} {\bibfnamefont {B.~J.~P.}\ \bibnamefont {Jones}}, \bibinfo {author} {\bibfnamefont {J.~R.}\ \bibnamefont {Jordan}}, \bibinfo {author} {\bibfnamefont {N.~W.}\ \bibnamefont {Kamp}}, \bibinfo {author} {\bibfnamefont {T.}~\bibnamefont {Katori}}, \bibinfo {author} {\bibfnamefont {S.~B.}\ \bibnamefont {Kim}}, \bibinfo {author} {\bibfnamefont {L.~W.}\ \bibnamefont {Koerner}}, \bibinfo {author} {\bibfnamefont {M.}~\bibnamefont {Lamoureux}}, \bibinfo {author} {\bibfnamefont {T.}~\bibnamefont {Lasserre}}, \bibinfo {author} {\bibfnamefont {K.~G.}\ \bibnamefont {Leach}}, \bibinfo {author} {\bibfnamefont {J.}~\bibnamefont {Learned}}, \bibinfo {author} {\bibfnamefont {Y.~F.}\ \bibnamefont {Li}}, \bibinfo {author} {\bibfnamefont {J.~M.}\ \bibnamefont {Link}}, \bibinfo {author} {\bibfnamefont {W.~C.}\ \bibnamefont {Louis}}, \bibinfo {author} {\bibfnamefont {K.}~\bibnamefont {Mahn}}, \bibinfo {author} {\bibfnamefont {P.~D.}\ \bibnamefont {Meyers}}, \bibinfo
  {author} {\bibfnamefont {J.}~\bibnamefont {Maricic}}, \bibinfo {author} {\bibfnamefont {D.}~\bibnamefont {Marko}}, \bibinfo {author} {\bibfnamefont {T.}~\bibnamefont {Maruyama}}, \bibinfo {author} {\bibfnamefont {S.}~\bibnamefont {Mertens}}, \bibinfo {author} {\bibfnamefont {H.}~\bibnamefont {Minakata}}, \bibinfo {author} {\bibfnamefont {I.}~\bibnamefont {Mocioiu}}, \bibinfo {author} {\bibfnamefont {M.}~\bibnamefont {Mooney}}, \bibinfo {author} {\bibfnamefont {M.~H.}\ \bibnamefont {Moulai}}, \bibinfo {author} {\bibfnamefont {H.}~\bibnamefont {Nunokawa}}, \bibinfo {author} {\bibfnamefont {J.~P.}\ \bibnamefont {Ochoa-Ricoux}}, \bibinfo {author} {\bibfnamefont {Y.~M.}\ \bibnamefont {Oh}}, \bibinfo {author} {\bibfnamefont {T.}~\bibnamefont {Ohlsson}}, \bibinfo {author} {\bibfnamefont {H.}~\bibnamefont {P{\"{a}}s}}, \bibinfo {author} {\bibfnamefont {D.}~\bibnamefont {Pershey}}, \bibinfo {author} {\bibfnamefont {R.~G.~H.}\ \bibnamefont {Robertson}}, \bibinfo {author} {\bibfnamefont {S.}~\bibnamefont
  {Rosauro-Alcaraz}}, \bibinfo {author} {\bibfnamefont {C.}~\bibnamefont {Rott}}, \bibinfo {author} {\bibfnamefont {S.}~\bibnamefont {Roy}}, \bibinfo {author} {\bibfnamefont {J.}~\bibnamefont {Salvado}}, \bibinfo {author} {\bibfnamefont {M.}~\bibnamefont {Scott}}, \bibinfo {author} {\bibfnamefont {S.~H.}\ \bibnamefont {Seo}}, \bibinfo {author} {\bibfnamefont {M.~H.}\ \bibnamefont {Shaevitz}}, \bibinfo {author} {\bibfnamefont {M.}~\bibnamefont {Smiley}}, \bibinfo {author} {\bibfnamefont {J.}~\bibnamefont {Spitz}}, \bibinfo {author} {\bibfnamefont {J.}~\bibnamefont {Stachurska}}, \bibinfo {author} {\bibfnamefont {T.}~\bibnamefont {Thakore}}, \bibinfo {author} {\bibfnamefont {C.~A.}\ \bibnamefont {Ternes}}, \bibinfo {author} {\bibfnamefont {A.}~\bibnamefont {Thompson}}, \bibinfo {author} {\bibfnamefont {S.}~\bibnamefont {Tseng}}, \bibinfo {author} {\bibfnamefont {B.}~\bibnamefont {Vogelaar}}, \bibinfo {author} {\bibfnamefont {T.}~\bibnamefont {Weiss}}, \bibinfo {author} {\bibfnamefont {R.~A.}\ \bibnamefont
  {Wendell}}, \bibinfo {author} {\bibfnamefont {T.}~\bibnamefont {Wright}}, \bibinfo {author} {\bibfnamefont {Z.}~\bibnamefont {Xin}}, \bibinfo {author} {\bibfnamefont {B.~S.}\ \bibnamefont {Yang}}, \bibinfo {author} {\bibfnamefont {J.}~\bibnamefont {Yoo}}, \bibinfo {author} {\bibfnamefont {J.}~\bibnamefont {Zennamo}}, \bibinfo {author} {\bibfnamefont {J.}~\bibnamefont {Zettlemoyer}}, \bibinfo {author} {\bibfnamefont {J.~D.}\ \bibnamefont {Zornoza}}, \bibinfo {author} {\bibfnamefont {S.}~\bibnamefont {Ahmad}}, \bibinfo {author} {\bibfnamefont {V.~S.}\ \bibnamefont {Basto-Gonzalez}}, \bibinfo {author} {\bibfnamefont {N.~S.}\ \bibnamefont {Bowden}}, \bibinfo {author} {\bibfnamefont {B.~C.}\ \bibnamefont {Ca{\~{n}}as}}, \bibinfo {author} {\bibfnamefont {D.}~\bibnamefont {Caratelli}}, \bibinfo {author} {\bibfnamefont {C.~V.}\ \bibnamefont {Chang}}, \bibinfo {author} {\bibfnamefont {C.}~\bibnamefont {Chen}}, \bibinfo {author} {\bibfnamefont {T.}~\bibnamefont {Classen}}, \bibinfo {author} {\bibfnamefont
  {M.}~\bibnamefont {Convery}}, \bibinfo {author} {\bibfnamefont {G.~S.}\ \bibnamefont {Davies}}, \bibinfo {author} {\bibfnamefont {S.~R.}\ \bibnamefont {Dennis}}, \bibinfo {author} {\bibfnamefont {Z.}~\bibnamefont {Djurcic}}, \bibinfo {author} {\bibfnamefont {R.}~\bibnamefont {Dorrill}}, \bibinfo {author} {\bibfnamefont {Y.}~\bibnamefont {Du}}, \bibinfo {author} {\bibfnamefont {J.~J.}\ \bibnamefont {Evans}}, \bibinfo {author} {\bibfnamefont {U.}~\bibnamefont {Fahrendholz}}, \bibinfo {author} {\bibfnamefont {J.~A.}\ \bibnamefont {Formaggio}}, \bibinfo {author} {\bibfnamefont {B.~T.}\ \bibnamefont {Foust}}, \bibinfo {author} {\bibfnamefont {H.~F.}\ \bibnamefont {Gatti}}, \bibinfo {author} {\bibfnamefont {D.}~\bibnamefont {Garcia-Gamez}}, \bibinfo {author} {\bibfnamefont {S.}~\bibnamefont {Gariazzo}}, \bibinfo {author} {\bibfnamefont {J.}~\bibnamefont {Gehrlein}}, \bibinfo {author} {\bibfnamefont {C.}~\bibnamefont {Grant}}, \bibinfo {author} {\bibfnamefont {R.~A.}\ \bibnamefont {Gomes}}, \bibinfo {author}
  {\bibfnamefont {A.~B.}\ \bibnamefont {Hansell}}, \bibinfo {author} {\bibfnamefont {F.}~\bibnamefont {Halzen}}, \bibinfo {author} {\bibfnamefont {S.}~\bibnamefont {Ho}}, \bibinfo {author} {\bibfnamefont {J.~H.}\ \bibnamefont {Zink}}, \bibinfo {author} {\bibfnamefont {R.~S.}\ \bibnamefont {Jones}}, \bibinfo {author} {\bibfnamefont {P.}~\bibnamefont {Kunkle}}, \bibinfo {author} {\bibfnamefont {J.~Y.}\ \bibnamefont {Li}}, \bibinfo {author} {\bibfnamefont {S.~C.}\ \bibnamefont {Li}}, \bibinfo {author} {\bibfnamefont {X.}~\bibnamefont {Luo}}, \bibinfo {author} {\bibfnamefont {Y.}~\bibnamefont {Malyshkin}}, \bibinfo {author} {\bibfnamefont {D.}~\bibnamefont {Massaro}}, \bibinfo {author} {\bibfnamefont {A.}~\bibnamefont {Mastbaum}}, \bibinfo {author} {\bibfnamefont {R.}~\bibnamefont {Mohanta}}, \bibinfo {author} {\bibfnamefont {H.~P.}\ \bibnamefont {Mumm}}, \bibinfo {author} {\bibfnamefont {M.}~\bibnamefont {Nebot-Guinot}}, \bibinfo {author} {\bibfnamefont {R.}~\bibnamefont {Neilson}}, \bibinfo {author}
  {\bibfnamefont {K.}~\bibnamefont {Ni}}, \bibinfo {author} {\bibfnamefont {J.}~\bibnamefont {Nieves}}, \bibinfo {author} {\bibfnamefont {G.~D.~O.}\ \bibnamefont {Gann}}, \bibinfo {author} {\bibfnamefont {V.}~\bibnamefont {Pandey}}, \bibinfo {author} {\bibfnamefont {S.}~\bibnamefont {Pascoli}}, \bibinfo {author} {\bibfnamefont {X.}~\bibnamefont {Qian}}, \bibinfo {author} {\bibfnamefont {M.}~\bibnamefont {Rajaoalisoa}}, \bibinfo {author} {\bibfnamefont {C.}~\bibnamefont {Roca}}, \bibinfo {author} {\bibfnamefont {B.}~\bibnamefont {Roskovec}}, \bibinfo {author} {\bibfnamefont {E.}~\bibnamefont {Saul-Sala}}, \bibinfo {author} {\bibfnamefont {L.}~\bibnamefont {Salda{\~{n}}a}}, \bibinfo {author} {\bibfnamefont {K.}~\bibnamefont {Scholberg}}, \bibinfo {author} {\bibfnamefont {B.}~\bibnamefont {Shakya}}, \bibinfo {author} {\bibfnamefont {P.~L.}\ \bibnamefont {Slocum}}, \bibinfo {author} {\bibfnamefont {E.~L.}\ \bibnamefont {Snider}}, \bibinfo {author} {\bibfnamefont {H.~T.~J.}\ \bibnamefont {Steiger}}, \bibinfo
  {author} {\bibfnamefont {A.~F.}\ \bibnamefont {Steklain}}, \bibinfo {author} {\bibfnamefont {M.~R.}\ \bibnamefont {Stock}}, \bibinfo {author} {\bibfnamefont {F.}~\bibnamefont {Sutanto}}, \bibinfo {author} {\bibfnamefont {V.}~\bibnamefont {Takhistov}}, \bibinfo {author} {\bibfnamefont {Y.~D.}\ \bibnamefont {Tsai}}, \bibinfo {author} {\bibfnamefont {Y.~T.}\ \bibnamefont {Tsai}}, \bibinfo {author} {\bibfnamefont {D.}~\bibnamefont {Venegas-Vargas}}, \bibinfo {author} {\bibfnamefont {M.}~\bibnamefont {Wallbank}}, \bibinfo {author} {\bibfnamefont {E.}~\bibnamefont {Wang}}, \bibinfo {author} {\bibfnamefont {P.}~\bibnamefont {Weatherly}}, \bibinfo {author} {\bibfnamefont {S.}~\bibnamefont {Westerdale}}, \bibinfo {author} {\bibfnamefont {E.}~\bibnamefont {Worcester}}, \bibinfo {author} {\bibfnamefont {W.}~\bibnamefont {Wu}}, \bibinfo {author} {\bibfnamefont {G.}~\bibnamefont {Yang}}, \ and\ \bibinfo {author} {\bibfnamefont {B.}~\bibnamefont {Zamorano}},\ }\href {http://arxiv.org/abs/2203.07323} {\  (\bibinfo {year}
  {2022})},\ \Eprint {http://arxiv.org/abs/2203.07323} {arXiv:2203.07323} \BibitemShut {NoStop}%
\bibitem [{\citenamefont {Giunti}\ and\ \citenamefont {Lasserre}(2019)}]{Giunti2019}%
  \BibitemOpen
  \bibfield  {author} {\bibinfo {author} {\bibfnamefont {C.}~\bibnamefont {Giunti}}\ and\ \bibinfo {author} {\bibfnamefont {T.}~\bibnamefont {Lasserre}},\ }\href {\doibase 10.1146/annurev-nucl-101918-023755} {\bibfield  {journal} {\bibinfo  {journal} {Annual Review of Nuclear and Particle Science}\ }\textbf {\bibinfo {volume} {69}},\ \bibinfo {pages} {163} (\bibinfo {year} {2019})}\BibitemShut {NoStop}%
\bibitem [{\citenamefont {Qian}\ and\ \citenamefont {Peng}(2019)}]{Qian2019}%
  \BibitemOpen
  \bibfield  {author} {\bibinfo {author} {\bibfnamefont {X.}~\bibnamefont {Qian}}\ and\ \bibinfo {author} {\bibfnamefont {J.-C.}\ \bibnamefont {Peng}},\ }\href {\doibase 10.1088/1361-6633/aae881} {\bibfield  {journal} {\bibinfo  {journal} {Reports on Progress in Physics}\ }\textbf {\bibinfo {volume} {82}},\ \bibinfo {pages} {036201} (\bibinfo {year} {2019})}\BibitemShut {NoStop}%
\bibitem [{\citenamefont {Ang}\ \emph {et~al.}(2023)\citenamefont {Ang}, \citenamefont {Lee},\ and\ \citenamefont {Prasad}}]{Ang2023}%
  \BibitemOpen
  \bibfield  {author} {\bibinfo {author} {\bibfnamefont {W.~E.}\ \bibnamefont {Ang}}, \bibinfo {author} {\bibfnamefont {S.}~\bibnamefont {Lee}}, \ and\ \bibinfo {author} {\bibfnamefont {S.}~\bibnamefont {Prasad}},\ }\href {\doibase 10.1080/00295639.2022.2103348} {\bibfield  {journal} {\bibinfo  {journal} {Nuclear Science and Engineering}\ }\textbf {\bibinfo {volume} {197}},\ \bibinfo {pages} {443} (\bibinfo {year} {2023})}\BibitemShut {NoStop}%
\bibitem [{\citenamefont {Estienne}\ \emph {et~al.}(2019)\citenamefont {Estienne}, \citenamefont {Fallot}, \citenamefont {Algora}, \citenamefont {Briz-Monago}, \citenamefont {Bui}, \citenamefont {Cormon}, \citenamefont {Gelletly}, \citenamefont {Giot}, \citenamefont {Guadilla}, \citenamefont {Jordan}, \citenamefont {Meur}, \citenamefont {Porta}, \citenamefont {Rice}, \citenamefont {Rubio}, \citenamefont {Tain}, \citenamefont {Valencia},\ and\ \citenamefont {Zakari-Issoufou}}]{Estienne2019}%
  \BibitemOpen
  \bibfield  {author} {\bibinfo {author} {\bibfnamefont {M.}~\bibnamefont {Estienne}}, \bibinfo {author} {\bibfnamefont {M.}~\bibnamefont {Fallot}}, \bibinfo {author} {\bibfnamefont {A.}~\bibnamefont {Algora}}, \bibinfo {author} {\bibfnamefont {J.}~\bibnamefont {Briz-Monago}}, \bibinfo {author} {\bibfnamefont {V.~M.}\ \bibnamefont {Bui}}, \bibinfo {author} {\bibfnamefont {S.}~\bibnamefont {Cormon}}, \bibinfo {author} {\bibfnamefont {W.}~\bibnamefont {Gelletly}}, \bibinfo {author} {\bibfnamefont {L.}~\bibnamefont {Giot}}, \bibinfo {author} {\bibfnamefont {V.}~\bibnamefont {Guadilla}}, \bibinfo {author} {\bibfnamefont {D.}~\bibnamefont {Jordan}}, \bibinfo {author} {\bibfnamefont {L.~L.}\ \bibnamefont {Meur}}, \bibinfo {author} {\bibfnamefont {A.}~\bibnamefont {Porta}}, \bibinfo {author} {\bibfnamefont {S.}~\bibnamefont {Rice}}, \bibinfo {author} {\bibfnamefont {B.}~\bibnamefont {Rubio}}, \bibinfo {author} {\bibfnamefont {J.~L.}\ \bibnamefont {Tain}}, \bibinfo {author} {\bibfnamefont {E.}~\bibnamefont
  {Valencia}}, \ and\ \bibinfo {author} {\bibfnamefont {A.-A.}\ \bibnamefont {Zakari-Issoufou}},\ }\href {\doibase 10.1103/PhysRevLett.123.022502} {\bibfield  {journal} {\bibinfo  {journal} {Physical Review Letters}\ }\textbf {\bibinfo {volume} {123}},\ \bibinfo {pages} {022502} (\bibinfo {year} {2019})},\ \Eprint {http://arxiv.org/abs/1904.09358} {arXiv:1904.09358} \BibitemShut {NoStop}%
\bibitem [{\citenamefont {Schreckenbach}\ \emph {et~al.}(1981)\citenamefont {Schreckenbach}, \citenamefont {Faust}, \citenamefont {von Feilitzsch}, \citenamefont {Hahn}, \citenamefont {Hawerkamp},\ and\ \citenamefont {Vuilleumier}}]{Schreckenbach1981}%
  \BibitemOpen
  \bibfield  {author} {\bibinfo {author} {\bibfnamefont {K.}~\bibnamefont {Schreckenbach}}, \bibinfo {author} {\bibfnamefont {H.}~\bibnamefont {Faust}}, \bibinfo {author} {\bibfnamefont {F.}~\bibnamefont {von Feilitzsch}}, \bibinfo {author} {\bibfnamefont {A.}~\bibnamefont {Hahn}}, \bibinfo {author} {\bibfnamefont {K.}~\bibnamefont {Hawerkamp}}, \ and\ \bibinfo {author} {\bibfnamefont {J.}~\bibnamefont {Vuilleumier}},\ }\href {\doibase 10.1016/0370-2693(81)91120-5} {\bibfield  {journal} {\bibinfo  {journal} {Physics Letters B}\ }\textbf {\bibinfo {volume} {99}},\ \bibinfo {pages} {251} (\bibinfo {year} {1981})}\BibitemShut {NoStop}%
\bibitem [{\citenamefont {Schreckenbach}\ \emph {et~al.}(1985)\citenamefont {Schreckenbach}, \citenamefont {Colvin}, \citenamefont {Gelletly},\ and\ \citenamefont {{Von Feilitzsch}}}]{Schreckenbach1985}%
  \BibitemOpen
  \bibfield  {author} {\bibinfo {author} {\bibfnamefont {K.}~\bibnamefont {Schreckenbach}}, \bibinfo {author} {\bibfnamefont {G.}~\bibnamefont {Colvin}}, \bibinfo {author} {\bibfnamefont {W.}~\bibnamefont {Gelletly}}, \ and\ \bibinfo {author} {\bibfnamefont {F.}~\bibnamefont {{Von Feilitzsch}}},\ }\href {\doibase 10.1016/0370-2693(85)91337-1} {\bibfield  {journal} {\bibinfo  {journal} {Physics Letters B}\ }\textbf {\bibinfo {volume} {160}},\ \bibinfo {pages} {325} (\bibinfo {year} {1985})}\BibitemShut {NoStop}%
\bibitem [{\citenamefont {Haag}\ \emph {et~al.}(2014)\citenamefont {Haag}, \citenamefont {Gelletly}, \citenamefont {von Feilitzsch}, \citenamefont {Oberauer}, \citenamefont {Potzel}, \citenamefont {Schreckenbach},\ and\ \citenamefont {Sonzogni}}]{Haag2014}%
  \BibitemOpen
  \bibfield  {author} {\bibinfo {author} {\bibfnamefont {N.}~\bibnamefont {Haag}}, \bibinfo {author} {\bibfnamefont {W.}~\bibnamefont {Gelletly}}, \bibinfo {author} {\bibfnamefont {F.}~\bibnamefont {von Feilitzsch}}, \bibinfo {author} {\bibfnamefont {L.}~\bibnamefont {Oberauer}}, \bibinfo {author} {\bibfnamefont {W.}~\bibnamefont {Potzel}}, \bibinfo {author} {\bibfnamefont {K.}~\bibnamefont {Schreckenbach}}, \ and\ \bibinfo {author} {\bibfnamefont {A.~A.}\ \bibnamefont {Sonzogni}},\ }\href {http://arxiv.org/abs/1405.3501} {\  (\bibinfo {year} {2014})},\ \Eprint {http://arxiv.org/abs/1405.3501} {arXiv:1405.3501} \BibitemShut {NoStop}%
\bibitem [{\citenamefont {Kopeikin}\ \emph {et~al.}(2021)\citenamefont {Kopeikin}, \citenamefont {Skorokhvatov},\ and\ \citenamefont {Titov}}]{Kopeikin2021}%
  \BibitemOpen
  \bibfield  {author} {\bibinfo {author} {\bibfnamefont {V.}~\bibnamefont {Kopeikin}}, \bibinfo {author} {\bibfnamefont {M.}~\bibnamefont {Skorokhvatov}}, \ and\ \bibinfo {author} {\bibfnamefont {O.}~\bibnamefont {Titov}},\ }\href {\doibase 10.1103/PhysRevD.104.L071301} {\bibfield  {journal} {\bibinfo  {journal} {Physical Review D}\ }\textbf {\bibinfo {volume} {104}},\ \bibinfo {pages} {L071301} (\bibinfo {year} {2021})},\ \Eprint {http://arxiv.org/abs/2103.01684} {arXiv:2103.01684} \BibitemShut {NoStop}%
\bibitem [{\citenamefont {Lindner}\ \emph {et~al.}(2017)\citenamefont {Lindner}, \citenamefont {Rodejohann},\ and\ \citenamefont {Xu}}]{Lindner2017}%
  \BibitemOpen
  \bibfield  {author} {\bibinfo {author} {\bibfnamefont {M.}~\bibnamefont {Lindner}}, \bibinfo {author} {\bibfnamefont {W.}~\bibnamefont {Rodejohann}}, \ and\ \bibinfo {author} {\bibfnamefont {X.~J.}\ \bibnamefont {Xu}},\ }\href {\doibase 10.1007/JHEP03(2017)097} {\bibfield  {journal} {\bibinfo  {journal} {Journal of High Energy Physics}\ }\textbf {\bibinfo {volume} {2017}} (\bibinfo {year} {2017}),\ 10.1007/JHEP03(2017)097}\BibitemShut {NoStop}%
\bibitem [{\citenamefont {Workman}\ \emph {et~al.}(2022)\citenamefont {Workman}, \citenamefont {Burkert}, \citenamefont {Crede}, \citenamefont {Klempt}, \citenamefont {Thoma}, \citenamefont {Tiator}, \citenamefont {Agashe}, \citenamefont {Aielli}, \citenamefont {Allanach}, \citenamefont {Amsler}, \citenamefont {Antonelli}, \citenamefont {Aschenauer}, \citenamefont {Asner}, \citenamefont {Baer}, \citenamefont {Banerjee}, \citenamefont {Barnett}, \citenamefont {Baudis}, \citenamefont {Bauer}, \citenamefont {Beatty}, \citenamefont {Belousov}, \citenamefont {Beringer}, \citenamefont {Bettini}, \citenamefont {Biebel}, \citenamefont {Black}, \citenamefont {Blucher}, \citenamefont {Bonventre}, \citenamefont {Bryzgalov}, \citenamefont {Buchmuller}, \citenamefont {Bychkov}, \citenamefont {Cahn}, \citenamefont {Carena}, \citenamefont {Ceccucci}, \citenamefont {Cerri}, \citenamefont {Chivukula}, \citenamefont {Cowan}, \citenamefont {Cranmer}, \citenamefont {Cremonesi}, \citenamefont {D'Ambrosio}, \citenamefont {Damour},
  \citenamefont {de~Florian}, \citenamefont {de~Gouv{\^{e}}a}, \citenamefont {DeGrand}, \citenamefont {de~Jong}, \citenamefont {Demers}, \citenamefont {Dobrescu}, \citenamefont {D'Onofrio}, \citenamefont {Doser}, \citenamefont {Dreiner}, \citenamefont {Eerola}, \citenamefont {Egede}, \citenamefont {Eidelman}, \citenamefont {El-Khadra}, \citenamefont {Ellis}, \citenamefont {Eno}, \citenamefont {Erler}, \citenamefont {Ezhela}, \citenamefont {Fetscher}, \citenamefont {Fields}, \citenamefont {Freitas}, \citenamefont {Gallagher}, \citenamefont {Gershtein}, \citenamefont {Gherghetta}, \citenamefont {Gonzalez-Garcia}, \citenamefont {Goodman}, \citenamefont {Grab}, \citenamefont {Gritsan}, \citenamefont {Grojean}, \citenamefont {Groom}, \citenamefont {Gr{\"{u}}newald}, \citenamefont {Gurtu}, \citenamefont {Gutsche}, \citenamefont {Haber}, \citenamefont {Hamel}, \citenamefont {Hanhart}, \citenamefont {Hashimoto}, \citenamefont {Hayato}, \citenamefont {Hebecker}, \citenamefont {Heinemeyer}, \citenamefont
  {Hern{\'{a}}ndez-Rey}, \citenamefont {Hikasa}, \citenamefont {Hisano}, \citenamefont {H{\"{o}}cker}, \citenamefont {Holder}, \citenamefont {Hsu}, \citenamefont {Huston}, \citenamefont {Hyodo}, \citenamefont {Ianni}, \citenamefont {Kado}, \citenamefont {Karliner}, \citenamefont {Katz}, \citenamefont {Kenzie}, \citenamefont {Khoze}, \citenamefont {Klein}, \citenamefont {Krauss}, \citenamefont {Kreps}, \citenamefont {Kri{\v{z}}an}, \citenamefont {Krusche}, \citenamefont {Kwon}, \citenamefont {Lahav}, \citenamefont {Laiho}, \citenamefont {Lellouch}, \citenamefont {Lesgourgues}, \citenamefont {Liddle}, \citenamefont {Ligeti}, \citenamefont {Lin}, \citenamefont {Lippmann}, \citenamefont {Liss}, \citenamefont {Littenberg}, \citenamefont {Louren{\c{c}}o}, \citenamefont {Lugovsky}, \citenamefont {Lugovsky}, \citenamefont {Lusiani}, \citenamefont {Makida}, \citenamefont {Maltoni}, \citenamefont {Mannel}, \citenamefont {Manohar}, \citenamefont {Marciano}, \citenamefont {Masoni}, \citenamefont {Matthews}, \citenamefont
  {Mei{\ss}ner}, \citenamefont {Melzer-Pellmann}, \citenamefont {Mikhasenko}, \citenamefont {Miller}, \citenamefont {Milstead}, \citenamefont {Mitchell}, \citenamefont {M{\"{o}}nig}, \citenamefont {Molaro}, \citenamefont {Moortgat}, \citenamefont {Moskovic}, \citenamefont {Nakamura}, \citenamefont {Narain}, \citenamefont {Nason}, \citenamefont {Navas}, \citenamefont {Nelles}, \citenamefont {Neubert}, \citenamefont {Nevski}, \citenamefont {Nir}, \citenamefont {Olive}, \citenamefont {Patrignani}, \citenamefont {Peacock}, \citenamefont {Petrov}, \citenamefont {Pianori}, \citenamefont {Pich}, \citenamefont {Piepke}, \citenamefont {Pietropaolo}, \citenamefont {Pomarol}, \citenamefont {Pordes}, \citenamefont {Profumo}, \citenamefont {Quadt}, \citenamefont {Rabbertz}, \citenamefont {Rademacker}, \citenamefont {Raffelt}, \citenamefont {Ramsey-Musolf}, \citenamefont {Ratcliff}, \citenamefont {Richardson}, \citenamefont {Ringwald}, \citenamefont {Robinson}, \citenamefont {Roesler}, \citenamefont {Rolli}, \citenamefont
  {Romaniouk}, \citenamefont {Rosenberg}, \citenamefont {Rosner}, \citenamefont {Rybka}, \citenamefont {Ryskin}, \citenamefont {Ryutin}, \citenamefont {Sakai}, \citenamefont {Sarkar}, \citenamefont {Sauli}, \citenamefont {Schneider}, \citenamefont {Sch{\"{o}}nert}, \citenamefont {Scholberg}, \citenamefont {Schwartz}, \citenamefont {Schwiening}, \citenamefont {Scott}, \citenamefont {Sefkow}, \citenamefont {Seljak}, \citenamefont {Sharma}, \citenamefont {Sharpe}, \citenamefont {Shiltsev}, \citenamefont {Signorelli}, \citenamefont {Silari}, \citenamefont {Simon}, \citenamefont {Sj{\"{o}}strand}, \citenamefont {Skands}, \citenamefont {Skwarnicki}, \citenamefont {Smoot}, \citenamefont {Soffer}, \citenamefont {Sozzi}, \citenamefont {Spanier}, \citenamefont {Spiering}, \citenamefont {Stahl}, \citenamefont {Stone}, \citenamefont {Sumino}, \citenamefont {Syphers}, \citenamefont {Takahashi}, \citenamefont {Tanabashi}, \citenamefont {Tanaka}, \citenamefont {Ta{\v{s}}evsk{\'{y}}}, \citenamefont {Terao}, \citenamefont
  {Terashi}, \citenamefont {Terning}, \citenamefont {Thorne}, \citenamefont {Titov}, \citenamefont {Tkachenko}, \citenamefont {Tovey}, \citenamefont {Trabelsi}, \citenamefont {Urquijo}, \citenamefont {Valencia}, \citenamefont {{Van de Water}}, \citenamefont {Varelas}, \citenamefont {Venanzoni}, \citenamefont {Verde}, \citenamefont {Vivarelli}, \citenamefont {Vogel}, \citenamefont {Vogelsang}, \citenamefont {Vorobyev}, \citenamefont {Wakely}, \citenamefont {Walkowiak}, \citenamefont {Walter}, \citenamefont {Wands}, \citenamefont {Weinberg}, \citenamefont {Weinberg}, \citenamefont {Wermes}, \citenamefont {White}, \citenamefont {Wiencke}, \citenamefont {Willocq}, \citenamefont {Wohl}, \citenamefont {Woody}, \citenamefont {Yao}, \citenamefont {Yokoyama}, \citenamefont {Yoshida}, \citenamefont {Zanderighi}, \citenamefont {Zeller}, \citenamefont {Zenin}, \citenamefont {Zhu}, \citenamefont {Zhu}, \citenamefont {Zimmermann},\ and\ \citenamefont {Zyla}}]{Workman2022}%
  \BibitemOpen
  \bibfield  {author} {\bibinfo {author} {\bibfnamefont {R.~L.}\ \bibnamefont {Workman}}, \bibinfo {author} {\bibfnamefont {V.~D.}\ \bibnamefont {Burkert}}, \bibinfo {author} {\bibfnamefont {V.}~\bibnamefont {Crede}}, \bibinfo {author} {\bibfnamefont {E.}~\bibnamefont {Klempt}}, \bibinfo {author} {\bibfnamefont {U.}~\bibnamefont {Thoma}}, \bibinfo {author} {\bibfnamefont {L.}~\bibnamefont {Tiator}}, \bibinfo {author} {\bibfnamefont {K.}~\bibnamefont {Agashe}}, \bibinfo {author} {\bibfnamefont {G.}~\bibnamefont {Aielli}}, \bibinfo {author} {\bibfnamefont {B.~C.}\ \bibnamefont {Allanach}}, \bibinfo {author} {\bibfnamefont {C.}~\bibnamefont {Amsler}}, \bibinfo {author} {\bibfnamefont {M.}~\bibnamefont {Antonelli}}, \bibinfo {author} {\bibfnamefont {E.~C.}\ \bibnamefont {Aschenauer}}, \bibinfo {author} {\bibfnamefont {D.~M.}\ \bibnamefont {Asner}}, \bibinfo {author} {\bibfnamefont {H.}~\bibnamefont {Baer}}, \bibinfo {author} {\bibfnamefont {S.}~\bibnamefont {Banerjee}}, \bibinfo {author} {\bibfnamefont {R.~M.}\
  \bibnamefont {Barnett}}, \bibinfo {author} {\bibfnamefont {L.}~\bibnamefont {Baudis}}, \bibinfo {author} {\bibfnamefont {C.~W.}\ \bibnamefont {Bauer}}, \bibinfo {author} {\bibfnamefont {J.~J.}\ \bibnamefont {Beatty}}, \bibinfo {author} {\bibfnamefont {V.~I.}\ \bibnamefont {Belousov}}, \bibinfo {author} {\bibfnamefont {J.}~\bibnamefont {Beringer}}, \bibinfo {author} {\bibfnamefont {A.}~\bibnamefont {Bettini}}, \bibinfo {author} {\bibfnamefont {O.}~\bibnamefont {Biebel}}, \bibinfo {author} {\bibfnamefont {K.~M.}\ \bibnamefont {Black}}, \bibinfo {author} {\bibfnamefont {E.}~\bibnamefont {Blucher}}, \bibinfo {author} {\bibfnamefont {R.}~\bibnamefont {Bonventre}}, \bibinfo {author} {\bibfnamefont {V.~V.}\ \bibnamefont {Bryzgalov}}, \bibinfo {author} {\bibfnamefont {O.}~\bibnamefont {Buchmuller}}, \bibinfo {author} {\bibfnamefont {M.~A.}\ \bibnamefont {Bychkov}}, \bibinfo {author} {\bibfnamefont {R.~N.}\ \bibnamefont {Cahn}}, \bibinfo {author} {\bibfnamefont {M.}~\bibnamefont {Carena}}, \bibinfo {author}
  {\bibfnamefont {A.}~\bibnamefont {Ceccucci}}, \bibinfo {author} {\bibfnamefont {A.}~\bibnamefont {Cerri}}, \bibinfo {author} {\bibfnamefont {R.~S.}\ \bibnamefont {Chivukula}}, \bibinfo {author} {\bibfnamefont {G.}~\bibnamefont {Cowan}}, \bibinfo {author} {\bibfnamefont {K.}~\bibnamefont {Cranmer}}, \bibinfo {author} {\bibfnamefont {O.}~\bibnamefont {Cremonesi}}, \bibinfo {author} {\bibfnamefont {G.}~\bibnamefont {D'Ambrosio}}, \bibinfo {author} {\bibfnamefont {T.}~\bibnamefont {Damour}}, \bibinfo {author} {\bibfnamefont {D.}~\bibnamefont {de~Florian}}, \bibinfo {author} {\bibfnamefont {A.}~\bibnamefont {de~Gouv{\^{e}}a}}, \bibinfo {author} {\bibfnamefont {T.}~\bibnamefont {DeGrand}}, \bibinfo {author} {\bibfnamefont {P.}~\bibnamefont {de~Jong}}, \bibinfo {author} {\bibfnamefont {S.}~\bibnamefont {Demers}}, \bibinfo {author} {\bibfnamefont {B.~A.}\ \bibnamefont {Dobrescu}}, \bibinfo {author} {\bibfnamefont {M.}~\bibnamefont {D'Onofrio}}, \bibinfo {author} {\bibfnamefont {M.}~\bibnamefont {Doser}}, \bibinfo
  {author} {\bibfnamefont {H.~K.}\ \bibnamefont {Dreiner}}, \bibinfo {author} {\bibfnamefont {P.}~\bibnamefont {Eerola}}, \bibinfo {author} {\bibfnamefont {U.}~\bibnamefont {Egede}}, \bibinfo {author} {\bibfnamefont {S.}~\bibnamefont {Eidelman}}, \bibinfo {author} {\bibfnamefont {A.~X.}\ \bibnamefont {El-Khadra}}, \bibinfo {author} {\bibfnamefont {J.}~\bibnamefont {Ellis}}, \bibinfo {author} {\bibfnamefont {S.~C.}\ \bibnamefont {Eno}}, \bibinfo {author} {\bibfnamefont {J.}~\bibnamefont {Erler}}, \bibinfo {author} {\bibfnamefont {V.~V.}\ \bibnamefont {Ezhela}}, \bibinfo {author} {\bibfnamefont {W.}~\bibnamefont {Fetscher}}, \bibinfo {author} {\bibfnamefont {B.~D.}\ \bibnamefont {Fields}}, \bibinfo {author} {\bibfnamefont {A.}~\bibnamefont {Freitas}}, \bibinfo {author} {\bibfnamefont {H.}~\bibnamefont {Gallagher}}, \bibinfo {author} {\bibfnamefont {Y.}~\bibnamefont {Gershtein}}, \bibinfo {author} {\bibfnamefont {T.}~\bibnamefont {Gherghetta}}, \bibinfo {author} {\bibfnamefont {M.~C.}\ \bibnamefont
  {Gonzalez-Garcia}}, \bibinfo {author} {\bibfnamefont {M.}~\bibnamefont {Goodman}}, \bibinfo {author} {\bibfnamefont {C.}~\bibnamefont {Grab}}, \bibinfo {author} {\bibfnamefont {A.~V.}\ \bibnamefont {Gritsan}}, \bibinfo {author} {\bibfnamefont {C.}~\bibnamefont {Grojean}}, \bibinfo {author} {\bibfnamefont {D.~E.}\ \bibnamefont {Groom}}, \bibinfo {author} {\bibfnamefont {M.}~\bibnamefont {Gr{\"{u}}newald}}, \bibinfo {author} {\bibfnamefont {A.}~\bibnamefont {Gurtu}}, \bibinfo {author} {\bibfnamefont {T.}~\bibnamefont {Gutsche}}, \bibinfo {author} {\bibfnamefont {H.~E.}\ \bibnamefont {Haber}}, \bibinfo {author} {\bibfnamefont {M.}~\bibnamefont {Hamel}}, \bibinfo {author} {\bibfnamefont {C.}~\bibnamefont {Hanhart}}, \bibinfo {author} {\bibfnamefont {S.}~\bibnamefont {Hashimoto}}, \bibinfo {author} {\bibfnamefont {Y.}~\bibnamefont {Hayato}}, \bibinfo {author} {\bibfnamefont {A.}~\bibnamefont {Hebecker}}, \bibinfo {author} {\bibfnamefont {S.}~\bibnamefont {Heinemeyer}}, \bibinfo {author} {\bibfnamefont {J.~J.}\
  \bibnamefont {Hern{\'{a}}ndez-Rey}}, \bibinfo {author} {\bibfnamefont {K.}~\bibnamefont {Hikasa}}, \bibinfo {author} {\bibfnamefont {J.}~\bibnamefont {Hisano}}, \bibinfo {author} {\bibfnamefont {A.}~\bibnamefont {H{\"{o}}cker}}, \bibinfo {author} {\bibfnamefont {J.}~\bibnamefont {Holder}}, \bibinfo {author} {\bibfnamefont {L.}~\bibnamefont {Hsu}}, \bibinfo {author} {\bibfnamefont {J.}~\bibnamefont {Huston}}, \bibinfo {author} {\bibfnamefont {T.}~\bibnamefont {Hyodo}}, \bibinfo {author} {\bibfnamefont {A.}~\bibnamefont {Ianni}}, \bibinfo {author} {\bibfnamefont {M.}~\bibnamefont {Kado}}, \bibinfo {author} {\bibfnamefont {M.}~\bibnamefont {Karliner}}, \bibinfo {author} {\bibfnamefont {U.~F.}\ \bibnamefont {Katz}}, \bibinfo {author} {\bibfnamefont {M.}~\bibnamefont {Kenzie}}, \bibinfo {author} {\bibfnamefont {V.~A.}\ \bibnamefont {Khoze}}, \bibinfo {author} {\bibfnamefont {S.~R.}\ \bibnamefont {Klein}}, \bibinfo {author} {\bibfnamefont {F.}~\bibnamefont {Krauss}}, \bibinfo {author} {\bibfnamefont
  {M.}~\bibnamefont {Kreps}}, \bibinfo {author} {\bibfnamefont {P.}~\bibnamefont {Kri{\v{z}}an}}, \bibinfo {author} {\bibfnamefont {B.}~\bibnamefont {Krusche}}, \bibinfo {author} {\bibfnamefont {Y.}~\bibnamefont {Kwon}}, \bibinfo {author} {\bibfnamefont {O.}~\bibnamefont {Lahav}}, \bibinfo {author} {\bibfnamefont {J.}~\bibnamefont {Laiho}}, \bibinfo {author} {\bibfnamefont {L.~P.}\ \bibnamefont {Lellouch}}, \bibinfo {author} {\bibfnamefont {J.}~\bibnamefont {Lesgourgues}}, \bibinfo {author} {\bibfnamefont {A.~R.}\ \bibnamefont {Liddle}}, \bibinfo {author} {\bibfnamefont {Z.}~\bibnamefont {Ligeti}}, \bibinfo {author} {\bibfnamefont {C.-J.}\ \bibnamefont {Lin}}, \bibinfo {author} {\bibfnamefont {C.}~\bibnamefont {Lippmann}}, \bibinfo {author} {\bibfnamefont {T.~M.}\ \bibnamefont {Liss}}, \bibinfo {author} {\bibfnamefont {L.}~\bibnamefont {Littenberg}}, \bibinfo {author} {\bibfnamefont {C.}~\bibnamefont {Louren{\c{c}}o}}, \bibinfo {author} {\bibfnamefont {K.~S.}\ \bibnamefont {Lugovsky}}, \bibinfo {author}
  {\bibfnamefont {S.~B.}\ \bibnamefont {Lugovsky}}, \bibinfo {author} {\bibfnamefont {A.}~\bibnamefont {Lusiani}}, \bibinfo {author} {\bibfnamefont {Y.}~\bibnamefont {Makida}}, \bibinfo {author} {\bibfnamefont {F.}~\bibnamefont {Maltoni}}, \bibinfo {author} {\bibfnamefont {T.}~\bibnamefont {Mannel}}, \bibinfo {author} {\bibfnamefont {A.~V.}\ \bibnamefont {Manohar}}, \bibinfo {author} {\bibfnamefont {W.~J.}\ \bibnamefont {Marciano}}, \bibinfo {author} {\bibfnamefont {A.}~\bibnamefont {Masoni}}, \bibinfo {author} {\bibfnamefont {J.}~\bibnamefont {Matthews}}, \bibinfo {author} {\bibfnamefont {U.-G.}\ \bibnamefont {Mei{\ss}ner}}, \bibinfo {author} {\bibfnamefont {I.-A.}\ \bibnamefont {Melzer-Pellmann}}, \bibinfo {author} {\bibfnamefont {M.}~\bibnamefont {Mikhasenko}}, \bibinfo {author} {\bibfnamefont {D.~J.}\ \bibnamefont {Miller}}, \bibinfo {author} {\bibfnamefont {D.}~\bibnamefont {Milstead}}, \bibinfo {author} {\bibfnamefont {R.~E.}\ \bibnamefont {Mitchell}}, \bibinfo {author} {\bibfnamefont {K.}~\bibnamefont
  {M{\"{o}}nig}}, \bibinfo {author} {\bibfnamefont {P.}~\bibnamefont {Molaro}}, \bibinfo {author} {\bibfnamefont {F.}~\bibnamefont {Moortgat}}, \bibinfo {author} {\bibfnamefont {M.}~\bibnamefont {Moskovic}}, \bibinfo {author} {\bibfnamefont {K.}~\bibnamefont {Nakamura}}, \bibinfo {author} {\bibfnamefont {M.}~\bibnamefont {Narain}}, \bibinfo {author} {\bibfnamefont {P.}~\bibnamefont {Nason}}, \bibinfo {author} {\bibfnamefont {S.}~\bibnamefont {Navas}}, \bibinfo {author} {\bibfnamefont {A.}~\bibnamefont {Nelles}}, \bibinfo {author} {\bibfnamefont {M.}~\bibnamefont {Neubert}}, \bibinfo {author} {\bibfnamefont {P.}~\bibnamefont {Nevski}}, \bibinfo {author} {\bibfnamefont {Y.}~\bibnamefont {Nir}}, \bibinfo {author} {\bibfnamefont {K.~A.}\ \bibnamefont {Olive}}, \bibinfo {author} {\bibfnamefont {C.}~\bibnamefont {Patrignani}}, \bibinfo {author} {\bibfnamefont {J.~A.}\ \bibnamefont {Peacock}}, \bibinfo {author} {\bibfnamefont {V.~A.}\ \bibnamefont {Petrov}}, \bibinfo {author} {\bibfnamefont {E.}~\bibnamefont
  {Pianori}}, \bibinfo {author} {\bibfnamefont {A.}~\bibnamefont {Pich}}, \bibinfo {author} {\bibfnamefont {A.}~\bibnamefont {Piepke}}, \bibinfo {author} {\bibfnamefont {F.}~\bibnamefont {Pietropaolo}}, \bibinfo {author} {\bibfnamefont {A.}~\bibnamefont {Pomarol}}, \bibinfo {author} {\bibfnamefont {S.}~\bibnamefont {Pordes}}, \bibinfo {author} {\bibfnamefont {S.}~\bibnamefont {Profumo}}, \bibinfo {author} {\bibfnamefont {A.}~\bibnamefont {Quadt}}, \bibinfo {author} {\bibfnamefont {K.}~\bibnamefont {Rabbertz}}, \bibinfo {author} {\bibfnamefont {J.}~\bibnamefont {Rademacker}}, \bibinfo {author} {\bibfnamefont {G.}~\bibnamefont {Raffelt}}, \bibinfo {author} {\bibfnamefont {M.}~\bibnamefont {Ramsey-Musolf}}, \bibinfo {author} {\bibfnamefont {B.~N.}\ \bibnamefont {Ratcliff}}, \bibinfo {author} {\bibfnamefont {P.}~\bibnamefont {Richardson}}, \bibinfo {author} {\bibfnamefont {A.}~\bibnamefont {Ringwald}}, \bibinfo {author} {\bibfnamefont {D.~J.}\ \bibnamefont {Robinson}}, \bibinfo {author} {\bibfnamefont
  {S.}~\bibnamefont {Roesler}}, \bibinfo {author} {\bibfnamefont {S.}~\bibnamefont {Rolli}}, \bibinfo {author} {\bibfnamefont {A.}~\bibnamefont {Romaniouk}}, \bibinfo {author} {\bibfnamefont {L.~J.}\ \bibnamefont {Rosenberg}}, \bibinfo {author} {\bibfnamefont {J.~L.}\ \bibnamefont {Rosner}}, \bibinfo {author} {\bibfnamefont {G.}~\bibnamefont {Rybka}}, \bibinfo {author} {\bibfnamefont {M.~G.}\ \bibnamefont {Ryskin}}, \bibinfo {author} {\bibfnamefont {R.~A.}\ \bibnamefont {Ryutin}}, \bibinfo {author} {\bibfnamefont {Y.}~\bibnamefont {Sakai}}, \bibinfo {author} {\bibfnamefont {S.}~\bibnamefont {Sarkar}}, \bibinfo {author} {\bibfnamefont {F.}~\bibnamefont {Sauli}}, \bibinfo {author} {\bibfnamefont {O.}~\bibnamefont {Schneider}}, \bibinfo {author} {\bibfnamefont {S.}~\bibnamefont {Sch{\"{o}}nert}}, \bibinfo {author} {\bibfnamefont {K.}~\bibnamefont {Scholberg}}, \bibinfo {author} {\bibfnamefont {A.~J.}\ \bibnamefont {Schwartz}}, \bibinfo {author} {\bibfnamefont {J.}~\bibnamefont {Schwiening}}, \bibinfo {author}
  {\bibfnamefont {D.}~\bibnamefont {Scott}}, \bibinfo {author} {\bibfnamefont {F.}~\bibnamefont {Sefkow}}, \bibinfo {author} {\bibfnamefont {U.}~\bibnamefont {Seljak}}, \bibinfo {author} {\bibfnamefont {V.}~\bibnamefont {Sharma}}, \bibinfo {author} {\bibfnamefont {S.~R.}\ \bibnamefont {Sharpe}}, \bibinfo {author} {\bibfnamefont {V.}~\bibnamefont {Shiltsev}}, \bibinfo {author} {\bibfnamefont {G.}~\bibnamefont {Signorelli}}, \bibinfo {author} {\bibfnamefont {M.}~\bibnamefont {Silari}}, \bibinfo {author} {\bibfnamefont {F.}~\bibnamefont {Simon}}, \bibinfo {author} {\bibfnamefont {T.}~\bibnamefont {Sj{\"{o}}strand}}, \bibinfo {author} {\bibfnamefont {P.}~\bibnamefont {Skands}}, \bibinfo {author} {\bibfnamefont {T.}~\bibnamefont {Skwarnicki}}, \bibinfo {author} {\bibfnamefont {G.~F.}\ \bibnamefont {Smoot}}, \bibinfo {author} {\bibfnamefont {A.}~\bibnamefont {Soffer}}, \bibinfo {author} {\bibfnamefont {M.~S.}\ \bibnamefont {Sozzi}}, \bibinfo {author} {\bibfnamefont {S.}~\bibnamefont {Spanier}}, \bibinfo {author}
  {\bibfnamefont {C.}~\bibnamefont {Spiering}}, \bibinfo {author} {\bibfnamefont {A.}~\bibnamefont {Stahl}}, \bibinfo {author} {\bibfnamefont {S.~L.}\ \bibnamefont {Stone}}, \bibinfo {author} {\bibfnamefont {Y.}~\bibnamefont {Sumino}}, \bibinfo {author} {\bibfnamefont {M.~J.}\ \bibnamefont {Syphers}}, \bibinfo {author} {\bibfnamefont {F.}~\bibnamefont {Takahashi}}, \bibinfo {author} {\bibfnamefont {M.}~\bibnamefont {Tanabashi}}, \bibinfo {author} {\bibfnamefont {J.}~\bibnamefont {Tanaka}}, \bibinfo {author} {\bibfnamefont {M.}~\bibnamefont {Ta{\v{s}}evsk{\'{y}}}}, \bibinfo {author} {\bibfnamefont {K.}~\bibnamefont {Terao}}, \bibinfo {author} {\bibfnamefont {K.}~\bibnamefont {Terashi}}, \bibinfo {author} {\bibfnamefont {J.}~\bibnamefont {Terning}}, \bibinfo {author} {\bibfnamefont {R.~S.}\ \bibnamefont {Thorne}}, \bibinfo {author} {\bibfnamefont {M.}~\bibnamefont {Titov}}, \bibinfo {author} {\bibfnamefont {N.~P.}\ \bibnamefont {Tkachenko}}, \bibinfo {author} {\bibfnamefont {D.~R.}\ \bibnamefont {Tovey}},
  \bibinfo {author} {\bibfnamefont {K.}~\bibnamefont {Trabelsi}}, \bibinfo {author} {\bibfnamefont {P.}~\bibnamefont {Urquijo}}, \bibinfo {author} {\bibfnamefont {G.}~\bibnamefont {Valencia}}, \bibinfo {author} {\bibfnamefont {R.}~\bibnamefont {{Van de Water}}}, \bibinfo {author} {\bibfnamefont {N.}~\bibnamefont {Varelas}}, \bibinfo {author} {\bibfnamefont {G.}~\bibnamefont {Venanzoni}}, \bibinfo {author} {\bibfnamefont {L.}~\bibnamefont {Verde}}, \bibinfo {author} {\bibfnamefont {I.}~\bibnamefont {Vivarelli}}, \bibinfo {author} {\bibfnamefont {P.}~\bibnamefont {Vogel}}, \bibinfo {author} {\bibfnamefont {W.}~\bibnamefont {Vogelsang}}, \bibinfo {author} {\bibfnamefont {V.}~\bibnamefont {Vorobyev}}, \bibinfo {author} {\bibfnamefont {S.~P.}\ \bibnamefont {Wakely}}, \bibinfo {author} {\bibfnamefont {W.}~\bibnamefont {Walkowiak}}, \bibinfo {author} {\bibfnamefont {C.~W.}\ \bibnamefont {Walter}}, \bibinfo {author} {\bibfnamefont {D.}~\bibnamefont {Wands}}, \bibinfo {author} {\bibfnamefont {D.~H.}\ \bibnamefont
  {Weinberg}}, \bibinfo {author} {\bibfnamefont {E.~J.}\ \bibnamefont {Weinberg}}, \bibinfo {author} {\bibfnamefont {N.}~\bibnamefont {Wermes}}, \bibinfo {author} {\bibfnamefont {M.}~\bibnamefont {White}}, \bibinfo {author} {\bibfnamefont {L.~R.}\ \bibnamefont {Wiencke}}, \bibinfo {author} {\bibfnamefont {S.}~\bibnamefont {Willocq}}, \bibinfo {author} {\bibfnamefont {C.~G.}\ \bibnamefont {Wohl}}, \bibinfo {author} {\bibfnamefont {C.~L.}\ \bibnamefont {Woody}}, \bibinfo {author} {\bibfnamefont {W.-M.}\ \bibnamefont {Yao}}, \bibinfo {author} {\bibfnamefont {M.}~\bibnamefont {Yokoyama}}, \bibinfo {author} {\bibfnamefont {R.}~\bibnamefont {Yoshida}}, \bibinfo {author} {\bibfnamefont {G.}~\bibnamefont {Zanderighi}}, \bibinfo {author} {\bibfnamefont {G.~P.}\ \bibnamefont {Zeller}}, \bibinfo {author} {\bibfnamefont {O.~V.}\ \bibnamefont {Zenin}}, \bibinfo {author} {\bibfnamefont {R.-Y.}\ \bibnamefont {Zhu}}, \bibinfo {author} {\bibfnamefont {S.-L.}\ \bibnamefont {Zhu}}, \bibinfo {author} {\bibfnamefont
  {F.}~\bibnamefont {Zimmermann}}, \ and\ \bibinfo {author} {\bibfnamefont {P.~A.}\ \bibnamefont {Zyla}},\ }\href {\doibase 10.1093/ptep/ptac097} {\bibfield  {journal} {\bibinfo  {journal} {Progress of Theoretical and Experimental Physics}\ }\textbf {\bibinfo {volume} {2022}} (\bibinfo {year} {2022}),\ 10.1093/ptep/ptac097}\BibitemShut {NoStop}%
\bibitem [{\citenamefont {Erler}\ and\ \citenamefont {Schott}(2019)}]{Erler2019}%
  \BibitemOpen
  \bibfield  {author} {\bibinfo {author} {\bibfnamefont {J.}~\bibnamefont {Erler}}\ and\ \bibinfo {author} {\bibfnamefont {M.}~\bibnamefont {Schott}},\ }\href {\doibase 10.1016/j.ppnp.2019.02.007} {\bibfield  {journal} {\bibinfo  {journal} {Progress in Particle and Nuclear Physics}\ }\textbf {\bibinfo {volume} {106}},\ \bibinfo {pages} {68} (\bibinfo {year} {2019})}\BibitemShut {NoStop}%
\bibitem [{\citenamefont {Valle}(1987)}]{Valle1987}%
  \BibitemOpen
  \bibfield  {author} {\bibinfo {author} {\bibfnamefont {J.}~\bibnamefont {Valle}},\ }\href {\doibase 10.1016/0370-2693(87)90947-6} {\bibfield  {journal} {\bibinfo  {journal} {Physics Letters B}\ }\textbf {\bibinfo {volume} {199}},\ \bibinfo {pages} {432} (\bibinfo {year} {1987})}\BibitemShut {NoStop}%
\bibitem [{\citenamefont {Vogel}\ and\ \citenamefont {Engel}(1989)}]{Vogel1989}%
  \BibitemOpen
  \bibfield  {author} {\bibinfo {author} {\bibfnamefont {P.}~\bibnamefont {Vogel}}\ and\ \bibinfo {author} {\bibfnamefont {J.}~\bibnamefont {Engel}},\ }\href {\doibase 10.1103/PhysRevD.39.3378} {\bibfield  {journal} {\bibinfo  {journal} {Physical Review D}\ }\textbf {\bibinfo {volume} {39}},\ \bibinfo {pages} {3378} (\bibinfo {year} {1989})}\BibitemShut {NoStop}%
\bibitem [{\citenamefont {Farzan}\ \emph {et~al.}(2018)\citenamefont {Farzan}, \citenamefont {Lindner}, \citenamefont {Rodejohann},\ and\ \citenamefont {Xu}}]{Farzan2018}%
  \BibitemOpen
  \bibfield  {author} {\bibinfo {author} {\bibfnamefont {Y.}~\bibnamefont {Farzan}}, \bibinfo {author} {\bibfnamefont {M.}~\bibnamefont {Lindner}}, \bibinfo {author} {\bibfnamefont {W.}~\bibnamefont {Rodejohann}}, \ and\ \bibinfo {author} {\bibfnamefont {X.-J.}\ \bibnamefont {Xu}},\ }\href {\doibase 10.1007/JHEP05(2018)066} {\bibfield  {journal} {\bibinfo  {journal} {Journal of High Energy Physics}\ }\textbf {\bibinfo {volume} {2018}},\ \bibinfo {pages} {66} (\bibinfo {year} {2018})}\BibitemShut {NoStop}%
\bibitem [{\citenamefont {Cerdeno}\ \emph {et~al.}(2016)\citenamefont {Cerdeno}, \citenamefont {Fairbairn}, \citenamefont {Jubb}, \citenamefont {Machado}, \citenamefont {Vincent},\ and\ \citenamefont {Boehm}}]{Cerdeno2016}%
  \BibitemOpen
  \bibfield  {author} {\bibinfo {author} {\bibfnamefont {D.~G.}\ \bibnamefont {Cerdeno}}, \bibinfo {author} {\bibfnamefont {M.}~\bibnamefont {Fairbairn}}, \bibinfo {author} {\bibfnamefont {T.}~\bibnamefont {Jubb}}, \bibinfo {author} {\bibfnamefont {P.~A.~N.}\ \bibnamefont {Machado}}, \bibinfo {author} {\bibfnamefont {A.~C.}\ \bibnamefont {Vincent}}, \ and\ \bibinfo {author} {\bibfnamefont {C.}~\bibnamefont {Boehm}},\ }\href {\doibase 10.1007/JHEP05(2016)118} {\bibfield  {journal} {\bibinfo  {journal} {Journal of High Energy Physics}\ }\textbf {\bibinfo {volume} {2016}},\ \bibinfo {pages} {118} (\bibinfo {year} {2016})}\BibitemShut {NoStop}%
\bibitem [{\citenamefont {Coloma}\ \emph {et~al.}(2022)\citenamefont {Coloma}, \citenamefont {Esteban}, \citenamefont {Gonzalez-Garcia}, \citenamefont {Larizgoitia}, \citenamefont {Monrabal},\ and\ \citenamefont {Palomares-Ruiz}}]{Coloma2022}%
  \BibitemOpen
  \bibfield  {author} {\bibinfo {author} {\bibfnamefont {P.}~\bibnamefont {Coloma}}, \bibinfo {author} {\bibfnamefont {I.}~\bibnamefont {Esteban}}, \bibinfo {author} {\bibfnamefont {M.~C.}\ \bibnamefont {Gonzalez-Garcia}}, \bibinfo {author} {\bibfnamefont {L.}~\bibnamefont {Larizgoitia}}, \bibinfo {author} {\bibfnamefont {F.}~\bibnamefont {Monrabal}}, \ and\ \bibinfo {author} {\bibfnamefont {S.}~\bibnamefont {Palomares-Ruiz}},\ }\href {\doibase 10.1007/JHEP05(2022)037} {\bibfield  {journal} {\bibinfo  {journal} {Journal of High Energy Physics}\ }\textbf {\bibinfo {volume} {2022}},\ \bibinfo {pages} {37} (\bibinfo {year} {2022})},\ \Eprint {http://arxiv.org/abs/2202.10829} {arXiv:2202.10829} \BibitemShut {NoStop}%
\bibitem [{\citenamefont {Majumdar}\ \emph {et~al.}(2022)\citenamefont {Majumdar}, \citenamefont {Papoulias}, \citenamefont {Srivastava},\ and\ \citenamefont {Valle}}]{Majumdar2022}%
  \BibitemOpen
  \bibfield  {author} {\bibinfo {author} {\bibfnamefont {A.}~\bibnamefont {Majumdar}}, \bibinfo {author} {\bibfnamefont {D.~K.}\ \bibnamefont {Papoulias}}, \bibinfo {author} {\bibfnamefont {R.}~\bibnamefont {Srivastava}}, \ and\ \bibinfo {author} {\bibfnamefont {J.~W.}\ \bibnamefont {Valle}},\ }\href {\doibase 10.1103/PhysRevD.106.093010} {\bibfield  {journal} {\bibinfo  {journal} {Physical Review D}\ }\textbf {\bibinfo {volume} {106}},\ \bibinfo {pages} {093010} (\bibinfo {year} {2022})},\ \Eprint {http://arxiv.org/abs/2208.13262} {arXiv:2208.13262} \BibitemShut {NoStop}%
\bibitem [{\citenamefont {Boucenna}\ \emph {et~al.}(2014)\citenamefont {Boucenna}, \citenamefont {Morisi},\ and\ \citenamefont {Valle}}]{Boucenna2014}%
  \BibitemOpen
  \bibfield  {author} {\bibinfo {author} {\bibfnamefont {S.~M.}\ \bibnamefont {Boucenna}}, \bibinfo {author} {\bibfnamefont {S.}~\bibnamefont {Morisi}}, \ and\ \bibinfo {author} {\bibfnamefont {J.~W.~F.}\ \bibnamefont {Valle}},\ }\href {\doibase 10.1155/2014/831598} {\bibfield  {journal} {\bibinfo  {journal} {Advances in High Energy Physics}\ }\textbf {\bibinfo {volume} {2014}},\ \bibinfo {pages} {1} (\bibinfo {year} {2014})}\BibitemShut {NoStop}%
\bibitem [{\citenamefont {Barranco}\ \emph {et~al.}(2005)\citenamefont {Barranco}, \citenamefont {Miranda},\ and\ \citenamefont {Rashba}}]{Barranco2005}%
  \BibitemOpen
  \bibfield  {author} {\bibinfo {author} {\bibfnamefont {J.}~\bibnamefont {Barranco}}, \bibinfo {author} {\bibfnamefont {O.~G.}\ \bibnamefont {Miranda}}, \ and\ \bibinfo {author} {\bibfnamefont {T.~I.}\ \bibnamefont {Rashba}},\ }\href {\doibase 10.1088/1126-6708/2005/12/021} {\bibfield  {journal} {\bibinfo  {journal} {Journal of High Energy Physics}\ }\textbf {\bibinfo {volume} {2005}},\ \bibinfo {pages} {021} (\bibinfo {year} {2005})}\BibitemShut {NoStop}%
\bibitem [{\citenamefont {Lee}\ and\ \citenamefont {Yang}(1956)}]{Lee1956}%
  \BibitemOpen
  \bibfield  {author} {\bibinfo {author} {\bibfnamefont {T.~D.}\ \bibnamefont {Lee}}\ and\ \bibinfo {author} {\bibfnamefont {C.~N.}\ \bibnamefont {Yang}},\ }\href {\doibase 10.1103/PhysRev.104.254} {\bibfield  {journal} {\bibinfo  {journal} {Physical Review}\ }\textbf {\bibinfo {volume} {104}},\ \bibinfo {pages} {254} (\bibinfo {year} {1956})}\BibitemShut {NoStop}%
\bibitem [{\citenamefont {Sierra}\ \emph {et~al.}(2018)\citenamefont {Sierra}, \citenamefont {Romeri},\ and\ \citenamefont {Rojas}}]{Aristizabal-Sierra2018}%
  \BibitemOpen
  \bibfield  {author} {\bibinfo {author} {\bibfnamefont {D.~A.}\ \bibnamefont {Sierra}}, \bibinfo {author} {\bibfnamefont {V.~D.}\ \bibnamefont {Romeri}}, \ and\ \bibinfo {author} {\bibfnamefont {N.}~\bibnamefont {Rojas}},\ }\href {\doibase 10.1103/PhysRevD.98.075018} {\bibfield  {journal} {\bibinfo  {journal} {Physical Review D}\ }\textbf {\bibinfo {volume} {98}},\ \bibinfo {pages} {075018} (\bibinfo {year} {2018})}\BibitemShut {NoStop}%
\bibitem [{\citenamefont {Sierra}\ \emph {et~al.}(2019)\citenamefont {Sierra}, \citenamefont {Liao},\ and\ \citenamefont {Marfatia}}]{Aristizabal-Sierra2019}%
  \BibitemOpen
  \bibfield  {author} {\bibinfo {author} {\bibfnamefont {D.~A.}\ \bibnamefont {Sierra}}, \bibinfo {author} {\bibfnamefont {J.}~\bibnamefont {Liao}}, \ and\ \bibinfo {author} {\bibfnamefont {D.}~\bibnamefont {Marfatia}},\ }\href {\doibase 10.1007/JHEP06(2019)141} {\bibfield  {journal} {\bibinfo  {journal} {Journal of High Energy Physics}\ }\textbf {\bibinfo {volume} {2019}},\ \bibinfo {pages} {141} (\bibinfo {year} {2019})}\BibitemShut {NoStop}%
\bibitem [{\citenamefont {WONG}\ and\ \citenamefont {LI}(2005)}]{Wong2005}%
  \BibitemOpen
  \bibfield  {author} {\bibinfo {author} {\bibfnamefont {H.~T.}\ \bibnamefont {WONG}}\ and\ \bibinfo {author} {\bibfnamefont {H.-B.}\ \bibnamefont {LI}},\ }\href {\doibase 10.1142/S0217732305017482} {\bibfield  {journal} {\bibinfo  {journal} {Modern Physics Letters A}\ }\textbf {\bibinfo {volume} {20}},\ \bibinfo {pages} {1103} (\bibinfo {year} {2005})}\BibitemShut {NoStop}%
\bibitem [{\citenamefont {Dvornikov}\ and\ \citenamefont {Studenikin}(2004)}]{Dvornikov2004}%
  \BibitemOpen
  \bibfield  {author} {\bibinfo {author} {\bibfnamefont {M.}~\bibnamefont {Dvornikov}}\ and\ \bibinfo {author} {\bibfnamefont {A.}~\bibnamefont {Studenikin}},\ }\href {\doibase 10.1103/PhysRevD.69.073001} {\bibfield  {journal} {\bibinfo  {journal} {Physical Review D}\ }\textbf {\bibinfo {volume} {69}},\ \bibinfo {pages} {073001} (\bibinfo {year} {2004})}\BibitemShut {NoStop}%
\bibitem [{\citenamefont {Fija{\l}kowska}\ \emph {et~al.}(2017)\citenamefont {Fija{\l}kowska}, \citenamefont {Karny}, \citenamefont {Rykaczewski}, \citenamefont {Rasco}, \citenamefont {Grzywacz}, \citenamefont {Gross}, \citenamefont {Woli{\'{n}}ska-Cichocka}, \citenamefont {Goetz}, \citenamefont {Stracener}, \citenamefont {Bielewski}, \citenamefont {Goans}, \citenamefont {Hamilton}, \citenamefont {Johnson}, \citenamefont {Jost}, \citenamefont {Madurga}, \citenamefont {Miernik}, \citenamefont {Miller}, \citenamefont {Padgett}, \citenamefont {Paulauskas}, \citenamefont {Ramayya},\ and\ \citenamefont {Zganjar}}]{Fijakowska2017}%
  \BibitemOpen
  \bibfield  {author} {\bibinfo {author} {\bibfnamefont {A.}~\bibnamefont {Fija{\l}kowska}}, \bibinfo {author} {\bibfnamefont {M.}~\bibnamefont {Karny}}, \bibinfo {author} {\bibfnamefont {K.~P.}\ \bibnamefont {Rykaczewski}}, \bibinfo {author} {\bibfnamefont {B.~C.}\ \bibnamefont {Rasco}}, \bibinfo {author} {\bibfnamefont {R.}~\bibnamefont {Grzywacz}}, \bibinfo {author} {\bibfnamefont {C.~J.}\ \bibnamefont {Gross}}, \bibinfo {author} {\bibfnamefont {M.}~\bibnamefont {Woli{\'{n}}ska-Cichocka}}, \bibinfo {author} {\bibfnamefont {K.~C.}\ \bibnamefont {Goetz}}, \bibinfo {author} {\bibfnamefont {D.~W.}\ \bibnamefont {Stracener}}, \bibinfo {author} {\bibfnamefont {W.}~\bibnamefont {Bielewski}}, \bibinfo {author} {\bibfnamefont {R.}~\bibnamefont {Goans}}, \bibinfo {author} {\bibfnamefont {J.~H.}\ \bibnamefont {Hamilton}}, \bibinfo {author} {\bibfnamefont {J.~W.}\ \bibnamefont {Johnson}}, \bibinfo {author} {\bibfnamefont {C.}~\bibnamefont {Jost}}, \bibinfo {author} {\bibfnamefont {M.}~\bibnamefont {Madurga}}, \bibinfo
  {author} {\bibfnamefont {K.}~\bibnamefont {Miernik}}, \bibinfo {author} {\bibfnamefont {D.}~\bibnamefont {Miller}}, \bibinfo {author} {\bibfnamefont {S.~W.}\ \bibnamefont {Padgett}}, \bibinfo {author} {\bibfnamefont {S.~V.}\ \bibnamefont {Paulauskas}}, \bibinfo {author} {\bibfnamefont {A.~V.}\ \bibnamefont {Ramayya}}, \ and\ \bibinfo {author} {\bibfnamefont {E.~F.}\ \bibnamefont {Zganjar}},\ }\href {\doibase 10.1103/PhysRevLett.119.052503} {\bibfield  {journal} {\bibinfo  {journal} {Physical Review Letters}\ }\textbf {\bibinfo {volume} {119}},\ \bibinfo {pages} {052503} (\bibinfo {year} {2017})}\BibitemShut {NoStop}%
\bibitem [{\citenamefont {Guadilla}\ \emph {et~al.}(2019)\citenamefont {Guadilla}, \citenamefont {Algora}, \citenamefont {Tain}, \citenamefont {Agramunt}, \citenamefont {Aysto}, \citenamefont {Briz}, \citenamefont {Cucoanes}, \citenamefont {Eronen}, \citenamefont {Estienne}, \citenamefont {Fallot}, \citenamefont {Fraile}, \citenamefont {Ganioglu}, \citenamefont {Gelletly}, \citenamefont {Gorelov}, \citenamefont {Hakala}, \citenamefont {Jokinen}, \citenamefont {Jordan}, \citenamefont {Kankainen}, \citenamefont {Kolhinen}, \citenamefont {Koponen}, \citenamefont {Lebois}, \citenamefont {{Le Meur}}, \citenamefont {Martinez}, \citenamefont {Monserrate}, \citenamefont {Montaner-Piz{\'{a}}}, \citenamefont {Moore}, \citenamefont {N{\'{a}}cher}, \citenamefont {Orrigo}, \citenamefont {Penttil{\"{a}}}, \citenamefont {Pohjalainen}, \citenamefont {Porta}, \citenamefont {Reinikainen}, \citenamefont {Reponen}, \citenamefont {Rinta-Antila}, \citenamefont {Rubio}, \citenamefont {Rytk{\"{o}}nen}, \citenamefont {Sarriguren},
  \citenamefont {Shiba}, \citenamefont {Sonnenschein}, \citenamefont {Sonzogni}, \citenamefont {Valencia}, \citenamefont {Vedia}, \citenamefont {Voss}, \citenamefont {Wilson},\ and\ \citenamefont {Zakari-Issoufou}}]{Guadilla2019}%
  \BibitemOpen
  \bibfield  {author} {\bibinfo {author} {\bibfnamefont {V.}~\bibnamefont {Guadilla}}, \bibinfo {author} {\bibfnamefont {A.}~\bibnamefont {Algora}}, \bibinfo {author} {\bibfnamefont {J.~L.}\ \bibnamefont {Tain}}, \bibinfo {author} {\bibfnamefont {J.}~\bibnamefont {Agramunt}}, \bibinfo {author} {\bibfnamefont {J.}~\bibnamefont {Aysto}}, \bibinfo {author} {\bibfnamefont {J.~A.}\ \bibnamefont {Briz}}, \bibinfo {author} {\bibfnamefont {A.}~\bibnamefont {Cucoanes}}, \bibinfo {author} {\bibfnamefont {T.}~\bibnamefont {Eronen}}, \bibinfo {author} {\bibfnamefont {M.}~\bibnamefont {Estienne}}, \bibinfo {author} {\bibfnamefont {M.}~\bibnamefont {Fallot}}, \bibinfo {author} {\bibfnamefont {L.~M.}\ \bibnamefont {Fraile}}, \bibinfo {author} {\bibfnamefont {E.}~\bibnamefont {Ganioglu}}, \bibinfo {author} {\bibfnamefont {W.}~\bibnamefont {Gelletly}}, \bibinfo {author} {\bibfnamefont {D.}~\bibnamefont {Gorelov}}, \bibinfo {author} {\bibfnamefont {J.}~\bibnamefont {Hakala}}, \bibinfo {author} {\bibfnamefont {A.}~\bibnamefont
  {Jokinen}}, \bibinfo {author} {\bibfnamefont {D.}~\bibnamefont {Jordan}}, \bibinfo {author} {\bibfnamefont {A.}~\bibnamefont {Kankainen}}, \bibinfo {author} {\bibfnamefont {V.}~\bibnamefont {Kolhinen}}, \bibinfo {author} {\bibfnamefont {J.}~\bibnamefont {Koponen}}, \bibinfo {author} {\bibfnamefont {M.}~\bibnamefont {Lebois}}, \bibinfo {author} {\bibfnamefont {L.}~\bibnamefont {{Le Meur}}}, \bibinfo {author} {\bibfnamefont {T.}~\bibnamefont {Martinez}}, \bibinfo {author} {\bibfnamefont {M.}~\bibnamefont {Monserrate}}, \bibinfo {author} {\bibfnamefont {A.}~\bibnamefont {Montaner-Piz{\'{a}}}}, \bibinfo {author} {\bibfnamefont {I.}~\bibnamefont {Moore}}, \bibinfo {author} {\bibfnamefont {E.}~\bibnamefont {N{\'{a}}cher}}, \bibinfo {author} {\bibfnamefont {S.~E.~A.}\ \bibnamefont {Orrigo}}, \bibinfo {author} {\bibfnamefont {H.}~\bibnamefont {Penttil{\"{a}}}}, \bibinfo {author} {\bibfnamefont {I.}~\bibnamefont {Pohjalainen}}, \bibinfo {author} {\bibfnamefont {A.}~\bibnamefont {Porta}}, \bibinfo {author}
  {\bibfnamefont {J.}~\bibnamefont {Reinikainen}}, \bibinfo {author} {\bibfnamefont {M.}~\bibnamefont {Reponen}}, \bibinfo {author} {\bibfnamefont {S.}~\bibnamefont {Rinta-Antila}}, \bibinfo {author} {\bibfnamefont {B.}~\bibnamefont {Rubio}}, \bibinfo {author} {\bibfnamefont {K.}~\bibnamefont {Rytk{\"{o}}nen}}, \bibinfo {author} {\bibfnamefont {P.}~\bibnamefont {Sarriguren}}, \bibinfo {author} {\bibfnamefont {T.}~\bibnamefont {Shiba}}, \bibinfo {author} {\bibfnamefont {V.}~\bibnamefont {Sonnenschein}}, \bibinfo {author} {\bibfnamefont {A.~A.}\ \bibnamefont {Sonzogni}}, \bibinfo {author} {\bibfnamefont {E.}~\bibnamefont {Valencia}}, \bibinfo {author} {\bibfnamefont {V.}~\bibnamefont {Vedia}}, \bibinfo {author} {\bibfnamefont {A.}~\bibnamefont {Voss}}, \bibinfo {author} {\bibfnamefont {J.~N.}\ \bibnamefont {Wilson}}, \ and\ \bibinfo {author} {\bibfnamefont {A.-A.}\ \bibnamefont {Zakari-Issoufou}},\ }\href {\doibase 10.1103/PhysRevC.100.024311} {\bibfield  {journal} {\bibinfo  {journal} {Physical Review C}\
  }\textbf {\bibinfo {volume} {100}},\ \bibinfo {pages} {024311} (\bibinfo {year} {2019})},\ \Eprint {http://arxiv.org/abs/1904.07036} {arXiv:1904.07036} \BibitemShut {NoStop}%
\bibitem [{\citenamefont {Zakari-Issoufou}\ \emph {et~al.}(2015)\citenamefont {Zakari-Issoufou}, \citenamefont {Fallot}, \citenamefont {Porta}, \citenamefont {Algora}, \citenamefont {Tain}, \citenamefont {Valencia}, \citenamefont {Rice}, \citenamefont {Bui}, \citenamefont {Cormon}, \citenamefont {Estienne}, \citenamefont {Agramunt}, \citenamefont {AystO}, \citenamefont {Bowry}, \citenamefont {Briz}, \citenamefont {Caballero-Folch}, \citenamefont {Cano-Ott}, \citenamefont {Cucoanes}, \citenamefont {Elomaa}, \citenamefont {Eronen}, \citenamefont {Estévez}, \citenamefont {Farrelly}, \citenamefont {Garcia}, \citenamefont {Gelletly}, \citenamefont {Gomez-Hornillos}, \citenamefont {Gorlychev}, \citenamefont {Hakala}, \citenamefont {Jokinen}, \citenamefont {Jordan}, \citenamefont {Kankainen}, \citenamefont {Karvonen}, \citenamefont {Kolhinen}, \citenamefont {Kondev}, \citenamefont {Martinez}, \citenamefont {Mendoza}, \citenamefont {Molina}, \citenamefont {Moore}, \citenamefont {Perez-Cerdan}, \citenamefont
  {Podolyak}, \citenamefont {Penttila}, \citenamefont {Regan}, \citenamefont {Reponen}, \citenamefont {Rissanen}, \citenamefont {Rubio}, \citenamefont {Shiba}, \citenamefont {Sonzogni},\ and\ \citenamefont {Weber}}]{Zakari-Issoufou2015}%
  \BibitemOpen
  \bibfield  {author} {\bibinfo {author} {\bibfnamefont {A.-A.}\ \bibnamefont {Zakari-Issoufou}}, \bibinfo {author} {\bibfnamefont {M.}~\bibnamefont {Fallot}}, \bibinfo {author} {\bibfnamefont {A.}~\bibnamefont {Porta}}, \bibinfo {author} {\bibfnamefont {A.}~\bibnamefont {Algora}}, \bibinfo {author} {\bibfnamefont {J.~L.}\ \bibnamefont {Tain}}, \bibinfo {author} {\bibfnamefont {E.}~\bibnamefont {Valencia}}, \bibinfo {author} {\bibfnamefont {S.}~\bibnamefont {Rice}}, \bibinfo {author} {\bibfnamefont {V.~M.}\ \bibnamefont {Bui}}, \bibinfo {author} {\bibfnamefont {S.}~\bibnamefont {Cormon}}, \bibinfo {author} {\bibfnamefont {M.}~\bibnamefont {Estienne}}, \bibinfo {author} {\bibfnamefont {J.}~\bibnamefont {Agramunt}}, \bibinfo {author} {\bibfnamefont {J.}~\bibnamefont {AystO}}, \bibinfo {author} {\bibfnamefont {M.}~\bibnamefont {Bowry}}, \bibinfo {author} {\bibfnamefont {J.~A.}\ \bibnamefont {Briz}}, \bibinfo {author} {\bibfnamefont {R.}~\bibnamefont {Caballero-Folch}}, \bibinfo {author} {\bibfnamefont
  {D.}~\bibnamefont {Cano-Ott}}, \bibinfo {author} {\bibfnamefont {A.}~\bibnamefont {Cucoanes}}, \bibinfo {author} {\bibfnamefont {V.-V.}\ \bibnamefont {Elomaa}}, \bibinfo {author} {\bibfnamefont {T.}~\bibnamefont {Eronen}}, \bibinfo {author} {\bibfnamefont {E.}~\bibnamefont {Estévez}}, \bibinfo {author} {\bibfnamefont {G.~F.}\ \bibnamefont {Farrelly}}, \bibinfo {author} {\bibfnamefont {A.~R.}\ \bibnamefont {Garcia}}, \bibinfo {author} {\bibfnamefont {W.}~\bibnamefont {Gelletly}}, \bibinfo {author} {\bibfnamefont {M.~B.}\ \bibnamefont {Gomez-Hornillos}}, \bibinfo {author} {\bibfnamefont {V.}~\bibnamefont {Gorlychev}}, \bibinfo {author} {\bibfnamefont {J.}~\bibnamefont {Hakala}}, \bibinfo {author} {\bibfnamefont {A.}~\bibnamefont {Jokinen}}, \bibinfo {author} {\bibfnamefont {M.~D.}\ \bibnamefont {Jordan}}, \bibinfo {author} {\bibfnamefont {A.}~\bibnamefont {Kankainen}}, \bibinfo {author} {\bibfnamefont {P.}~\bibnamefont {Karvonen}}, \bibinfo {author} {\bibfnamefont {V.~S.}\ \bibnamefont {Kolhinen}}, \bibinfo
  {author} {\bibfnamefont {F.~G.}\ \bibnamefont {Kondev}}, \bibinfo {author} {\bibfnamefont {T.}~\bibnamefont {Martinez}}, \bibinfo {author} {\bibfnamefont {E.}~\bibnamefont {Mendoza}}, \bibinfo {author} {\bibfnamefont {F.}~\bibnamefont {Molina}}, \bibinfo {author} {\bibfnamefont {I.}~\bibnamefont {Moore}}, \bibinfo {author} {\bibfnamefont {A.~B.}\ \bibnamefont {Perez-Cerdan}}, \bibinfo {author} {\bibfnamefont {Z.}~\bibnamefont {Podolyak}}, \bibinfo {author} {\bibfnamefont {H.}~\bibnamefont {Penttila}}, \bibinfo {author} {\bibfnamefont {P.~H.}\ \bibnamefont {Regan}}, \bibinfo {author} {\bibfnamefont {M.}~\bibnamefont {Reponen}}, \bibinfo {author} {\bibfnamefont {J.}~\bibnamefont {Rissanen}}, \bibinfo {author} {\bibfnamefont {B.}~\bibnamefont {Rubio}}, \bibinfo {author} {\bibfnamefont {T.}~\bibnamefont {Shiba}}, \bibinfo {author} {\bibfnamefont {A.~A.}\ \bibnamefont {Sonzogni}}, \ and\ \bibinfo {author} {\bibfnamefont {C.}~\bibnamefont {Weber}},\ }\href {\doibase 10.1103/PhysRevLett.115.102503} {\bibfield
  {journal} {\bibinfo  {journal} {Physical Review Letters}\ }\textbf {\bibinfo {volume} {115}},\ \bibinfo {pages} {102503} (\bibinfo {year} {2015})},\ \Eprint {http://arxiv.org/abs/1504.05812} {arXiv:1504.05812} \BibitemShut {NoStop}%
\bibitem [{\citenamefont {Fallot}\ \emph {et~al.}(2019)\citenamefont {Fallot}, \citenamefont {Littlejohn},\ and\ \citenamefont {Dimitriou}}]{Fallot2019}%
  \BibitemOpen
  \bibfield  {author} {\bibinfo {author} {\bibfnamefont {M.}~\bibnamefont {Fallot}}, \bibinfo {author} {\bibfnamefont {B.}~\bibnamefont {Littlejohn}}, \ and\ \bibinfo {author} {\bibfnamefont {P.}~\bibnamefont {Dimitriou}},\ }\href {https://www-nds.iaea.org/publications/indc/indc-nds-0786.pdf} {\emph {\bibinfo {title} {{Antineutrino spectra and their applications}}}},\ \bibinfo {type} {Tech. Rep.}\ \bibinfo {number} {INDC(NDS)-0786}\ (\bibinfo  {institution} {International Atomic Energy Agency},\ \bibinfo {year} {2019})\BibitemShut {NoStop}%
\bibitem [{\citenamefont {Hayen}\ \emph {et~al.}(2018)\citenamefont {Hayen}, \citenamefont {Severijns}, \citenamefont {Bodek}, \citenamefont {Rozpedzik},\ and\ \citenamefont {Mougeot}}]{Hayen2018}%
  \BibitemOpen
  \bibfield  {author} {\bibinfo {author} {\bibfnamefont {L.}~\bibnamefont {Hayen}}, \bibinfo {author} {\bibfnamefont {N.}~\bibnamefont {Severijns}}, \bibinfo {author} {\bibfnamefont {K.}~\bibnamefont {Bodek}}, \bibinfo {author} {\bibfnamefont {D.}~\bibnamefont {Rozpedzik}}, \ and\ \bibinfo {author} {\bibfnamefont {X.}~\bibnamefont {Mougeot}},\ }\href {\doibase 10.1103/RevModPhys.90.015008} {\bibfield  {journal} {\bibinfo  {journal} {Reviews of Modern Physics}\ }\textbf {\bibinfo {volume} {90}},\ \bibinfo {pages} {015008} (\bibinfo {year} {2018})},\ \Eprint {http://arxiv.org/abs/1709.07530} {arXiv:1709.07530} \BibitemShut {NoStop}%
\bibitem [{\citenamefont {Sirlin}(2011)}]{Sirlin2011}%
  \BibitemOpen
  \bibfield  {author} {\bibinfo {author} {\bibfnamefont {A.}~\bibnamefont {Sirlin}},\ }\href {\doibase 10.1103/PhysRevD.84.014021} {\bibfield  {journal} {\bibinfo  {journal} {Physical Review D}\ }\textbf {\bibinfo {volume} {84}},\ \bibinfo {pages} {014021} (\bibinfo {year} {2011})},\ \Eprint {http://arxiv.org/abs/arXiv:1105.2842v2} {arXiv:arXiv:1105.2842v2} \BibitemShut {NoStop}%
\bibitem [{\citenamefont {Hayen}\ \emph {et~al.}(2019{\natexlab{a}})\citenamefont {Hayen}, \citenamefont {Kostensalo}, \citenamefont {Severijns},\ and\ \citenamefont {Suhonen}}]{Hayen2019}%
  \BibitemOpen
  \bibfield  {author} {\bibinfo {author} {\bibfnamefont {L.}~\bibnamefont {Hayen}}, \bibinfo {author} {\bibfnamefont {J.}~\bibnamefont {Kostensalo}}, \bibinfo {author} {\bibfnamefont {N.}~\bibnamefont {Severijns}}, \ and\ \bibinfo {author} {\bibfnamefont {J.}~\bibnamefont {Suhonen}},\ }\href {\doibase 10.1103/PhysRevC.99.031301} {\bibfield  {journal} {\bibinfo  {journal} {Physical Review C}\ }\textbf {\bibinfo {volume} {99}},\ \bibinfo {pages} {031301(R)} (\bibinfo {year} {2019}{\natexlab{a}})}\BibitemShut {NoStop}%
\bibitem [{\citenamefont {Hayen}\ \emph {et~al.}(2019{\natexlab{b}})\citenamefont {Hayen}, \citenamefont {Kostensalo}, \citenamefont {Severijns},\ and\ \citenamefont {Suhonen}}]{Hayen2019b}%
  \BibitemOpen
  \bibfield  {author} {\bibinfo {author} {\bibfnamefont {L.}~\bibnamefont {Hayen}}, \bibinfo {author} {\bibfnamefont {J.}~\bibnamefont {Kostensalo}}, \bibinfo {author} {\bibfnamefont {N.}~\bibnamefont {Severijns}}, \ and\ \bibinfo {author} {\bibfnamefont {J.}~\bibnamefont {Suhonen}},\ }\href {\doibase 10.1103/PhysRevC.100.054323} {\bibfield  {journal} {\bibinfo  {journal} {Physical Review C}\ }\textbf {\bibinfo {volume} {100}},\ \bibinfo {pages} {054323} (\bibinfo {year} {2019}{\natexlab{b}})},\ \Eprint {http://arxiv.org/abs/1805.12259} {arXiv:1805.12259} \BibitemShut {NoStop}%
\bibitem [{\citenamefont {Bonhomme}\ \emph {et~al.}(2022)\citenamefont {Bonhomme}, \citenamefont {Bonet}, \citenamefont {Buck}, \citenamefont {Hakenm{\"{u}}ller}, \citenamefont {Heusser}, \citenamefont {Hugle}, \citenamefont {Lindner}, \citenamefont {Maneschg}, \citenamefont {Nolte}, \citenamefont {Rink}, \citenamefont {Pirovano},\ and\ \citenamefont {Strecker}}]{Bonhomme2022}%
  \BibitemOpen
  \bibfield  {author} {\bibinfo {author} {\bibfnamefont {A.}~\bibnamefont {Bonhomme}}, \bibinfo {author} {\bibfnamefont {H.}~\bibnamefont {Bonet}}, \bibinfo {author} {\bibfnamefont {C.}~\bibnamefont {Buck}}, \bibinfo {author} {\bibfnamefont {J.}~\bibnamefont {Hakenm{\"{u}}ller}}, \bibinfo {author} {\bibfnamefont {G.}~\bibnamefont {Heusser}}, \bibinfo {author} {\bibfnamefont {T.}~\bibnamefont {Hugle}}, \bibinfo {author} {\bibfnamefont {M.}~\bibnamefont {Lindner}}, \bibinfo {author} {\bibfnamefont {W.}~\bibnamefont {Maneschg}}, \bibinfo {author} {\bibfnamefont {R.}~\bibnamefont {Nolte}}, \bibinfo {author} {\bibfnamefont {T.}~\bibnamefont {Rink}}, \bibinfo {author} {\bibfnamefont {E.}~\bibnamefont {Pirovano}}, \ and\ \bibinfo {author} {\bibfnamefont {H.}~\bibnamefont {Strecker}},\ }\href {\doibase 10.1140/epjc/s10052-022-10768-1} {\bibfield  {journal} {\bibinfo  {journal} {The European Physical Journal C}\ }\textbf {\bibinfo {volume} {82}},\ \bibinfo {pages} {815} (\bibinfo {year} {2022})},\ \Eprint
  {http://arxiv.org/abs/2202.03754} {arXiv:2202.03754} \BibitemShut {NoStop}%
\bibitem [{\citenamefont {Khan}\ and\ \citenamefont {Rodejohann}(2019)}]{Khan2019}%
  \BibitemOpen
  \bibfield  {author} {\bibinfo {author} {\bibfnamefont {A.~N.}\ \bibnamefont {Khan}}\ and\ \bibinfo {author} {\bibfnamefont {W.}~\bibnamefont {Rodejohann}},\ }\href {\doibase 10.1103/PhysRevD.100.113003} {\bibfield  {journal} {\bibinfo  {journal} {Physical Review D}\ }\textbf {\bibinfo {volume} {100}},\ \bibinfo {pages} {113003} (\bibinfo {year} {2019})}\BibitemShut {NoStop}%
\bibitem [{\citenamefont {{Daya Bay collaboration}}\ \emph {et~al.}(2021)\citenamefont {{Daya Bay collaboration}}, \citenamefont {An}, \citenamefont {Balantekin}, \citenamefont {Band}, \citenamefont {Bishai}, \citenamefont {Blyth}, \citenamefont {Cao}, \citenamefont {Cao}, \citenamefont {Chang}, \citenamefont {Chang}, \citenamefont {Chen}, \citenamefont {Chen}, \citenamefont {Chen}, \citenamefont {Chen}, \citenamefont {Cheng}, \citenamefont {Cheng}, \citenamefont {Cherwinka}, \citenamefont {Chu}, \citenamefont {Cummings}, \citenamefont {Dalager}, \citenamefont {Deng}, \citenamefont {Ding}, \citenamefont {Diwan}, \citenamefont {Dohnal}, \citenamefont {Dove}, \citenamefont {Dvorak}, \citenamefont {Dwyer}, \citenamefont {Gallo}, \citenamefont {Gonchar}, \citenamefont {Gong}, \citenamefont {Gong}, \citenamefont {Gu}, \citenamefont {Guo}, \citenamefont {Guo}, \citenamefont {Guo}, \citenamefont {Guo}, \citenamefont {Guo}, \citenamefont {Hackenburg}, \citenamefont {Hans}, \citenamefont {He}, \citenamefont {Heeger},
  \citenamefont {Heng}, \citenamefont {Higuera}, \citenamefont {Hor}, \citenamefont {Hsiung}, \citenamefont {Hu}, \citenamefont {Hu}, \citenamefont {Hu}, \citenamefont {Hu}, \citenamefont {Huang}, \citenamefont {Huang}, \citenamefont {Huang}, \citenamefont {Huber}, \citenamefont {Jaffe}, \citenamefont {Jen}, \citenamefont {Ji}, \citenamefont {Ji}, \citenamefont {Johnson}, \citenamefont {Jones}, \citenamefont {Kang}, \citenamefont {Kettell}, \citenamefont {Kohn}, \citenamefont {Kramer}, \citenamefont {Langford}, \citenamefont {Lee}, \citenamefont {Lee}, \citenamefont {Lei}, \citenamefont {Leitner}, \citenamefont {Leung}, \citenamefont {Li}, \citenamefont {Li}, \citenamefont {Li}, \citenamefont {Li}, \citenamefont {Li}, \citenamefont {Li}, \citenamefont {Li}, \citenamefont {Li}, \citenamefont {Li}, \citenamefont {Li}, \citenamefont {Liang}, \citenamefont {Lin}, \citenamefont {Lin}, \citenamefont {Lin}, \citenamefont {Ling}, \citenamefont {Link}, \citenamefont {Littenberg}, \citenamefont {Littlejohn},
  \citenamefont {Liu}, \citenamefont {Liu}, \citenamefont {Lu}, \citenamefont {Lu}, \citenamefont {Lu}, \citenamefont {Luk}, \citenamefont {Ma}, \citenamefont {Ma}, \citenamefont {Ma}, \citenamefont {Marshall}, \citenamefont {Caicedo}, \citenamefont {McDonald}, \citenamefont {McKeown}, \citenamefont {Meng}, \citenamefont {Napolitano}, \citenamefont {Naumov}, \citenamefont {Naumova}, \citenamefont {Ochoa-Ricoux}, \citenamefont {Olshevskiy}, \citenamefont {Pan}, \citenamefont {Park}, \citenamefont {Patton}, \citenamefont {Peng}, \citenamefont {Pun}, \citenamefont {Qi}, \citenamefont {Qi}, \citenamefont {Qian}, \citenamefont {Raper}, \citenamefont {Ren}, \citenamefont {Reveco}, \citenamefont {Rosero}, \citenamefont {Roskovec}, \citenamefont {Ruan}, \citenamefont {Steiner}, \citenamefont {Sun}, \citenamefont {Tmej}, \citenamefont {Treskov}, \citenamefont {Tse}, \citenamefont {Tull}, \citenamefont {Viren}, \citenamefont {Vorobel}, \citenamefont {Wang}, \citenamefont {Wang}, \citenamefont {Wang}, \citenamefont
  {Wang}, \citenamefont {Wang}, \citenamefont {Wang}, \citenamefont {Wang}, \citenamefont {Wang}, \citenamefont {Wang}, \citenamefont {Wang}, \citenamefont {Wang}, \citenamefont {Wang}, \citenamefont {Wang}, \citenamefont {Wei}, \citenamefont {Wei}, \citenamefont {Wen}, \citenamefont {Whisnant}, \citenamefont {White}, \citenamefont {Wong}, \citenamefont {Worcester}, \citenamefont {Wu}, \citenamefont {Wu}, \citenamefont {Wu}, \citenamefont {Wu}, \citenamefont {Xia}, \citenamefont {Xie}, \citenamefont {Xing}, \citenamefont {Xu}, \citenamefont {Xu}, \citenamefont {Xue}, \citenamefont {Yang}, \citenamefont {Yang}, \citenamefont {Yang}, \citenamefont {Yao}, \citenamefont {Ye}, \citenamefont {Yeh}, \citenamefont {Young}, \citenamefont {Yu}, \citenamefont {Yu}, \citenamefont {Yue}, \citenamefont {Zeng}, \citenamefont {Zeng}, \citenamefont {Zhan}, \citenamefont {Zhang}, \citenamefont {Zhang}, \citenamefont {Zhang}, \citenamefont {Zhang}, \citenamefont {Zhang}, \citenamefont {Zhang}, \citenamefont {Zhang},
  \citenamefont {Zhang}, \citenamefont {Zhang}, \citenamefont {Zhang}, \citenamefont {Zhang}, \citenamefont {Zhang}, \citenamefont {Zhao}, \citenamefont {Zhou}, \citenamefont {Zhuang},\ and\ \citenamefont {Zou}}]{DayaBaycollaboration2021}%
  \BibitemOpen
  \bibfield  {author} {\bibinfo {author} {\bibnamefont {{Daya Bay collaboration}}}, \bibinfo {author} {\bibfnamefont {F.~P.}\ \bibnamefont {An}}, \bibinfo {author} {\bibfnamefont {A.~B.}\ \bibnamefont {Balantekin}}, \bibinfo {author} {\bibfnamefont {H.~R.}\ \bibnamefont {Band}}, \bibinfo {author} {\bibfnamefont {M.}~\bibnamefont {Bishai}}, \bibinfo {author} {\bibfnamefont {S.}~\bibnamefont {Blyth}}, \bibinfo {author} {\bibfnamefont {G.~F.}\ \bibnamefont {Cao}}, \bibinfo {author} {\bibfnamefont {J.}~\bibnamefont {Cao}}, \bibinfo {author} {\bibfnamefont {J.~F.}\ \bibnamefont {Chang}}, \bibinfo {author} {\bibfnamefont {Y.}~\bibnamefont {Chang}}, \bibinfo {author} {\bibfnamefont {H.~S.}\ \bibnamefont {Chen}}, \bibinfo {author} {\bibfnamefont {S.~M.}\ \bibnamefont {Chen}}, \bibinfo {author} {\bibfnamefont {Y.}~\bibnamefont {Chen}}, \bibinfo {author} {\bibfnamefont {Y.~X.}\ \bibnamefont {Chen}}, \bibinfo {author} {\bibfnamefont {J.}~\bibnamefont {Cheng}}, \bibinfo {author} {\bibfnamefont {Z.~K.}\ \bibnamefont
  {Cheng}}, \bibinfo {author} {\bibfnamefont {J.~J.}\ \bibnamefont {Cherwinka}}, \bibinfo {author} {\bibfnamefont {M.~C.}\ \bibnamefont {Chu}}, \bibinfo {author} {\bibfnamefont {J.~P.}\ \bibnamefont {Cummings}}, \bibinfo {author} {\bibfnamefont {O.}~\bibnamefont {Dalager}}, \bibinfo {author} {\bibfnamefont {F.~S.}\ \bibnamefont {Deng}}, \bibinfo {author} {\bibfnamefont {Y.~Y.}\ \bibnamefont {Ding}}, \bibinfo {author} {\bibfnamefont {M.~V.}\ \bibnamefont {Diwan}}, \bibinfo {author} {\bibfnamefont {T.}~\bibnamefont {Dohnal}}, \bibinfo {author} {\bibfnamefont {J.}~\bibnamefont {Dove}}, \bibinfo {author} {\bibfnamefont {M.}~\bibnamefont {Dvorak}}, \bibinfo {author} {\bibfnamefont {D.~A.}\ \bibnamefont {Dwyer}}, \bibinfo {author} {\bibfnamefont {J.~P.}\ \bibnamefont {Gallo}}, \bibinfo {author} {\bibfnamefont {M.}~\bibnamefont {Gonchar}}, \bibinfo {author} {\bibfnamefont {G.~H.}\ \bibnamefont {Gong}}, \bibinfo {author} {\bibfnamefont {H.}~\bibnamefont {Gong}}, \bibinfo {author} {\bibfnamefont {W.~Q.}\ \bibnamefont
  {Gu}}, \bibinfo {author} {\bibfnamefont {J.~Y.}\ \bibnamefont {Guo}}, \bibinfo {author} {\bibfnamefont {L.}~\bibnamefont {Guo}}, \bibinfo {author} {\bibfnamefont {X.~H.}\ \bibnamefont {Guo}}, \bibinfo {author} {\bibfnamefont {Y.~H.}\ \bibnamefont {Guo}}, \bibinfo {author} {\bibfnamefont {Z.}~\bibnamefont {Guo}}, \bibinfo {author} {\bibfnamefont {R.~W.}\ \bibnamefont {Hackenburg}}, \bibinfo {author} {\bibfnamefont {S.}~\bibnamefont {Hans}}, \bibinfo {author} {\bibfnamefont {M.}~\bibnamefont {He}}, \bibinfo {author} {\bibfnamefont {K.~M.}\ \bibnamefont {Heeger}}, \bibinfo {author} {\bibfnamefont {Y.~K.}\ \bibnamefont {Heng}}, \bibinfo {author} {\bibfnamefont {A.}~\bibnamefont {Higuera}}, \bibinfo {author} {\bibfnamefont {Y.~K.}\ \bibnamefont {Hor}}, \bibinfo {author} {\bibfnamefont {Y.~B.}\ \bibnamefont {Hsiung}}, \bibinfo {author} {\bibfnamefont {B.~Z.}\ \bibnamefont {Hu}}, \bibinfo {author} {\bibfnamefont {J.~R.}\ \bibnamefont {Hu}}, \bibinfo {author} {\bibfnamefont {T.}~\bibnamefont {Hu}}, \bibinfo
  {author} {\bibfnamefont {Z.~J.}\ \bibnamefont {Hu}}, \bibinfo {author} {\bibfnamefont {H.~X.}\ \bibnamefont {Huang}}, \bibinfo {author} {\bibfnamefont {X.~T.}\ \bibnamefont {Huang}}, \bibinfo {author} {\bibfnamefont {Y.~B.}\ \bibnamefont {Huang}}, \bibinfo {author} {\bibfnamefont {P.}~\bibnamefont {Huber}}, \bibinfo {author} {\bibfnamefont {D.~E.}\ \bibnamefont {Jaffe}}, \bibinfo {author} {\bibfnamefont {K.~L.}\ \bibnamefont {Jen}}, \bibinfo {author} {\bibfnamefont {X.~L.}\ \bibnamefont {Ji}}, \bibinfo {author} {\bibfnamefont {X.~P.}\ \bibnamefont {Ji}}, \bibinfo {author} {\bibfnamefont {R.~A.}\ \bibnamefont {Johnson}}, \bibinfo {author} {\bibfnamefont {D.}~\bibnamefont {Jones}}, \bibinfo {author} {\bibfnamefont {L.}~\bibnamefont {Kang}}, \bibinfo {author} {\bibfnamefont {S.~H.}\ \bibnamefont {Kettell}}, \bibinfo {author} {\bibfnamefont {S.}~\bibnamefont {Kohn}}, \bibinfo {author} {\bibfnamefont {M.}~\bibnamefont {Kramer}}, \bibinfo {author} {\bibfnamefont {T.~J.}\ \bibnamefont {Langford}}, \bibinfo
  {author} {\bibfnamefont {J.}~\bibnamefont {Lee}}, \bibinfo {author} {\bibfnamefont {J.~H.~C.}\ \bibnamefont {Lee}}, \bibinfo {author} {\bibfnamefont {R.~T.}\ \bibnamefont {Lei}}, \bibinfo {author} {\bibfnamefont {R.}~\bibnamefont {Leitner}}, \bibinfo {author} {\bibfnamefont {J.~K.~C.}\ \bibnamefont {Leung}}, \bibinfo {author} {\bibfnamefont {F.}~\bibnamefont {Li}}, \bibinfo {author} {\bibfnamefont {J.~J.}\ \bibnamefont {Li}}, \bibinfo {author} {\bibfnamefont {Q.~J.}\ \bibnamefont {Li}}, \bibinfo {author} {\bibfnamefont {S.}~\bibnamefont {Li}}, \bibinfo {author} {\bibfnamefont {S.~C.}\ \bibnamefont {Li}}, \bibinfo {author} {\bibfnamefont {W.~D.}\ \bibnamefont {Li}}, \bibinfo {author} {\bibfnamefont {X.~N.}\ \bibnamefont {Li}}, \bibinfo {author} {\bibfnamefont {X.~Q.}\ \bibnamefont {Li}}, \bibinfo {author} {\bibfnamefont {Y.~F.}\ \bibnamefont {Li}}, \bibinfo {author} {\bibfnamefont {Z.~B.}\ \bibnamefont {Li}}, \bibinfo {author} {\bibfnamefont {H.}~\bibnamefont {Liang}}, \bibinfo {author} {\bibfnamefont
  {C.~J.}\ \bibnamefont {Lin}}, \bibinfo {author} {\bibfnamefont {G.~L.}\ \bibnamefont {Lin}}, \bibinfo {author} {\bibfnamefont {S.}~\bibnamefont {Lin}}, \bibinfo {author} {\bibfnamefont {J.~J.}\ \bibnamefont {Ling}}, \bibinfo {author} {\bibfnamefont {J.~M.}\ \bibnamefont {Link}}, \bibinfo {author} {\bibfnamefont {L.}~\bibnamefont {Littenberg}}, \bibinfo {author} {\bibfnamefont {B.~R.}\ \bibnamefont {Littlejohn}}, \bibinfo {author} {\bibfnamefont {J.~C.}\ \bibnamefont {Liu}}, \bibinfo {author} {\bibfnamefont {J.~L.}\ \bibnamefont {Liu}}, \bibinfo {author} {\bibfnamefont {C.}~\bibnamefont {Lu}}, \bibinfo {author} {\bibfnamefont {H.~Q.}\ \bibnamefont {Lu}}, \bibinfo {author} {\bibfnamefont {J.~S.}\ \bibnamefont {Lu}}, \bibinfo {author} {\bibfnamefont {K.~B.}\ \bibnamefont {Luk}}, \bibinfo {author} {\bibfnamefont {X.~B.}\ \bibnamefont {Ma}}, \bibinfo {author} {\bibfnamefont {X.~Y.}\ \bibnamefont {Ma}}, \bibinfo {author} {\bibfnamefont {Y.~Q.}\ \bibnamefont {Ma}}, \bibinfo {author} {\bibfnamefont
  {C.}~\bibnamefont {Marshall}}, \bibinfo {author} {\bibfnamefont {D.~A.~M.}\ \bibnamefont {Caicedo}}, \bibinfo {author} {\bibfnamefont {K.~T.}\ \bibnamefont {McDonald}}, \bibinfo {author} {\bibfnamefont {R.~D.}\ \bibnamefont {McKeown}}, \bibinfo {author} {\bibfnamefont {Y.}~\bibnamefont {Meng}}, \bibinfo {author} {\bibfnamefont {J.}~\bibnamefont {Napolitano}}, \bibinfo {author} {\bibfnamefont {D.}~\bibnamefont {Naumov}}, \bibinfo {author} {\bibfnamefont {E.}~\bibnamefont {Naumova}}, \bibinfo {author} {\bibfnamefont {J.~P.}\ \bibnamefont {Ochoa-Ricoux}}, \bibinfo {author} {\bibfnamefont {A.}~\bibnamefont {Olshevskiy}}, \bibinfo {author} {\bibfnamefont {H.~R.}\ \bibnamefont {Pan}}, \bibinfo {author} {\bibfnamefont {J.}~\bibnamefont {Park}}, \bibinfo {author} {\bibfnamefont {S.}~\bibnamefont {Patton}}, \bibinfo {author} {\bibfnamefont {J.~C.}\ \bibnamefont {Peng}}, \bibinfo {author} {\bibfnamefont {C.~S.~J.}\ \bibnamefont {Pun}}, \bibinfo {author} {\bibfnamefont {F.~Z.}\ \bibnamefont {Qi}}, \bibinfo {author}
  {\bibfnamefont {M.}~\bibnamefont {Qi}}, \bibinfo {author} {\bibfnamefont {X.}~\bibnamefont {Qian}}, \bibinfo {author} {\bibfnamefont {N.}~\bibnamefont {Raper}}, \bibinfo {author} {\bibfnamefont {J.}~\bibnamefont {Ren}}, \bibinfo {author} {\bibfnamefont {C.~M.}\ \bibnamefont {Reveco}}, \bibinfo {author} {\bibfnamefont {R.}~\bibnamefont {Rosero}}, \bibinfo {author} {\bibfnamefont {B.}~\bibnamefont {Roskovec}}, \bibinfo {author} {\bibfnamefont {X.~C.}\ \bibnamefont {Ruan}}, \bibinfo {author} {\bibfnamefont {H.}~\bibnamefont {Steiner}}, \bibinfo {author} {\bibfnamefont {J.~L.}\ \bibnamefont {Sun}}, \bibinfo {author} {\bibfnamefont {T.}~\bibnamefont {Tmej}}, \bibinfo {author} {\bibfnamefont {K.}~\bibnamefont {Treskov}}, \bibinfo {author} {\bibfnamefont {W.~H.}\ \bibnamefont {Tse}}, \bibinfo {author} {\bibfnamefont {C.~E.}\ \bibnamefont {Tull}}, \bibinfo {author} {\bibfnamefont {B.}~\bibnamefont {Viren}}, \bibinfo {author} {\bibfnamefont {V.}~\bibnamefont {Vorobel}}, \bibinfo {author} {\bibfnamefont {C.~H.}\
  \bibnamefont {Wang}}, \bibinfo {author} {\bibfnamefont {J.}~\bibnamefont {Wang}}, \bibinfo {author} {\bibfnamefont {M.}~\bibnamefont {Wang}}, \bibinfo {author} {\bibfnamefont {N.~Y.}\ \bibnamefont {Wang}}, \bibinfo {author} {\bibfnamefont {R.~G.}\ \bibnamefont {Wang}}, \bibinfo {author} {\bibfnamefont {W.}~\bibnamefont {Wang}}, \bibinfo {author} {\bibfnamefont {W.}~\bibnamefont {Wang}}, \bibinfo {author} {\bibfnamefont {X.}~\bibnamefont {Wang}}, \bibinfo {author} {\bibfnamefont {Y.}~\bibnamefont {Wang}}, \bibinfo {author} {\bibfnamefont {Y.~F.}\ \bibnamefont {Wang}}, \bibinfo {author} {\bibfnamefont {Z.}~\bibnamefont {Wang}}, \bibinfo {author} {\bibfnamefont {Z.}~\bibnamefont {Wang}}, \bibinfo {author} {\bibfnamefont {Z.~M.}\ \bibnamefont {Wang}}, \bibinfo {author} {\bibfnamefont {H.~Y.}\ \bibnamefont {Wei}}, \bibinfo {author} {\bibfnamefont {L.~H.}\ \bibnamefont {Wei}}, \bibinfo {author} {\bibfnamefont {L.~J.}\ \bibnamefont {Wen}}, \bibinfo {author} {\bibfnamefont {K.}~\bibnamefont {Whisnant}}, \bibinfo
  {author} {\bibfnamefont {C.~G.}\ \bibnamefont {White}}, \bibinfo {author} {\bibfnamefont {H.~L.~H.}\ \bibnamefont {Wong}}, \bibinfo {author} {\bibfnamefont {E.}~\bibnamefont {Worcester}}, \bibinfo {author} {\bibfnamefont {D.~R.}\ \bibnamefont {Wu}}, \bibinfo {author} {\bibfnamefont {F.~L.}\ \bibnamefont {Wu}}, \bibinfo {author} {\bibfnamefont {Q.}~\bibnamefont {Wu}}, \bibinfo {author} {\bibfnamefont {W.~J.}\ \bibnamefont {Wu}}, \bibinfo {author} {\bibfnamefont {D.~M.}\ \bibnamefont {Xia}}, \bibinfo {author} {\bibfnamefont {Z.~Q.}\ \bibnamefont {Xie}}, \bibinfo {author} {\bibfnamefont {Z.~Z.}\ \bibnamefont {Xing}}, \bibinfo {author} {\bibfnamefont {J.~L.}\ \bibnamefont {Xu}}, \bibinfo {author} {\bibfnamefont {T.}~\bibnamefont {Xu}}, \bibinfo {author} {\bibfnamefont {T.}~\bibnamefont {Xue}}, \bibinfo {author} {\bibfnamefont {C.~G.}\ \bibnamefont {Yang}}, \bibinfo {author} {\bibfnamefont {L.}~\bibnamefont {Yang}}, \bibinfo {author} {\bibfnamefont {Y.~Z.}\ \bibnamefont {Yang}}, \bibinfo {author} {\bibfnamefont
  {H.~F.}\ \bibnamefont {Yao}}, \bibinfo {author} {\bibfnamefont {M.}~\bibnamefont {Ye}}, \bibinfo {author} {\bibfnamefont {M.}~\bibnamefont {Yeh}}, \bibinfo {author} {\bibfnamefont {B.~L.}\ \bibnamefont {Young}}, \bibinfo {author} {\bibfnamefont {H.~Z.}\ \bibnamefont {Yu}}, \bibinfo {author} {\bibfnamefont {Z.~Y.}\ \bibnamefont {Yu}}, \bibinfo {author} {\bibfnamefont {B.~B.}\ \bibnamefont {Yue}}, \bibinfo {author} {\bibfnamefont {S.}~\bibnamefont {Zeng}}, \bibinfo {author} {\bibfnamefont {Y.}~\bibnamefont {Zeng}}, \bibinfo {author} {\bibfnamefont {L.}~\bibnamefont {Zhan}}, \bibinfo {author} {\bibfnamefont {C.}~\bibnamefont {Zhang}}, \bibinfo {author} {\bibfnamefont {F.~Y.}\ \bibnamefont {Zhang}}, \bibinfo {author} {\bibfnamefont {H.~H.}\ \bibnamefont {Zhang}}, \bibinfo {author} {\bibfnamefont {J.~W.}\ \bibnamefont {Zhang}}, \bibinfo {author} {\bibfnamefont {Q.~M.}\ \bibnamefont {Zhang}}, \bibinfo {author} {\bibfnamefont {X.~T.}\ \bibnamefont {Zhang}}, \bibinfo {author} {\bibfnamefont {Y.~M.}\ \bibnamefont
  {Zhang}}, \bibinfo {author} {\bibfnamefont {Y.~X.}\ \bibnamefont {Zhang}}, \bibinfo {author} {\bibfnamefont {Y.~Y.}\ \bibnamefont {Zhang}}, \bibinfo {author} {\bibfnamefont {Z.~J.}\ \bibnamefont {Zhang}}, \bibinfo {author} {\bibfnamefont {Z.~P.}\ \bibnamefont {Zhang}}, \bibinfo {author} {\bibfnamefont {Z.~Y.}\ \bibnamefont {Zhang}}, \bibinfo {author} {\bibfnamefont {J.}~\bibnamefont {Zhao}}, \bibinfo {author} {\bibfnamefont {L.}~\bibnamefont {Zhou}}, \bibinfo {author} {\bibfnamefont {H.~L.}\ \bibnamefont {Zhuang}}, \ and\ \bibinfo {author} {\bibfnamefont {J.~H.}\ \bibnamefont {Zou}},\ }\href {http://arxiv.org/abs/2102.04614} {\ ,\ \bibinfo {pages} {1} (\bibinfo {year} {2021})},\ \Eprint {http://arxiv.org/abs/2102.04614} {arXiv:2102.04614} \BibitemShut {NoStop}%
\bibitem [{\citenamefont {Letourneau}\ \emph {et~al.}(2023)\citenamefont {Letourneau}, \citenamefont {Savu}, \citenamefont {Lhuillier}, \citenamefont {Lasserre}, \citenamefont {Materna}, \citenamefont {Mention}, \citenamefont {Mougeot}, \citenamefont {Onillon}, \citenamefont {Perisse},\ and\ \citenamefont {Vivier}}]{Letourneau2023}%
  \BibitemOpen
  \bibfield  {author} {\bibinfo {author} {\bibfnamefont {A.}~\bibnamefont {Letourneau}}, \bibinfo {author} {\bibfnamefont {V.}~\bibnamefont {Savu}}, \bibinfo {author} {\bibfnamefont {D.}~\bibnamefont {Lhuillier}}, \bibinfo {author} {\bibfnamefont {T.}~\bibnamefont {Lasserre}}, \bibinfo {author} {\bibfnamefont {T.}~\bibnamefont {Materna}}, \bibinfo {author} {\bibfnamefont {G.}~\bibnamefont {Mention}}, \bibinfo {author} {\bibfnamefont {X.}~\bibnamefont {Mougeot}}, \bibinfo {author} {\bibfnamefont {A.}~\bibnamefont {Onillon}}, \bibinfo {author} {\bibfnamefont {L.}~\bibnamefont {Perisse}}, \ and\ \bibinfo {author} {\bibfnamefont {M.}~\bibnamefont {Vivier}},\ }\href {\doibase 10.1103/PhysRevLett.130.021801} {\bibfield  {journal} {\bibinfo  {journal} {Physical Review Letters}\ }\textbf {\bibinfo {volume} {130}} (\bibinfo {year} {2023}),\ 10.1103/PhysRevLett.130.021801}\BibitemShut {NoStop}%
\bibitem [{\citenamefont {Behrens}\ and\ \citenamefont {J{\"{a}}necke}(1969)}]{Behrens1969}%
  \BibitemOpen
  \bibfield  {author} {\bibinfo {author} {\bibfnamefont {H.}~\bibnamefont {Behrens}}\ and\ \bibinfo {author} {\bibfnamefont {J.}~\bibnamefont {J{\"{a}}necke}},\ }\href@noop {} {\emph {\bibinfo {title} {{Landolt-B{\"{o}}rnstein Tables, Gruppe I, Band 4}}}}\ (\bibinfo  {publisher} {Springer},\ \bibinfo {year} {1969})\BibitemShut {NoStop}%
\bibitem [{\citenamefont {Behrens}\ and\ \citenamefont {B{\"{u}}hring}(1982)}]{Behrens1982}%
  \BibitemOpen
  \bibfield  {author} {\bibinfo {author} {\bibfnamefont {H.}~\bibnamefont {Behrens}}\ and\ \bibinfo {author} {\bibfnamefont {W.}~\bibnamefont {B{\"{u}}hring}},\ }\href@noop {} {\emph {\bibinfo {title} {{Electron radial wave functions and nuclear beta-decay}}}}\ (\bibinfo  {publisher} {Clarendon Press, Oxford},\ \bibinfo {year} {1982})\BibitemShut {NoStop}%
\bibitem [{\citenamefont {Nasteva}(2021)}]{Nasteva2021}%
  \BibitemOpen
  \bibfield  {author} {\bibinfo {author} {\bibfnamefont {I.}~\bibnamefont {Nasteva}},\ }\href {\doibase 10.1088/1742-6596/2156/1/012115} {\bibfield  {journal} {\bibinfo  {journal} {Journal of Physics: Conference Series}\ }\textbf {\bibinfo {volume} {2156}},\ \bibinfo {pages} {012115} (\bibinfo {year} {2021})}\BibitemShut {NoStop}%
\bibitem [{\citenamefont {Beda}\ \emph {et~al.}(2013)\citenamefont {Beda}, \citenamefont {Brudanin}, \citenamefont {Egorov}, \citenamefont {Medvedev}, \citenamefont {Pogosov}, \citenamefont {Shevchik}, \citenamefont {Shirchenko}, \citenamefont {Starostin},\ and\ \citenamefont {Zhitnikov}}]{Beda2013}%
  \BibitemOpen
  \bibfield  {author} {\bibinfo {author} {\bibfnamefont {A.~G.}\ \bibnamefont {Beda}}, \bibinfo {author} {\bibfnamefont {V.~B.}\ \bibnamefont {Brudanin}}, \bibinfo {author} {\bibfnamefont {V.~G.}\ \bibnamefont {Egorov}}, \bibinfo {author} {\bibfnamefont {D.~V.}\ \bibnamefont {Medvedev}}, \bibinfo {author} {\bibfnamefont {V.~S.}\ \bibnamefont {Pogosov}}, \bibinfo {author} {\bibfnamefont {E.~A.}\ \bibnamefont {Shevchik}}, \bibinfo {author} {\bibfnamefont {M.~V.}\ \bibnamefont {Shirchenko}}, \bibinfo {author} {\bibfnamefont {A.~S.}\ \bibnamefont {Starostin}}, \ and\ \bibinfo {author} {\bibfnamefont {I.~V.}\ \bibnamefont {Zhitnikov}},\ }\href {\doibase 10.1134/S1547477113020027} {\bibfield  {journal} {\bibinfo  {journal} {Physics of Particles and Nuclei Letters}\ }\textbf {\bibinfo {volume} {10}},\ \bibinfo {pages} {139} (\bibinfo {year} {2013})}\BibitemShut {NoStop}%
\bibitem [{\citenamefont {Agostini}\ \emph {et~al.}(2017)\citenamefont {Agostini}, \citenamefont {Altenmueller}, \citenamefont {Appel}, \citenamefont {Atroshchenko}, \citenamefont {Bagdasarian}, \citenamefont {Basilico}, \citenamefont {Bellini}, \citenamefont {Benziger}, \citenamefont {Bick}, \citenamefont {Bonfini}, \citenamefont {Bravo}, \citenamefont {Caccianiga}, \citenamefont {Calaprice}, \citenamefont {Caminata}, \citenamefont {Caprioli}, \citenamefont {Carlini}, \citenamefont {Cavalcante}, \citenamefont {Chepurnov}, \citenamefont {Choi}, \citenamefont {Collica}, \citenamefont {D'Angelo}, \citenamefont {Davini}, \citenamefont {Derbin}, \citenamefont {Ding}, \citenamefont {Ludovico}, \citenamefont {Noto}, \citenamefont {Drachnev}, \citenamefont {Fomenko}, \citenamefont {Formozov}, \citenamefont {Franco}, \citenamefont {Froborg}, \citenamefont {Gabriele}, \citenamefont {Galbiati}, \citenamefont {Ghiano}, \citenamefont {Giammarchi}, \citenamefont {Goretti}, \citenamefont {Gromov}, \citenamefont {Guffanti},
  \citenamefont {Hagner}, \citenamefont {Houdy}, \citenamefont {Hungerford}, \citenamefont {Ianni}, \citenamefont {Ianni}, \citenamefont {Jany}, \citenamefont {Jeschke}, \citenamefont {Kobychev}, \citenamefont {Korablev}, \citenamefont {Korga}, \citenamefont {Kryn}, \citenamefont {Laubenstein}, \citenamefont {Litvinovich}, \citenamefont {Lombardi}, \citenamefont {Lombardi}, \citenamefont {Ludhova}, \citenamefont {Lukyanchenko}, \citenamefont {Lukyanchenko}, \citenamefont {Machulin}, \citenamefont {Manuzio}, \citenamefont {Marcocci}, \citenamefont {Martyn}, \citenamefont {Meroni}, \citenamefont {Meyer}, \citenamefont {Miramonti}, \citenamefont {Misiaszek}, \citenamefont {Muratova}, \citenamefont {Neumair}, \citenamefont {Oberauer}, \citenamefont {Opitz}, \citenamefont {Orekhov}, \citenamefont {Ortica}, \citenamefont {Pallavicini}, \citenamefont {Papp}, \citenamefont {Penek}, \citenamefont {Pilipenko}, \citenamefont {Pocar}, \citenamefont {Porcelli}, \citenamefont {Ranucci}, \citenamefont {Razeto},
  \citenamefont {Re}, \citenamefont {Redchuk}, \citenamefont {Romani}, \citenamefont {Roncin}, \citenamefont {Rossi}, \citenamefont {Schoenert}, \citenamefont {Semenov}, \citenamefont {Skorokhvatov}, \citenamefont {Smirnov}, \citenamefont {Sotnikov}, \citenamefont {Stokes}, \citenamefont {Suvorov}, \citenamefont {Tartaglia}, \citenamefont {Testera}, \citenamefont {Thurn}, \citenamefont {Toropova}, \citenamefont {Unzhakov}, \citenamefont {Vishneva}, \citenamefont {Vogelaar}, \citenamefont {von Feilitzsch}, \citenamefont {Wang}, \citenamefont {Weinz}, \citenamefont {Wojcik}, \citenamefont {Wurm}, \citenamefont {Yokley}, \citenamefont {Zaimidoroga}, \citenamefont {Zavatarelli}, \citenamefont {Zuber},\ and\ \citenamefont {Zuzel}}]{Agostini2017a}%
  \BibitemOpen
  \bibfield  {author} {\bibinfo {author} {\bibfnamefont {M.}~\bibnamefont {Agostini}}, \bibinfo {author} {\bibfnamefont {K.}~\bibnamefont {Altenmueller}}, \bibinfo {author} {\bibfnamefont {S.}~\bibnamefont {Appel}}, \bibinfo {author} {\bibfnamefont {V.}~\bibnamefont {Atroshchenko}}, \bibinfo {author} {\bibfnamefont {Z.}~\bibnamefont {Bagdasarian}}, \bibinfo {author} {\bibfnamefont {D.}~\bibnamefont {Basilico}}, \bibinfo {author} {\bibfnamefont {G.}~\bibnamefont {Bellini}}, \bibinfo {author} {\bibfnamefont {J.}~\bibnamefont {Benziger}}, \bibinfo {author} {\bibfnamefont {D.}~\bibnamefont {Bick}}, \bibinfo {author} {\bibfnamefont {G.}~\bibnamefont {Bonfini}}, \bibinfo {author} {\bibfnamefont {D.}~\bibnamefont {Bravo}}, \bibinfo {author} {\bibfnamefont {B.}~\bibnamefont {Caccianiga}}, \bibinfo {author} {\bibfnamefont {F.}~\bibnamefont {Calaprice}}, \bibinfo {author} {\bibfnamefont {A.}~\bibnamefont {Caminata}}, \bibinfo {author} {\bibfnamefont {S.}~\bibnamefont {Caprioli}}, \bibinfo {author} {\bibfnamefont
  {M.}~\bibnamefont {Carlini}}, \bibinfo {author} {\bibfnamefont {P.}~\bibnamefont {Cavalcante}}, \bibinfo {author} {\bibfnamefont {A.}~\bibnamefont {Chepurnov}}, \bibinfo {author} {\bibfnamefont {K.}~\bibnamefont {Choi}}, \bibinfo {author} {\bibfnamefont {L.}~\bibnamefont {Collica}}, \bibinfo {author} {\bibfnamefont {D.}~\bibnamefont {D'Angelo}}, \bibinfo {author} {\bibfnamefont {S.}~\bibnamefont {Davini}}, \bibinfo {author} {\bibfnamefont {A.}~\bibnamefont {Derbin}}, \bibinfo {author} {\bibfnamefont {X.}~\bibnamefont {Ding}}, \bibinfo {author} {\bibfnamefont {A.~D.}\ \bibnamefont {Ludovico}}, \bibinfo {author} {\bibfnamefont {L.~D.}\ \bibnamefont {Noto}}, \bibinfo {author} {\bibfnamefont {I.}~\bibnamefont {Drachnev}}, \bibinfo {author} {\bibfnamefont {K.}~\bibnamefont {Fomenko}}, \bibinfo {author} {\bibfnamefont {A.}~\bibnamefont {Formozov}}, \bibinfo {author} {\bibfnamefont {D.}~\bibnamefont {Franco}}, \bibinfo {author} {\bibfnamefont {F.}~\bibnamefont {Froborg}}, \bibinfo {author} {\bibfnamefont
  {F.}~\bibnamefont {Gabriele}}, \bibinfo {author} {\bibfnamefont {C.}~\bibnamefont {Galbiati}}, \bibinfo {author} {\bibfnamefont {C.}~\bibnamefont {Ghiano}}, \bibinfo {author} {\bibfnamefont {M.}~\bibnamefont {Giammarchi}}, \bibinfo {author} {\bibfnamefont {A.}~\bibnamefont {Goretti}}, \bibinfo {author} {\bibfnamefont {M.}~\bibnamefont {Gromov}}, \bibinfo {author} {\bibfnamefont {D.}~\bibnamefont {Guffanti}}, \bibinfo {author} {\bibfnamefont {C.}~\bibnamefont {Hagner}}, \bibinfo {author} {\bibfnamefont {T.}~\bibnamefont {Houdy}}, \bibinfo {author} {\bibfnamefont {E.}~\bibnamefont {Hungerford}}, \bibinfo {author} {\bibfnamefont {A.}~\bibnamefont {Ianni}}, \bibinfo {author} {\bibfnamefont {A.}~\bibnamefont {Ianni}}, \bibinfo {author} {\bibfnamefont {A.}~\bibnamefont {Jany}}, \bibinfo {author} {\bibfnamefont {D.}~\bibnamefont {Jeschke}}, \bibinfo {author} {\bibfnamefont {V.}~\bibnamefont {Kobychev}}, \bibinfo {author} {\bibfnamefont {D.}~\bibnamefont {Korablev}}, \bibinfo {author} {\bibfnamefont
  {G.}~\bibnamefont {Korga}}, \bibinfo {author} {\bibfnamefont {D.}~\bibnamefont {Kryn}}, \bibinfo {author} {\bibfnamefont {M.}~\bibnamefont {Laubenstein}}, \bibinfo {author} {\bibfnamefont {E.}~\bibnamefont {Litvinovich}}, \bibinfo {author} {\bibfnamefont {F.}~\bibnamefont {Lombardi}}, \bibinfo {author} {\bibfnamefont {P.}~\bibnamefont {Lombardi}}, \bibinfo {author} {\bibfnamefont {L.}~\bibnamefont {Ludhova}}, \bibinfo {author} {\bibfnamefont {G.}~\bibnamefont {Lukyanchenko}}, \bibinfo {author} {\bibfnamefont {L.}~\bibnamefont {Lukyanchenko}}, \bibinfo {author} {\bibfnamefont {I.}~\bibnamefont {Machulin}}, \bibinfo {author} {\bibfnamefont {G.}~\bibnamefont {Manuzio}}, \bibinfo {author} {\bibfnamefont {S.}~\bibnamefont {Marcocci}}, \bibinfo {author} {\bibfnamefont {J.}~\bibnamefont {Martyn}}, \bibinfo {author} {\bibfnamefont {E.}~\bibnamefont {Meroni}}, \bibinfo {author} {\bibfnamefont {M.}~\bibnamefont {Meyer}}, \bibinfo {author} {\bibfnamefont {L.}~\bibnamefont {Miramonti}}, \bibinfo {author} {\bibfnamefont
  {M.}~\bibnamefont {Misiaszek}}, \bibinfo {author} {\bibfnamefont {V.}~\bibnamefont {Muratova}}, \bibinfo {author} {\bibfnamefont {B.}~\bibnamefont {Neumair}}, \bibinfo {author} {\bibfnamefont {L.}~\bibnamefont {Oberauer}}, \bibinfo {author} {\bibfnamefont {B.}~\bibnamefont {Opitz}}, \bibinfo {author} {\bibfnamefont {V.}~\bibnamefont {Orekhov}}, \bibinfo {author} {\bibfnamefont {F.}~\bibnamefont {Ortica}}, \bibinfo {author} {\bibfnamefont {M.}~\bibnamefont {Pallavicini}}, \bibinfo {author} {\bibfnamefont {L.}~\bibnamefont {Papp}}, \bibinfo {author} {\bibfnamefont {O.}~\bibnamefont {Penek}}, \bibinfo {author} {\bibfnamefont {N.}~\bibnamefont {Pilipenko}}, \bibinfo {author} {\bibfnamefont {A.}~\bibnamefont {Pocar}}, \bibinfo {author} {\bibfnamefont {A.}~\bibnamefont {Porcelli}}, \bibinfo {author} {\bibfnamefont {G.}~\bibnamefont {Ranucci}}, \bibinfo {author} {\bibfnamefont {A.}~\bibnamefont {Razeto}}, \bibinfo {author} {\bibfnamefont {A.}~\bibnamefont {Re}}, \bibinfo {author} {\bibfnamefont {M.}~\bibnamefont
  {Redchuk}}, \bibinfo {author} {\bibfnamefont {A.}~\bibnamefont {Romani}}, \bibinfo {author} {\bibfnamefont {R.}~\bibnamefont {Roncin}}, \bibinfo {author} {\bibfnamefont {N.}~\bibnamefont {Rossi}}, \bibinfo {author} {\bibfnamefont {S.}~\bibnamefont {Schoenert}}, \bibinfo {author} {\bibfnamefont {D.}~\bibnamefont {Semenov}}, \bibinfo {author} {\bibfnamefont {M.}~\bibnamefont {Skorokhvatov}}, \bibinfo {author} {\bibfnamefont {O.}~\bibnamefont {Smirnov}}, \bibinfo {author} {\bibfnamefont {A.}~\bibnamefont {Sotnikov}}, \bibinfo {author} {\bibfnamefont {L.}~\bibnamefont {Stokes}}, \bibinfo {author} {\bibfnamefont {Y.}~\bibnamefont {Suvorov}}, \bibinfo {author} {\bibfnamefont {R.}~\bibnamefont {Tartaglia}}, \bibinfo {author} {\bibfnamefont {G.}~\bibnamefont {Testera}}, \bibinfo {author} {\bibfnamefont {J.}~\bibnamefont {Thurn}}, \bibinfo {author} {\bibfnamefont {M.}~\bibnamefont {Toropova}}, \bibinfo {author} {\bibfnamefont {E.}~\bibnamefont {Unzhakov}}, \bibinfo {author} {\bibfnamefont {A.}~\bibnamefont
  {Vishneva}}, \bibinfo {author} {\bibfnamefont {R.}~\bibnamefont {Vogelaar}}, \bibinfo {author} {\bibfnamefont {F.}~\bibnamefont {von Feilitzsch}}, \bibinfo {author} {\bibfnamefont {H.}~\bibnamefont {Wang}}, \bibinfo {author} {\bibfnamefont {S.}~\bibnamefont {Weinz}}, \bibinfo {author} {\bibfnamefont {M.}~\bibnamefont {Wojcik}}, \bibinfo {author} {\bibfnamefont {M.}~\bibnamefont {Wurm}}, \bibinfo {author} {\bibfnamefont {Z.}~\bibnamefont {Yokley}}, \bibinfo {author} {\bibfnamefont {O.}~\bibnamefont {Zaimidoroga}}, \bibinfo {author} {\bibfnamefont {S.}~\bibnamefont {Zavatarelli}}, \bibinfo {author} {\bibfnamefont {K.}~\bibnamefont {Zuber}}, \ and\ \bibinfo {author} {\bibfnamefont {G.}~\bibnamefont {Zuzel}},\ }\href {\doibase 10.1103/PhysRevD.96.091103} {\bibfield  {journal} {\bibinfo  {journal} {Physical Review D}\ }\textbf {\bibinfo {volume} {96}},\ \bibinfo {pages} {091103} (\bibinfo {year} {2017})}\BibitemShut {NoStop}%
\bibitem [{\citenamefont {Lattimer}\ and\ \citenamefont {Cooperstein}(1988)}]{Lattimer1988}%
  \BibitemOpen
  \bibfield  {author} {\bibinfo {author} {\bibfnamefont {J.~M.}\ \bibnamefont {Lattimer}}\ and\ \bibinfo {author} {\bibfnamefont {J.}~\bibnamefont {Cooperstein}},\ }\href {\doibase 10.1103/PhysRevLett.61.23} {\bibfield  {journal} {\bibinfo  {journal} {Physical Review Letters}\ }\textbf {\bibinfo {volume} {61}},\ \bibinfo {pages} {23} (\bibinfo {year} {1988})}\BibitemShut {NoStop}%
\end{thebibliography}%

\end{document}